\documentclass[english,12pt]{article}
\PassOptionsToPackage{natbib=true}{biblatex}

\usepackage[T1]{fontenc}
\usepackage[utf8]{inputenc}
\usepackage{array}
\usepackage{float}
\usepackage{multirow}
\usepackage{amsmath}
\usepackage{amsthm}
\usepackage{amssymb}
\usepackage{graphicx}
\usepackage{rotfloat}

\makeatletter

\providecommand{\tabularnewline}{\\}

\theoremstyle{plain}
\newtheorem{thm}{\protect\theoremname}
\theoremstyle{plain}
\newtheorem{lem}{\protect\lemmaname}

\usepackage[letterpaper,margin=1.0in]{geometry}
\usepackage{threeparttable}
\usepackage{hyperref}
\usepackage{bookmark}
\usepackage{bbm}
\usepackage{enumitem}
\usepackage{multicol}
\newtheorem{assumptionx}{Assumption}
\usepackage{footmisc}
\usepackage[explicit]{titlesec}
\titleformat{\part}{}{}{0pt}{}

\usepackage{setspace}
\makeatother

\usepackage{babel}
\usepackage[style=ext-authoryear-comp,articlein=false,uniquename=false,uniquelist=false,maxcitenames=2]{biblatex}
\providecommand{\lemmaname}{Lemma}
\providecommand{\theoremname}{Theorem}

\begin{document}
\title{Empirical Decomposition of the IV--OLS Gap with Heterogeneous and
Nonlinear Effects}
\author{Shoya Ishimaru\footnote{Hitotsubashi University, Department of Economics (email: shoya.ishimaru@r.hit-u.ac.jp). \newline I am grateful to Mark Colas, Chao Fu, John Kennan, Lance Lochner, Nobuhiko Nakazawa, Masayuki Sawada, Tymon Słoczyński, Yoichi Sugita, Christopher Taber, Matthew Wiswall, and many seminar participants for their comments. I greatly appreciate helpful suggestions from the editor and anonymous referees. All errors are mine.}}
\maketitle
\begin{abstract}
This study proposes an econometric framework to interpret and empirically
decompose the difference between IV and OLS estimates given by a linear
regression model when the true causal effects of the treatment are
nonlinear in treatment levels and heterogeneous across covariates.
I show that the IV--OLS coefficient gap consists of three estimable
components: the difference in weights on the covariates, the difference
in weights on the treatment levels, and the difference in identified
marginal effects that arises from endogeneity bias. Applications of
this framework to return-to-schooling estimates demonstrate the empirical
relevance of this distinction in properly interpreting the IV--OLS
gap.\medskip

\vfill
\end{abstract}

\part[Main Paper]{}

\begin{refsection}
\vspace{-6.5em}

\section{Introduction}

Instrumental variables (IV) regression is the most common approach
for estimating the causal effect of a potentially endogenous regressor.
A standard empirical approach specifies the following linear model:
\begin{equation}
Y=\beta X+W'\gamma+\varepsilon,\label{eq:linear_reg}
\end{equation}
where $Y$ is the outcome of interest, $X$ is a scalar (multivalued)
treatment, and $W$ is a vector of covariates. A standard econometric
textbook takes the linear regression equation (\ref{eq:linear_reg})
as the true causal relationship and interprets the gap between the
IV and ordinary least squares (OLS) coefficient estimates of $X$
as a consequence of endogeneity bias associated with omitted variables,
selection, or measurement error. 

If these interpretations fail to provide a plausible explanation for
the IV--OLS coefficient gap, empirical researchers often consider
the possibility that it instead arises because of how the IV and OLS
coefficients place different weights on different treatment margins
or groups of individuals. This interpretation treats the regression
equation (\ref{eq:linear_reg}) as a linear projection model rather
than a causal model, allowing for the treatment effects to be heterogeneous
and nonlinear in the true causal relationship. For example, \citet{card1995earnings,card1999causal,card2001estimating}
suggests that the positive IV--OLS gaps in many return-to-schooling
studies could be explained by higher returns among credit-constrained
individuals, who are more likely to be affected by cost-related instruments.
In addition, researchers sometimes perform an OLS regression restricted
to a sample of likely complier groups or complying treatment margins
to examine the robustness of the IV--OLS coefficient gap to the weight
difference. As a prominent example, \citet{angrist1991does} compute
an OLS estimate of the return to schooling by restricting their sample
to individuals with 9--12 years of schooling, who are expected to
be most influenced by quarter-of-birth instruments. However, despite
the widely recognized importance of the weight difference interpretation,
relatively few studies attempt to formally quantify how much it actually
matters for the IV--OLS coefficient gap.\footnote{Notable exceptions include \citet{kling2001interpreting}, \citet{lochner2001effect},
\citet{mogstad2010linearity}, \citet{loken2012linear}, and \citet{lochner2015estimating}.}

In this study, I propose an econometric framework to quantify the
sources of the IV--OLS coefficient gap. I begin my analysis by considering
the following causal model:
\begin{equation}
Y=g(X,W)+U,\label{eq:separable}
\end{equation}
with a valid instrument $Z$ that is uncorrelated with an unobservable
$U$ conditional on covariates $W$. With no restriction on the structural
function $g$, the equation allows for the treatment effects to be
heterogeneous across covariates $W$ and nonlinear in treatment levels
$X$ in any manner.\footnote{The separability restriction on the equation, which is relaxed in
Section \ref{subsec:Unobserved-Heterogeneity}, rules out unobserved
heterogeneity in the treatment effects.} I demonstrate that the OLS and IV estimates based on the linear regression
equation (\ref{eq:linear_reg}), when the true model is (\ref{eq:separable}),
represent different weighted averages of the marginal effects of the
treatment they identify. 

Using the weighted-average representations, I then decompose the IV--OLS
coefficient gap into three estimable components. These are: (i) \textbf{the
covariate weight difference}, being the difference in how the IV
and OLS coefficients place weights on the covariates $W$; (ii) \textbf{the
treatment-level weight difference}, as the difference in how they
place weights on treatment levels $X$; and (iii) \textbf{the endogeneity
bias} (or marginal effect difference), being the difference between
the IV- and OLS-identified marginal effects originating from the correlation
between the treatment $X$ and the unobservable $U$. For instance,
in the return-to-schooling context, (i) arises from heterogeneous
returns across observed personal backgrounds with different responses
to the instrument, which motivates the conjecture of \citet{card1995earnings,card1999causal,card2001estimating},
(ii) follows from nonlinear returns across schooling levels with different
sensitivity to the instrument, which motivates the robustness check
of \citet{angrist1991does}, and (iii) corresponds to endogeneity
bias associated with omitted unobserved ability.

To implement the decomposition, I propose a two-step approach to estimate
the ``IV-weighted OLS'' coefficients that serve as the intermediate
points between the IV and OLS coefficients.\footnote{The concept of an IV-weighted OLS coefficient originates in \citet*{mogstad2010linearity},
where they use the model $Y=g(X)+U$ and account for the treatment-level
weight difference.} The first step predicts the conditional mean of the outcome $Y$
given $(X,W$), and the second step uses the prediction to construct
a dependent variable and performs a quasi-IV regression. Depending
on the functional form restriction in the first step, the estimated
OLS coefficients in the second step have the IV weights on the covariates
or on both the covariates and treatment levels. Comparing the estimated
IV-weighted OLS coefficients with the IV and OLS coefficients reveals
the contributions of the weight difference components (i) and (ii)
and the endogeneity bias component (iii). Standard statistical packages
can compute the IV-weighted OLS estimates and perform statistical
inference based on them.\footnote{The Stata package that implements the decomposition, \textit{ivolsdec},
is available from the Boston College Statistical Software Components
(SSC) archive. Type ``ssc install ivolsdec'' in the Stata command
window to install it.}

In an extension, I consider a class of identification strategies that
use an instrument $Z$ deterministic in the covariates $W$, as in
difference-in-differences (DID) and regression discontinuity (RD)
designs.\footnote{Throughout this paper, the term “DID” refers to an identification
strategy that exploits DID variation in the instrument using a two-way
fixed effects regression, in which the treatment and the instrument
can be nonbinary. Examples include \citet{acemoglu2000large}, \citet{duflo2001schooling},
\citet{black2005apple}, and a fuzzy DID setup considered in \citet{de2018fuzzy}.
The term ``RD'' refers to a fuzzy RD design, which is usually implemented
as an IV regression.} I show that the weighted-average interpretation and the decomposition
approach can also be applied to these setups, with only a small modification
of the weight function.

It should be noted that my decomposition framework cannot fully isolate
endogeneity bias in the presence of unobserved heterogeneity in the
treatment effects. In an extension that relaxes the separability restriction
on the equation (\ref{eq:separable}), I demonstrate that the marginal
effect difference component (iii) captures not only endogeneity bias
but also the unobservable-driven discrepancy between the IV-identified
and the average marginal effects.\footnote{This corresponds to a classic impossibility result in the setting
with a binary treatment and no covariates, where endogeneity bias
cannot generally be separated from the difference between the local
and population average treatment effects.} While the weight difference components (i) and (ii) remain informative
about the implications of observed heterogeneity and nonlinearity,
it can be misleading to attribute the marginal effect difference component
(iii) entirely to endogeneity bias.

Using my framework, I examine the return-to-schooling estimates using
several common IV strategies. The first example employs geographic
variation in college costs \citep{cameron2004estimation,carneiro2011estimating}.
The second exploits a discontinuity in the minimum school-leaving
age across cohorts \citep{oreopoulos2006estimating}. The third and
final example uses DID variation in compulsory schooling laws across
cohorts and regions \citep{acemoglu2000large}. In these empirical
examples, the weight difference components are found to be as important
as the endogeneity bias component in explaining the IV--OLS coefficient
gap. The direction or extent of endogeneity bias implied by the estimated
IV--OLS gap differs entirely by taking into consideration how the
two coefficients place weights on the different observed personal
backgrounds and schooling margins.

\subsection*{Related Literature and Roadmap}

This paper advances the literature on the interpretation and decomposition
of linear regression coefficients by exploring a general and empirically
relevant setting in which the treatment effects are nonlinear in treatment
levels and heterogeneous across covariates. The local average treatment
effect (LATE) interpretation proposed by \citet{imbens1994identification}
in a binary treatment context originates the idea that the IV coefficient
is a weighted average of the marginal causal effects of the treatment.
\citet{angrist1995two} and \citet{angrist2000interpretation} extend
the basic insights of the LATE interpretation to a multivalued treatment
case. \citet{yitzhaki1996using} and \citet{angrist1999empirical}
suggest the analogous weighted-average interpretation of the OLS coefficient.\footnote{The OLS interpretation is also explored by \citet{angrist1998estimating},
\citet{aronow2016does}, and \citet{Sloczynski2020OLS} in a binary
treatment case with a focus on observed heterogeneity.} Much of the focus of the literature has been on a univariate model
with no or fixed covariates, which makes it difficult to immediately
apply these results to empirical settings. While drawing on these
existing results, my framework synthesizes them into an empirically
relevant format and provides an estimable decomposition of the IV--OLS
coefficient gap.

Motivated by the weighted-average interpretation developed in the
literature, \citet{mogstad2010linearity}, \citet*{loken2012linear},
and \citet*{lochner2015estimating} propose the empirical decomposition
of the IV--OLS coefficient gap into weight difference and endogeneity
bias components. They consider a model in which the treatment effects
are nonlinear in treatment levels but homogeneous across covariates.
My framework generalizes these previous works by allowing the treatment
effects to be heterogeneous across covariates. This is an empirically
meaningful generalization, as my empirical applications demonstrate
the relevance of the covariate weight difference in interpreting the
IV--OLS coefficient gap.

While the linear IV regression is often advocated for its transparency
\citep{angrist2010credibility}, it is also criticized for its lack
of a clear connection to an economic parameter of interest \citep{heckman2010comparing}.\footnote{For the linear OLS regression, \citet{Sloczynski2020OLS} shows that
the OLS coefficient of a binary treatment with covariates generally
does not represent the average treatment effect (ATE), the average
treatment effect on treated (ATT), or untreated (ATU).} A related strand of the literature based on the latter view aims
to develop alternatives to the linear IV regression, including the
policy-relevant treatment effect proposed by \citet{heckman2001policy}
and \citet{carneiro2010evaluating}. Nevertheless, many empirical
researchers use the linear regression for its simplicity. My framework
aims to provide useful diagnostics for empirical researchers who use
the linear regression while recognizing its potential limitations.

The rest of the paper proceeds as follows. Section 2 presents the
interpretation of the linear IV and OLS coefficients and proposes
the decomposition of the IV--OLS coefficient gap. Section 3 extends
these econometric results by exploring settings with alternative assumptions.
Section 4 proposes the estimators for the IV-weighted OLS coefficients
for empirically performing the decomposition. Section 5 presents the
results from the empirical applications, and Section 6 concludes.
Online Appendix presents the proofs of the theorems, explores additional
econometric results, and provides details for the empirical applications.

\section{Econometric Framework\label{sec:Econometric-Framework-for}}

\subsection{Setup and Assumptions}

I consider a random draw of $(Y,X,W,Z)$, where $Y$ is a scalar outcome
variable, $X$ is a scalar treatment variable, $W$ is a vector of
covariates, and $Z$ is a scalar instrument. Multi-instrument two-stage
least squares can fit into this setup by regarding the projection
of the treatment $X$ onto the instrument vector and covariates $W$
(i.e., the first-stage predicted value) as a synthetic scalar instrument.\footnote{\label{fn:multIV}If a regression of $X$ on a vector instrument $(Z_{1},\ldots,Z_{M})$
controlling for $W$ yields a first-stage coefficient $(\pi_{1},\ldots,\pi_{M})$,
I regard $Z=\sum_{m=1}^{M}\pi_{m}Z_{m}$ as a synthetic scalar instrument.
As my setting allows for negative IV weights, a partial monotonicity
condition for each individual instrument $Z_{m}$ as in \citet{mogstad2019causal}
is not required as long as the treatment $X$ and the synthetic instrument
$Z$ are correlated conditional on $W$.} 

Throughout this paper, I assume the existence of the first and second
moments of any random variable. I use $L_{w}(R)=w'E(WW')^{-1}E(WR)$
to denote a linear projection of a random variable $R$ onto $W$
evaluated at $W=w$ (i.e., a predicted value from a linear regression
of $R$ on $W$).\footnote{The covariate vector $W$ includes unity as one of its elements.}
I define $\widetilde{R}=R-L_{W}(R)$ to be a residual from the linear
projection. Let $m(x,w)=E[Y|X=x,W=w]$ be the conditional mean function
of $Y$ given $(X,W)$.
\global\long\def\ind{\mathbbm{1}}%

To assess the impact of the treatment $X$ on the outcome $Y$, a
standard approach specifies a linear regression model:
\begin{equation}
Y=\beta X+W'\gamma+\varepsilon,\,\,E(\varepsilon W)=0.\label{eq:linear}
\end{equation}
Additional moment conditions $E(\varepsilon Z)=0$ and $E(\varepsilon X)=0$,
respectively pin down the linear IV coefficient $\beta_{IV}$ and
OLS coefficient $\beta_{OLS}$ as
\begin{eqnarray}
\beta_{IV} & = & {\displaystyle {\displaystyle E(\widetilde{Y}\widetilde{Z})/E(\widetilde{X}\widetilde{Z})}},\label{eq:IV}\\
\beta_{OLS} & = & {\displaystyle E(\widetilde{Y}\widetilde{X})/E(\widetilde{X}^{2})}.\label{eq:OLS}
\end{eqnarray}
Note that I treat the equation (\ref{eq:linear}) merely as a statistical
model to characterize $\beta_{IV}$ and $\beta_{OLS}$, which does
not impose any assumption on the underlying causal relationship.

To consider the causal interpretation of the IV and OLS coefficients,
I define $Y(x)$ to be the potential outcome associated with the treatment
level $x$, which produces the observed outcome as $Y=Y(X)$. The
derivative $Y'(x)$ is the marginal effect of the treatment in a causal
sense. I make the following assumptions.

\renewcommand{\theassumptionx}{S}
\begin{assumptionx}

\label{Ass:sep}\textup{(Separability)} The potential outcome is
given by $Y(x)=g(x,W)+U$, $E(U|W)=0$.

\end{assumptionx}

\renewcommand{\theassumptionx}{C}
\begin{assumptionx}

\label{Ass:con}\textup{(Continuous Treatment)} The treatment $X$
is continuously distributed on support $(\underline{x},\overline{x})$,
with $-\infty\le\underline{x}<\overline{x}\le\infty$.

\end{assumptionx}

\renewcommand{\theassumptionx}{D}
\begin{assumptionx}

\label{Ass:deriv}\textup{(Regularity Conditions on Derivatives)}
Let $V_{a}^{b}(f,w)=\int_{\min\{a,b\}}^{\max\{a,b\}}|\frac{\partial}{\partial x}f(x,w)|dx$
be the total variation of a function $f(x,w)$ differentiable in $x$
between points $a$ and $b$.

\textup{(i) } $g(x,w)$ is differentiable in $x$ and $E\left[V_{x_{0}}^{X}(g,W)^{2}\right]<\infty$
for some $x_{0}\in(\underline{x},\overline{x})$;

\textup{(ii)} $m(x,w)$ is differentiable in $x$ and $E\left[V_{x_{0}}^{X}(m,W)^{2}\right]<\infty$
for some $x_{0}\in(\underline{x},\overline{x})$.

\end{assumptionx}

\renewcommand{\theassumptionx}{IV}
\begin{assumptionx}

\label{Ass:iv} The instrument $Z$ satisfies the conditions:

\textup{(i) (Exogeneity)} $E(U\widetilde{Z})=0$; \textup{(ii) (Relevance)}
$E(\widetilde{X}\widetilde{Z})\ne0$.

\end{assumptionx}

\renewcommand{\theassumptionx}{OLS}
\begin{assumptionx}

\label{Ass:ols} The treatment residual has positive variance, i.e.,
$E(\widetilde{X}^{2})>0$.

\end{assumptionx}

\renewcommand{\theassumptionx}{L}
\begin{assumptionx}

\label{Ass:linear}\textup{(Linearity of Conditional Means)} 

\textup{(i)} $E(Z|W)$ is linear in $W$; \textup{(ii)} $E(X|W)$
is linear in $W$.

\end{assumptionx}

Given Assumption \ref{Ass:sep}, the marginal causal effect of the
treatment is $Y'(x)=\frac{\partial}{\partial x}g(x,W)$, which can
be nonlinear in treatment levels $x$ and heterogeneous across covariates
$W$. However, it rules out unobserved heterogeneity in the effect.\footnote{The literature on the nonparametric IV approach typically makes a
similar separability assumption. See, for example, \citet{newey2003instrumental},
\citet{blundell2007semi}, and \citet{horowitz2011applied}.} Section \ref{subsec:Unobserved-Heterogeneity} relaxes this assumption. 

I make Assumption \ref{Ass:con} to focus on a continuous treatment
case, which is merely for expositional convenience.\footnote{As indicated in the assumption, the support $(\underline{x},\overline{x})$
can be unbounded. Any integral expression of $x$ in this paper is
taken over the support $(\underline{x},\overline{x})$, which is kept
implicit to simplify the exposition.} All econometric results can be applied to a discrete treatment case
by extending $g(x,w)$ to nonsupport points of the treatment.\footnote{For example, define $g(x,w)$ at nonsupport points by a linear interpolation
without loss of generality. Then, the derivatives and integrals of
$x$ in all econometric results can be replaced by the differences
and summations of $x$.} Assumption \ref{Ass:deriv} concerns the derivatives of the structural
function $g$ and the conditional mean function $m$. Assumption
\ref{Ass:iv} is a set of standard IV assumptions that require the
instrument to be exogenous and relevant after controlling for covariates.
Assumption \ref{Ass:ols} is a standard OLS assumption.

Assumption \ref{Ass:linear} is required for the exact weighted-average
interpretation of the regression coefficients. A similar linearity
assumption appears in, for example, \citet{angrist1999empirical},
\citet{lochner2015estimating}, and \citet{sloczynkski2020when}.
This assumption allows the analysis to abstract away from any omitted
variable bias associated with unaccounted nonlinear effects of covariates,
which is not as fundamental as endogeneity bias associated with unobservables.
While this assumption mechanically holds in a saturated model in which
a covariate vector $W$ consists of indicators for disjoint groups,
it can be restrictive with continuous covariates. An empirical researcher
may then want to choose elements of the vector $W$ based on a series
approximation and thereby flexibly account for the nonlinear and interaction
effects of the underlying observables.\footnote{Appendix \ref{subsec:Nonlinear} relaxes Assumption \ref{Ass:linear}
to explore what happens when a good linear approximation is not feasible
because of data limitations.} Assumption \ref{Ass:linear} does not hold by construction for identification
strategies based on DID or RD designs. Section \ref{subsec:DID_RD}
considers these cases.

\subsection{Weighted-Average Interpretation\label{subsec:Interpretation}}

I start by presenting a key theorem for interpreting the IV and OLS
coefficients. The OLS part of the theorem is shown by \citet{angrist1999empirical}
in a discrete treatment setting. I present it for completeness and
for reinterpretation in my setting.
\begin{thm}
\label{Thm:Interpret}The IV and OLS coefficients have a weighted-average
interpretation as below.
\begin{enumerate}
\item[(i)] With Assumptions \ref{Ass:sep}, \ref{Ass:con}, \ref{Ass:deriv}--(i),
\ref{Ass:iv}, and \ref{Ass:linear}--(i),
\begin{align*}
\beta_{IV} & =\int\int\frac{\partial}{\partial x}g(x,w)\omega_{Z}(x,w)dF_{W}(w)dx;
\end{align*}
\item[(ii)] \citep{angrist1999empirical} With Assumptions \ref{Ass:con}, \ref{Ass:deriv}--(ii),
\ref{Ass:ols}, and \ref{Ass:linear}--(ii),
\begin{align*}
\beta_{OLS} & =\int\int\frac{\partial}{\partial x}m(x,w)\omega_{X}(x,w)dF_{W}(w)dx.
\end{align*}
\end{enumerate}
The weight function is given by $\omega_{R}(x,w)=E[\widetilde{\ind}_{X\ge x}\widetilde{R}|W=w]/E(\widetilde{X}\widetilde{R})$
for $R=Z,X$, which satisfies $\int\int\omega_{R}(x,w)dF_{W}(w)dx=1$.
\end{thm}
Theorem \ref{Thm:Interpret} implies that the IV and OLS coefficients
are expressed as weighted averages of the marginal effects they identify.
While the IV coefficient identifies a weighted average of the causal
effects $\frac{\partial}{\partial x}g(x,w)$, the OLS coefficient
identifies a weighted average of the slopes of the conditional mean
function $\frac{\partial}{\partial x}m(x,w)$.

The relationship between the OLS-identified and IV-identified marginal
effects, $\frac{\partial}{\partial x}m(x,w)$ and $\frac{\partial}{\partial x}g(x,w)$,
can be expressed as
\begin{equation}
\frac{\partial}{\partial x}m(x,w)=\frac{\partial}{\partial x}g(x,w)+\frac{\partial}{\partial x}E[U|X=x,W=w].\label{eq:m-g}
\end{equation}
Therefore, the difference between the two marginal effects arises
from endogeneity bias, i.e., the correlation between the treatment
$X$ and the unobservable $U$. 

While endogeneity bias makes the IV and OLS coefficients differ, an
important implication from Theorem \ref{Thm:Interpret} is that the
difference in the weight functions, $\omega_{Z}$ and $\omega_{X}$,
also gives rise to the IV--OLS coefficient gap. To explore what makes
the IV weight $\omega_{Z}$ and the OLS weight $\omega_{X}$ differ,
let $\overline{\omega}_{R}(w)=\int\omega_{R}(x,w)dx$ be the marginal
weight on covariates $W=w$ for $R=Z,X$. The marginal IV and OLS
weights on $W=w$ are given by:
\begin{align}
\overline{\omega}_{Z}(w) & =Cov(X,Z|W=w)/E(\widetilde{X}\widetilde{Z}),\label{eq:IVweight_w}\\
\overline{\omega}_{X}(w) & =Var(X|W=w)/E(\widetilde{X}^{2}).\label{eq:OLSweight_w}
\end{align}
The IV weight $\overline{\omega}_{Z}(w)$ is proportional to the conditional
covariance $Cov(X,Z|W=w)$, which is the product of the regression
coefficient of $X$ on $Z$ given $W=w$ and the conditional variance
$Var(Z|W=w)$. This means that covariates $W$ with greater sensitivity
of the treatment to the instrument or larger variation in the instrument
are weighted more. In contrast, the OLS weight $\overline{\omega}_{X}(w)$
is proportional to the conditional variance $Var(X|W=w)$. This implies
that covariates $W$ with larger variation in the treatment are weighted
more.

Similarly, let $\overline{\omega}_{R}(x)=\int\omega_{R}(x,w)dF_{W}(w)$
be the marginal weight on the treatment level $X=x$ for $R=Z,X$.
The marginal IV and OLS weights on $X=x$ are given by:
\begin{align}
\overline{\omega}_{Z}(x) & =E(\widetilde{\ind}_{X\ge x}\widetilde{Z})/E(\widetilde{X}\widetilde{Z}),\label{eq:IVweight_x}\\
\overline{\omega}_{X}(x) & =E(\widetilde{\ind}_{X\ge x}\widetilde{X})/E(\widetilde{X}^{2}).
\end{align}
The IV weight $\overline{\omega}_{Z}(x)$ is proportional to a regression
coefficient of $\ind_{X\ge x}$ on the instrument $Z$, controlling
for $W$. This implies that the more the instrument $Z$ influences
the treatment $X$ at $x$, the more weighted the treatment level
$x$ is. For example, if a compulsory schooling instrument $Z$ increases
years of schooling $X$ through primary and secondary education but
does not influence college education, the IV weight is expected to
be positive with $x\le12$ and zero with $x>12$.\footnote{If $X$ is discrete, the weight on $x$ represents a change in treatment
levels between $x-1$ and $x$.} In general, the IV weight $\overline{\omega}_{Z}(x)$ is not guaranteed
to be positive because the instrument can have positive effects on
the treatment at some margins while having negative effects at others.

Interpreting the OLS weight expression is less straightforward, but
Appendix \ref{subsec:Derivation-of-the} shows
\begin{equation}
\overline{\omega}_{X}(x)\propto\int\int\int(x_{1}-x_{2})\ind_{x_{2}<x\le x_{1}}dF_{X|W}(x_{1}|w)dF_{X|W}(x_{2}|w)dF_{W}(w).\label{eq:OLSweightx}
\end{equation}
This implies that the OLS weight is proportional to the sum of differences
between the pairs of conditionally independent observations $X_{1},X_{2}\overset{i.i.d.}{\sim}F_{X|W}(\cdot|w)$
with $X_{1}\ge x>X_{2}$. Therefore, the treatment level $x$ is weighted
more if the treatment $X$ is densely distributed both above and below
$x$.\footnote{\citet{yitzhaki1996using} derives the OLS weight function in a simpler
case with no covariates and shows that the weight function is $\overline{\omega}_{X}(x)\propto(b-x)(x-a)$
if $X$ is uniformly distributed over $[a,b${]}. The weights are
zero at both ends of the support despite flat density because no pair
of observations can sandwich the endpoints.} As is evident from (\ref{eq:OLSweightx}), the OLS weight $\overline{\omega}_{X}(x)$
on each treatment level $x$ is nonnegative.

Theorem \ref{Thm:Interpret} can be considered as a generalization
of \citet{lochner2015estimating}, allowing for heterogeneity in the
marginal effects across covariates. In fact, restricting $g(x,w)$
to be additively separable in $x$ and $w$ yields a weighted-average
expression comparable to theirs. If $g(x,w)$ is additively separable,
$\frac{\partial}{\partial x}g(x,w)$ depends only on $x$. This yields
$\beta_{IV}=\int\frac{\partial}{\partial x}g(x,\cdot)\overline{\omega}_{Z}(x)dx$,
which matches the weighted-average expression provided by Proposition
1 in \citet{lochner2015estimating}.

\subsection{Related Work}

Theorem \ref{Thm:Interpret} closely relates to many existing results
in the literature. To provide further intuition for the weighted-average
interpretation, I explore the relationship between my results and
those in the literature. 

Denote the IV and OLS coefficients conditional on $W=w$ as
\begin{eqnarray}
b_{IV}(w) & = & Cov(Y,Z|W=w)/Cov(X,Z|W=w),\label{eq:IV_w}\\
b_{OLS}(w) & = & Cov(Y,X|W=w)/Var(X|W=w).
\end{eqnarray}
The following result from \citet{lochner2015estimating} shows that
the IV and OLS coefficients, $\beta_{IV}$ and $\beta_{OLS}$, can
be viewed as weighted averages of the covariate-specific coefficients,
$b_{IV}(w)$ and $b_{OLS}(w)$.
\begin{thm}
\label{Thm:avgw}\citep{lochner2015estimating}
\begin{enumerate}
\item[(i)] With Assumptions \ref{Ass:iv}--(ii) and \ref{Ass:linear}--(i),
\begin{align*}
\beta_{IV} & =\int b_{IV}(w)\overline{\omega}_{Z}(w)dF_{W}(w);
\end{align*}
\item[(ii)] With Assumptions \ref{Ass:ols} and \ref{Ass:linear}--(ii),
\begin{align*}
\beta_{OLS} & =\int b_{OLS}(w)\overline{\omega}_{X}(w)dF_{W}(w).
\end{align*}
\end{enumerate}
\end{thm}
Note that this theorem does not rely on assumptions about the causal
structure (Assumptions \ref{Ass:sep} and \ref{Ass:iv}--(i)) because
$b_{IV}(w)$ is defined in (\ref{eq:IV_w}) merely as a ratio of two
conditional covariances. The IV and OLS weights on covariates $W=w$
match the marginal weights $\overline{\omega}_{Z}(w)$ and $\overline{\omega}_{X}(w)$
defined in (\ref{eq:IVweight_w}) and (\ref{eq:OLSweight_w}). This
weighted-average interpretation is explored intensively in a binary
treatment case, in which the treatment effect is linear in treatment
levels $x\in\{0,1\}$ by construction and it is possible to focus
only on heterogeneity \citep{angrist1998estimating,aronow2016does,Sloczynski2020OLS,sloczynkski2020when}.

The weighted-average interpretation of the covariate-specific coefficients,
$b_{IV}(w)$ and $b_{OLS}(w)$, can be derived by applying Theorem
\ref{Thm:Interpret} conditional on $W=w$. This result originates
in \citet{yitzhaki1996using} and \citet{schechtman2004gini}.
\begin{thm}
\label{Thm:avgx_w}Let $\omega_{R}(x|w)=\omega_{R}(x,w)/\overline{\omega}_{R}(w)$
be the conditional weight on the treatment level $x$ given $W=w$.
\begin{enumerate}
\item[(i)]  \citep{schechtman2004gini} With Assumptions \ref{Ass:sep}, \ref{Ass:con},
\ref{Ass:deriv}--(i), $Cov(U,Z|W=w)=0$, and $Cov(X,Z|W=w)\ne0$,
\begin{align*}
b_{IV}(w) & =\int\frac{\partial}{\partial x}g(x,w)\omega_{Z}(x|w)dx;
\end{align*}
\item[(ii)]  \citep{yitzhaki1996using} With Assumptions \ref{Ass:con}, \ref{Ass:deriv}--(ii),
and $Var(X|W=w)>0$,
\begin{align*}
b_{OLS}(w) & =\int\frac{\partial}{\partial x}m(x,w)\omega_{X}(x|w)dx.
\end{align*}
\end{enumerate}
\end{thm}
The weighted-average expression for the covariate-specific IV coefficient
$b_{IV}(w)$ could also be derived as a special case of the results
from \citet{angrist1995two}, \citet{angrist2000interpretation},
and \citet{heckman2006understanding}, which consider a more general
setting that allows for unobserved heterogeneity in the treatment
effects.\footnote{\citet{angrist1995two} allow for covariates in a special case with
a ``saturated'' first stage, where $Z=E[X|Z_{1},\ldots,Z_{M},W]$
is generated from an instrument vector $(Z_{1},\ldots,Z_{M})$ and
a covariate vector $W$ that both consist of indicators for disjoint
groups.}

Synthesizing these existing results, Theorem \ref{Thm:Interpret}
can be divided into two components: the linear IV and OLS coefficients,
$\beta_{IV}$ and $\beta_{OLS}$, are weighted averages of the covariate-specific
coefficients, $b_{IV}(w)$ and $b_{OLS}(w)$, with different weights
on covariates (Theorem \ref{Thm:avgw}); and the covariate-specific
coefficients are weighted averages of the identified marginal effects,
$\frac{\partial}{\partial x}g(x,w)$ and $\frac{\partial}{\partial x}m(x,w)$,
with different weights on treatment levels (Theorem \ref{Thm:avgx_w}).

\subsection{Decomposing the IV--OLS Coefficient Gap\label{subsec:Decomposing-the-IV=002013OLS}}

The weighted-average interpretation of the IV and OLS coefficients
in Theorems \ref{Thm:Interpret}--\ref{Thm:avgx_w} motivates the
decomposition of the IV--OLS coefficient gap $\beta_{IV}-\beta_{OLS}$.
I decompose the gap into the following three components.
\begin{eqnarray}
\Delta_{CW} & = & \int b_{OLS}(w)\left(\overline{\omega}_{Z}(w)-\overline{\omega}_{X}(w)\right)dF_{W}(w),\label{eq:decom1}\\
\Delta_{TW} & = & \int\int\frac{\partial}{\partial x}m(x,w)\left(\omega_{Z}(x|w)-\omega_{X}(x|w)\right)\overline{\omega}_{Z}(w)dF_{W}(w)dx,\label{eq:decom2}\\
\Delta_{ME} & = & \int\int\left(\frac{\partial}{\partial x}g(x,w)-\frac{\partial}{\partial x}m(x,w)\right)\omega_{Z}(x,w)dF_{W}(w)dx.\label{eq:decom3}
\end{eqnarray}
The first component, $\Delta_{CW}$, which I call ``the covariate
weight difference,'' corresponds to how differently the IV and OLS
coefficients place weights on the covariates. The second component,
$\Delta_{TW}$, which I call ``the treatment-level weight difference,''
captures how differently the IV and OLS coefficients place weights
on the treatment levels, conditional on the covariates. The third
component, $\Delta_{ME}$, which I refer to as ``the endogeneity
bias'' or ``the marginal effect difference,'' captures the difference
between the IV- and OLS-identified marginal effects, which arises
from the endogeneity of the treatment as in (\ref{eq:m-g}). The sum
of the three components is the IV--OLS gap, i.e., $\beta_{IV}-\beta_{OLS}=\Delta_{CW}+\Delta_{TW}+\Delta_{ME}$.

This decomposition departs from the OLS coefficient $\beta_{OLS}$
and arrives at the IV coefficient $\beta_{IV}$ by first changing
the weights and then the marginal effects. This order follows an idea
of the ``IV-weighted OLS'' approach adopted by \citet{mogstad2010linearity}
and \citet{lochner2015estimating}. The advantage of this approach
is that the decomposition is always feasible. The decomposition requires
knowledge of the following IV-weighted OLS coefficients as intermediate
points:
\begin{eqnarray}
\beta_{C} & = & \int b_{OLS}(w)\overline{\omega}_{Z}(w)dF_{W}(w),\label{eq:beta_c}\\
\beta_{CT} & = & \int\int\frac{\partial}{\partial x}m(x,w)\omega_{Z}(x,w)dF_{W}(w)dx.\label{eq:beta_ct}
\end{eqnarray}
The first coefficient, $\beta_{C}$, is the OLS coefficient with the
IV weight on covariates, while the second, $\beta_{CT}$, is the OLS
coefficient with the IV weight on both covariates and treatment levels.
By construction, $\Delta_{CW}=\beta_{C}-\beta_{OLS}$, $\Delta_{TW}=\beta_{CT}-\beta_{C}$,
and $\Delta_{ME}=\beta_{IV}-\beta_{CT}$. These coefficients are always
well-defined because $\omega_{Z}(x,w)\ne0$ implies $\omega_{X}(x,w)>0$.

This is not a unique order in which the IV--OLS gap can be decomposed,
as is the case with Blinder--Oaxaca type decomposition methods. For
example, one can instead account for the marginal effect difference
first, then the weight differences. However, this alternative order
requires knowledge of the ``OLS-weighted IV'' coefficient, i.e.,
$\int\int\frac{\partial}{\partial x}g(x,w)\omega_{X}(x,w)dF_{W}(x)dx$,
as an intermediate point. This coefficient is not always identified
because the instrument may have no variation or no impact on the treatment
at some $(x,w)$, even with $\omega_{X}(x,w)>0$.\footnote{\citet{loken2012linear} propose a mixture of the IV-weighted OLS
and the OLS-weighted IV to make decomposition results independent
of whether the marginal effect or the weight difference is accounted
for first. This approach has the same identification issue as the
OLS-weighted IV approach.} This makes the OLS-weighted IV approach less practical despite its
potential theoretical appeal.\footnote{In all empirical examples in Section \ref{sec:Applications}, the
OLS-weighted IV coefficient cannot be identified because the instrument
influences only a subset of treatment margins.}

\section{Extensions\label{sec:more_general}}

I extend my econometric framework in several directions. Section \ref{subsec:DID_RD}
considers identification strategies based on DID and RD designs, in
which the instrument $Z$ is deterministic in the covariates $W$.
Section \ref{subsec:Unobserved-Heterogeneity} relaxes Assumption
\ref{Ass:sep} and allows for unobserved heterogeneity in the treatment
effects. Appendix \ref{sec:Additional-Extensions} explores additional
extensions that consider a setting without Assumption \ref{Ass:linear},
a setting with an invalid instrument, and a setting with DID or RD
designs in the presence of unobserved heterogeneity.

\subsection{Identification Based on DID or RD Designs\label{subsec:DID_RD}}

Assumption \ref{Ass:linear} does not hold by construction for two
important identification strategies: DID and RD designs. I explore
the weighted-average interpretation in these cases with an alternative
set of assumptions.

With a DID-based identification strategy, each observation belongs
to a particular group $g\in\{1,\ldots,G\}$ and period $t\in\{1,\ldots,T\}$,
and the instrument $Z$ is constant within each $(g,t)$. The regression
equation (\ref{eq:linear}) can be written as
\[
Y=\beta X+\sum_{g=1}^{G}\gamma_{g}d_{g}+\sum_{t=1}^{T-1}\delta_{t}D_{t}+\varepsilon,
\]
where $d_{g}$ indicates membership to a group $g$ and $D_{t}$ indicates
membership to a period $t$. By construction, the instrument $Z$
is a deterministic and nonlinear function of the covariate vector
$W=(d_{1},\ldots,d_{G},D_{1},\ldots,D_{T-1})$. This does not satisfy
the requirement by Assumption \ref{Ass:linear} that $E(Z|W)$ should
be linear in $W$. An empirical example from \citet{acemoglu2000large}
in Section \ref{subsec:CSL} fits into this setting.

With an RD-based identification strategy using a running variable
$C$ with a cutoff $c$, the instrument is $Z=\ind_{C\ge c}$. The
regression equation (\ref{eq:linear}) can be written as
\[
Y=\beta X+\sum_{k=1}^{K}\gamma_{k}p_{k}\left(C\right)+\varepsilon,
\]
where $(p_{1},\ldots,p_{K})$ is a set of basis functions with $p_{k}(c)=0$.\footnote{Given that an RD with local polynomials can be interpreted as a kernel-weighted
version of an RD with global polynomials, my description focuses on
a global polynomial case. The IV weight function should be multiplied
by a kernel weight in a local polynomial case.} The instrument $Z$ is a deterministic and nonlinear function of
the covariate vector $W=\left(p_{1}(C),\ldots,p_{K}(C)\right)$. An
empirical example from \citet{oreopoulos2006estimating} in Section
\ref{subsec:CSL_UK} fits into this setting.

The definition of the linear IV and OLS coefficients follows (\ref{eq:IV})
and (\ref{eq:OLS}). I rule out a ``sharp'' DID or RD setup with
$X=Z$, in which the IV and OLS coefficients are identical by construction.
Instead of Assumption \ref{Ass:linear}, I make the following assumption
to represent DID- and RD-based identification strategies.

\renewcommand{\theassumptionx}{LS}
\begin{assumptionx}

\label{Ass:DID_RD} \textup{(Linear Structural Function)} $g(x,w)$
is linear in $w$ for any $x\in(\underline{x},\overline{x})$.

\end{assumptionx}

In a DID, this assumption corresponds to a parallel trend assumption,
indicating that the group membership $d_{g}$ and the time membership
$D_{t}$ additively affect a potential outcome. In an RD, this assumption
implies that the relationship between a potential outcome and a running
variable $C$ is well approximated by a linear combination of the
basis functions $\left(p_{1}(C),\ldots,p_{K}(C)\right)$ around $C=c$,
which also implies continuity at $C=c$. In these settings, Assumption
\ref{Ass:iv}--(i) follows from $Var(Z|W)=0$, as $E(U\widetilde{Z})=E(E(U|W)\widetilde{Z})=0$.

With Assumption \ref{Ass:linear} replaced by Assumption \ref{Ass:DID_RD},
the weighted-average interpretation of the IV coefficient remains
valid with a modified weight function expression. 
\begin{thm}
\label{Thm:DID_RD} With Assumptions \ref{Ass:sep}, \ref{Ass:con},
\ref{Ass:deriv}--(i), \ref{Ass:iv}, and \ref{Ass:DID_RD}, the
IV coefficient $\beta_{IV}$ is given by
\begin{align*}
\beta_{IV} & =\int\int\frac{\partial}{\partial x}g(x,w)\omega_{Z}^{*}(x,w)dF_{W}(w)dx,
\end{align*}
where $\omega_{Z}^{*}(x,w)=L_{w}(\ind_{X\ge x}\widetilde{Z})/E(\widetilde{X}\widetilde{Z})$
satisfies $\int\int\omega_{Z}^{*}(x,w)dF_{W}(w)dx=1$.
\end{thm}
Note that there are infinitely many weight functions other than $\omega_{Z}^{*}$
that can make this weighted-average expression. In particular, $\omega_{Z}^{*}(x,w)+h(x,w)-L_{w}\left(h(x,W)\right)$
can also be a weight function, where $h(x,w)$ is any nonlinear function.
This property is a mere artifact of Assumption \ref{Ass:DID_RD},
which makes the marginal effect $\frac{\partial}{\partial x}g(x,w)$
linear in $w$ and orthogonal to any linear projection residual. Therefore,
it is most reasonable to use the linearly projected function $\omega_{Z}^{*}$
and omit the redundant variation in weights.\footnote{\citet{Chaisemartin2020} suggest the weight function expression for
a DID with a binary treatment, which is an important special case
of Theorem \ref{Thm:DID_RD}. Appendix \ref{subsec:Comparing-the-Weight}
discusses the relationship between Theorem \ref{Thm:DID_RD} and their
results.}

Given the weighted-average interpretation provided by Theorem \ref{Thm:DID_RD},
it is possible to define the weight difference and the marginal effect
difference components comparably to (\ref{eq:decom1}--\ref{eq:decom3}),
using $\omega_{Z}^{*}$ instead of $\omega_{Z}$ as the IV weight
function.\footnote{It still requires Assumption \ref{Ass:linear}--(ii) for the OLS
coefficient to have an exact weighted-average interpretation. In this
situation, a researcher may want to choose a more flexible specification
in performing the OLS regression, e.g., controlling for group--time
effects instead of additive group and time effects.} Note that the marginal weights on covariate $W=w$ and treatment
level $X=x$ are given by
\begin{align*}
\overline{\omega}_{Z}^{*}(w) & =\int\omega_{Z}^{*}(x,w)dx=L_{w}(X\widetilde{Z})/E(X\widetilde{Z}),\\
\overline{\omega}_{Z}^{*}(x) & =\int\omega_{Z}^{*}(x,w)dF_{W}(w)=E(\ind_{X\ge x}\widetilde{Z})/E(X\widetilde{Z}).
\end{align*}
While the treatment-level weight $\overline{\omega}_{Z}^{*}(x)$ is
identical to $\overline{\omega}_{Z}(x)$ defined in (\ref{eq:IVweight_x}),
the covariate weight $\overline{\omega}_{Z}^{*}(w)$ has a different
expression from $\overline{\omega}_{Z}(w)$ defined in (\ref{eq:IVweight_w}). 

\subsection{Unobserved Heterogeneity\label{subsec:Unobserved-Heterogeneity}}

Assumption \ref{Ass:sep} rules out unobserved heterogeneity in the
marginal effects of the treatment. This is a common but strong assumption.
Because the distinction between observables and unobservables arises
merely from data availability, it is natural to consider heterogeneity
in both observed and unobserved dimensions in the econometric model.
Allowing for unobserved heterogeneity in the marginal effects $Y'(x)$,
I define the average marginal effect (AME) as $\tau(x,w)=E\left[Y'(x)|W=w\right]$.
With Assumption \ref{Ass:sep}, the AME reduces to $\tau(x,w)=\frac{\partial}{\partial x}g(x,w)$.

Removing Assumptions \ref{Ass:sep}, \ref{Ass:deriv}--(i), and \ref{Ass:iv}--(i),
I make the following set of assumptions about the potential outcome
process $Y(x)$.

\renewcommand{\theassumptionx}{P}
\begin{assumptionx}

\label{Ass:pot}The potential outcome process $Y(x)$ satisfies the
following conditions:
\begin{enumerate}
\item[(i)] \textup{(Conditional Independence)} $Cov\left(Y(x),Z|W\right)=0$
for any $x\in(\underline{x},\overline{x})$;
\item[(ii)] \textup{(Differentiability)} $Y'(x)$ exists for any $x\in(\underline{x},\overline{x})$
and the total variation $V_{x_{0}}^{X}\left(Y(\cdot)\right)=\int_{\min\{x_{0},X\}}^{\max\{x_{0},X\}}\left|Y'(x)\right|dx$
satisfies $E\left[V_{x_{0}}^{X}\left(Y(\cdot)\right)^{2}\right]<\infty$
for some $x_{0}\in(\underline{x},\overline{x})$.
\end{enumerate}
\end{assumptionx}

Assumption \ref{Ass:pot}--(i) replaces \ref{Ass:iv}--(i), and
this is a standard exogeneity assumption.\footnote{This setting implicitly rules out any direct causal effects of the
instrument on the potential outcome. Some studies explicitly specify
the potential outcome $Y(x,z)$ to be a function of the potential
instrument assignment $z$ and then assume $Y(x,z')=Y(x,z)$ for any
$z\ne z'$.} Assumption \ref{Ass:pot}--(ii) replaces \ref{Ass:deriv}--(i).
Unobserved heterogeneity in the treatment effects arises when $Var\left(Y'(x)|W=w\right)>0$.
To rule out negative weights, some studies in the IV literature specify
the potential treatment process and assume it to be monotonic in the
instrument. On the other hand, I do not impose the monotonicity condition
and maintain Assumption \ref{Ass:iv}--(ii) (i.e., $E(\widetilde{X}\widetilde{Z})\ne0$),
thereby allowing for negative weights.\footnote{Appendix \ref{subsec:LATE_MTE} considers the weighted-average interpretation
under the monotonicity condition and explores its relationship with
the LATE interpretation in \citet{angrist1995two} and the marginal
treatment effect interpretation in \citet{heckman2006understanding}.}

The following theorem extends the weighted-average interpretation
of the IV coefficient provided by Theorem \ref{Thm:Interpret}.
\begin{thm}
\label{Thm:Unobs} With Assumptions \ref{Ass:pot}, \ref{Ass:con},
\ref{Ass:iv}--(ii), and \ref{Ass:linear}--(i), the IV coefficient
$\beta_{IV}$ is given by
\begin{align*}
\beta_{IV} & =\int\int\tau_{IV}(x,w)\omega_{Z}(x,w)dF_{W}(w)dx,
\end{align*}
\textup{where }the IV-identified marginal effect $\tau_{IV}(x,w)$
at each $(x,w)$ is given by
\[
\tau_{IV}(x,w)=E\left[Y'(x)\lambda\left(Y'(x)|x,w\right)|W=w\right]
\]
 with $\lambda(t|x,w)=\frac{Cov\left(\ind_{X\ge x},Z|Y'(x)=t,W=w\right)}{Cov\left(\ind_{X\ge x},Z|W=w\right)}$
and $E\left[\lambda(Y'(x)|x,w)|W=w\right]=1$.
\end{thm}
Theorem \ref{Thm:Unobs} implies that the IV coefficient is a weighted
average of the causal effects $Y'(x)$. Although treatment levels
$x$ and covariates $w$ are weighted exactly in the same manner as
in Theorem \ref{Thm:Interpret}, unobserved heterogeneity influences
how the effects $Y'(x)$ are weighted. In particular, the difference
between the IV-identified marginal effect $\tau_{IV}(x,w)$ and the
AME $\tau(x,w)$ can be written as
\[
\tau_{IV}(x,w)-\tau(x,w)=Cov\left(Y'(x),\lambda\left(Y'(x)|x,w\right)|W=w\right).
\]
This difference arises from unobservable-driven covariance between
the treatment effects $Y'(x)$ and the treatment responses to the
instrument. For example, suppose some unobservables positively influence
$Y'(x)$ and make the treatment more responsive to the instrument.
As the weight function $\lambda(t|x,w)$ represents how strongly the
instrument $Z$ induces a transition of the treatment $X$ from below
$x$ to above $x$ conditional on $Y'(x)=t$ and $W=w$, $\tau_{IV}(x,w)>\tau(x,w)$
results from a greater emphasis on a higher $Y'(x)$. This difference
corresponds to the discrepancy between the LATE and ATE in a binary
treatment setting with no covariates, although the difference is specific
to each treatment level $x$ and covariate value $w$ in this case.

Given the weighted-average interpretation, the decomposition in (\ref{eq:decom1}--\ref{eq:decom3})
remains valid by replacing the marginal effect difference component
in (\ref{eq:decom3}) with
\begin{equation}
\Delta_{ME}=\int\int\left(\tau_{IV}(x,w)-\frac{\partial}{\partial x}m(x,w)\right)\omega_{Z}(x,w)dF_{W}(w)dx.\label{eq:decom3u}
\end{equation}
However, this component no longer represents endogeneity bias alone.
In particular, (\ref{eq:decom3u}) can be further decomposed as
\begin{align*}
\Delta_{ME} & =\int\int\left(\tau(x,w)-\frac{\partial}{\partial x}m(x,w)\right)\omega_{Z}(x,w)dF_{W}(w)dx\\
 & +\int\int\left(\tau_{IV}(x,w)-\tau(x,w)\right)\omega_{Z}(x,w)dF_{W}(w)dx.
\end{align*}
The first term captures endogeneity bias through the difference between
the AME $\tau(x,w)$ and the slope of the conditional mean function
$\frac{\partial}{\partial x}m(x,w)$. The second term represents the
unobservable-driven weight difference discussed above.

In a nonbinary treatment setting, it is well known that the AME $\tau(x,w)$
or analogous causal objects cannot be identified without any restriction
on unobserved heterogeneity or causal structure.\footnote{Examples of restrictions that may enable this identification are:
one-dimensional unobservable in the first-stage relationship \citep{imbens2009identification};
one-dimensional unobservable in the potential outcome \citep{chernozhukov2007instrumental};
and the linear random coefficients model \citep{masten2016identification}.
One could estimate one of these models and recover the AME to further
decompose $\Delta_{ME}$ into endogeneity bias and the unobservable-driven
weight difference. Exploring the decomposition under these restrictions
is beyond the scope of this paper.} Thus, endogeneity bias and the unobservable-driven weight difference
cannot be identified separately in general. Nevertheless, it remains
helpful to isolate the covariate and treatment-level weight difference
components using the decomposition, rather than observing only the
raw IV--OLS gap.

\section{Estimation and Inference}

\subsection{IV-Weighted OLS Estimators\label{subsec:IV-weighted-OLS-Estimators}}

Performing the decomposition proposed in Section \ref{sec:Econometric-Framework-for}
requires estimators of the IV-weighted OLS coefficients defined in
(\ref{eq:beta_c}--\ref{eq:beta_ct}), which serve as intermediate
points between the IV and OLS estimates. To define the estimators,
let $(Y_{i},X_{i},Z_{i},W_{i})_{i=1}^{N}$ be an i.i.d. random sample
that satisfies the set of assumptions in Section \ref{sec:Econometric-Framework-for}.
The most natural estimators for the IV-weighted OLS coefficients are
their direct data counterparts:
\begin{align}
\widehat{\beta}_{C} & =\frac{1}{N}\sum_{i=1}^{N}\int\widehat{b}_{OLS}(W_{i})\widehat{\omega}_{Z}(x,W_{i})dx,\label{eq:betac_est}\\
\widehat{\beta}_{CT} & =\frac{1}{N}\sum_{i=1}^{N}\int\frac{\partial}{\partial x}\widehat{m}(x,W_{i})\widehat{\omega}_{Z}(x,W_{i})dx,\label{eq:betact_est}
\end{align}
using some estimators $(\widehat{b}_{OLS},\widehat{m},\widehat{\omega}_{Z})$
for $(b_{OLS},m,\omega_{Z}$).

While there are various choices for these estimators $(\widehat{b}_{OLS},\widehat{m},\widehat{\omega}_{Z})$,
I focus on the most practical setting. I assume that $\widehat{m}$
is a consistent estimator for $m$ estimated by minimizing $\frac{1}{N}\sum_{i=1}^{N}\left\{ Y_{i}-\widehat{m}(X_{i},W_{i})\right\} ^{2}$,
and that $\widehat{m}(x,w)$ is given in a series form 
\begin{equation}
\widehat{m}(x,w)=\sum_{k=1}^{K_{N}}\widehat{\alpha}_{k}(w)p_{k}(x),\,\,\,\widehat{\alpha}_{k}(w)=\sum_{\ell=1}^{L_{N}^{(k)}}\widehat{\theta}_{k\ell}q_{k\ell}(w),\label{eq:series}
\end{equation}
where $\{p_{k}(x),q_{k1}(w),\ldots,q_{kL_{N}^{(k)}}(w)\}_{k=1}^{K_{N}}$
is a set of basis functions chosen by the researcher. The numbers
of the basis functions $K_{N}$ and $(L_{N}^{(k)})_{k=1}^{K_{N}}$
are constant in a parametric approach, while they may increase with
the sample size $N$ with a nonparametric sieve estimation adopted.
While $\widehat{b}_{OLS}$ can be any consistent estimator for $b_{OLS}$,
for concreteness I assume that $\widehat{b}_{OLS}$ is also estimated
by the series form (\ref{eq:series}) with $K_{N}=2$ and $\left(p_{1}(x),p_{2}(x)\right)=\left(1,x\right)$,
in which $\widehat{\alpha}_{2}(w)$ corresponds to $\widehat{b}_{OLS}(w)$.
In addition, as the estimator for $\omega_{Z}(x,W_{i})$, I consider
its sample analogue 
\[
\widehat{\omega}_{Z}(x,W_{i})=\frac{\widetilde{\ind}_{X_{i}\ge x}\widetilde{Z}_{i}}{\frac{1}{N}\sum_{j=1}^{N}\widetilde{\ind}_{X_{j}\ge x}\widetilde{Z}_{j}}.
\]

Then, (\ref{eq:betac_est}--\ref{eq:betact_est}) can be rewritten
as
\begin{align}
\widehat{\beta}_{C} & =\frac{\sum_{i=1}^{N}\widehat{b}_{OLS}(W_{i})\widetilde{X}_{i}\widetilde{Z}_{i}}{\sum_{i=1}^{N}\widetilde{X}_{i}\widetilde{Z}_{i}},\label{eq:betac_est2}\\
\widehat{\beta}_{CT} & =\frac{\sum_{i=1}^{N}\left(\sum_{k=1}^{K_{N}}\widehat{\alpha}_{k}(W_{i})\widetilde{P}_{ik}\right)\widetilde{Z}_{i}}{\sum_{i=1}^{N}\widetilde{X}_{i}\widetilde{Z}_{i}},\label{eq:betact_est2}
\end{align}
where $P_{ik}=p_{k}(X_{i})$. Therefore, the following two-step procedure
gives the IV-weighted OLS estimates.\begin{description}[leftmargin=0em]

\item[Step 1:] Estimate $\widehat{b}_{OLS}(w)$ and $\left\{ \widehat{\alpha}_{k}(w)\right\} _{k=1}^{K_{N}}$
using the least-squares method with the series specification (\ref{eq:series}).

\item[Step 2:] Regress $Y_{2i}=\sum_{k=1}^{K_{N}}\widehat{\alpha}_{k}(W_{i})\widetilde{P}_{ik}$
on $(X_{i},W_{i})$ instrumenting $X_{i}$ with $Z_{i}$, which yields
$\widehat{\beta}_{CT}$ as the coefficient of $X_{i}$. To estimate
$\widehat{\beta}_{C}$, let $Y_{2i}=\widehat{b}_{OLS}(W_{i})\widetilde{X}_{i}$
and perform the same regression.

\end{description}

Note that this two-step approach naturally generalizes the one proposed
by \citet{lochner2015estimating} by allowing for a more general functional
form of $\widehat{m}(x,w)$. In fact, $\widehat{\beta}_{CT}$ given
in Step 2 is identical to the reweighted OLS estimator in \citet{lochner2015estimating}
if $\widehat{m}(x,w)$ is specified in Step 1 to be additively separable
in $x$ and $w$.

\subsubsection*{DID- or RD-type Instrument Case}

The setting considered in Section \ref{subsec:DID_RD} requires a
small modification of the estimators due to a difference in the weight
functions. Suppose that the weight function $\omega_{Z}^{*}(x,W_{i})$
defined in Theorem \ref{Thm:DID_RD} is estimated by its sample analogue
\[
\widehat{\omega}_{Z}^{*}(x,W_{i})=\frac{\widehat{L}_{W_{i}}\left(\ind_{X_{i}\ge x}\widetilde{Z}_{i}\right)}{\frac{1}{N}\sum_{j=1}^{N}\ind_{X_{j}\ge x}\widetilde{Z}_{j}},
\]
where $\widehat{L}_{w}\left(R_{i}\right)=w'\left(\sum_{j=1}^{N}W_{j}W_{j}'\right)^{-1}\left(\sum_{j=1}^{N}W_{j}R_{j}\right)$
is the linear projection estimate. Using $\widehat{\omega}_{Z}^{*}$
instead of $\widehat{\omega}_{Z}$ in deriving (\ref{eq:betac_est2}--\ref{eq:betact_est2})
yields
\begin{align}
\widehat{\beta}_{C} & =\frac{\sum_{i=1}^{N}\widehat{L}_{W_{i}}\left(\widehat{b}_{OLS}(W_{i})\right)X_{i}\widetilde{Z}_{i}}{\frac{1}{N}\sum_{i=1}^{N}\widetilde{X}_{i}\widetilde{Z}_{i}},\label{eq:betac_est3}\\
\widehat{\beta}_{CT} & =\frac{\sum_{i=1}^{N}\left(\sum_{k=1}^{K_{N}}\widehat{L}_{W_{i}}\left(\widehat{\alpha}_{k}(W_{i})\right)P_{ik}\right)\widetilde{Z}_{i}}{\frac{1}{N}\sum_{i=1}^{N}\widetilde{X}_{i}\widetilde{Z}_{i}}.\label{eq:betact_est3}
\end{align}
These expressions differ from (\ref{eq:betac_est2}--\ref{eq:betact_est2})
in two ways. First, $\widehat{b}_{OLS}(W_{i})$ and $\widehat{\alpha}_{k}(W_{i})$
are linearly projected onto $W_{i}$. In practice, $\widehat{b}_{OLS}(W_{i})$
and $\widehat{\alpha}_{k}(W_{i})$ may be specified to be linear at
the outset, rather than estimating them from a flexible nonlinear
specification first and then linearly predicting them. Second, $X_{i}$
and $P_{ik}$ instead of $\widetilde{X}_{i}$ and $\widetilde{P}_{ik}$
enter the numerators. These differences slightly change Step 2 in
estimating the IV-weighted OLS estimates as follows.\begin{description}[leftmargin=0em]

\item[Step 2 (DID- or RD-type Instrument):] Regress $Y_{2i}=\sum_{k=1}^{K_{N}}\widehat{L}_{W_{i}}\left(\widehat{\alpha}_{k}(W_{i})\right)P_{ik}$
on $(X_{i},W_{i})$ instrumenting $X_{i}$ with $Z_{i}$, which yields
$\widehat{\beta}_{CT}$ as the coefficient of $X_{i}$. To estimate
$\widehat{\beta}_{C}$, let $Y_{2i}=\widehat{L}_{W_{i}}\left(\widehat{b}_{OLS}(W_{i})\right)X_{i}$
and perform the same regression.\end{description}

\subsection{Asymptotic Properties}

Asymptotic properties of the IV-weighted OLS estimator can be derived
using the standard econometric results for a two-step estimator. The
following discussion focuses on the case in which the first step is
parametric, since \citet{ackerberg2012practical} illustrate that
a semiparametric two-step estimator that uses a series approximation
in the first step can be treated as if it were a parametric estimator
for the purpose of standard error computation.

Using the standard formula for a parametric two-step estimator \citep[Section 6]{newey1994large},
Appendix \ref{subsec:Statistical-Properties-of} derives the asymptotic
equivalence:
\begin{equation}
\sqrt{N}\left(\widehat{\beta}_{CT}-\beta_{CT}\right)\underset{p}{\to}\frac{\frac{1}{\sqrt{N}}\sum_{i=1}^{N}\left(v_{1i}\widehat{\widetilde{Z}_{i}}+v_{2i}\widetilde{Z}_{i}\right)}{E\left[\widetilde{X}_{i}\widetilde{Z}_{i}\right]},\label{eq:Std_Error}
\end{equation}
where $v_{1i}=Y_{i}-\sum_{k=1}^{K}\alpha_{k}(W_{i})P_{ik}$ is a residual
from the first step, $v_{2i}=\widetilde{Y}_{2i}-\beta_{CT}\widetilde{X}_{i}$
is a residual from the second step, and $\widehat{\widetilde{Z}_{i}}$
is a predicted value of $\widetilde{Z}_{i}$ given by the first step
using $\widetilde{Z}_{i}$ instead of $Y_{i}$ as the dependent variable.\footnote{Another correction term appears in the case with DID- or RD-based
instruments if $\widehat{b}_{OLS}(W_{i})$ and $\widehat{\alpha}_{k}(W_{i})$
are not specified to be linear, as presented in Appendix \ref{subsec:Statistical-Properties-of}.} Most statistical packages can estimate the standard error of $\widehat{\beta}_{CT}$
by estimating the standard error of the right-hand side of (\ref{eq:Std_Error})
under a certain distributional assumption (heteroscedasticity-robust,
clustered, etc.). For example, in an i.i.d. heteroscedastic case,
the asymptotic variance of $\sqrt{N}\left(\widehat{\beta}_{CT}-\beta_{CT}\right)$
is given by $V_{\beta_{CT}}=E\left[\widetilde{X}_{i}\widetilde{Z}_{i}\right]^{-1}E\left[(v_{1i}\widehat{\widetilde{Z}_{i}}+v_{2i}\widetilde{Z}_{i})^{2}\right]$
and the standard error of $\widehat{\beta}_{CT}$ can be estimated
using the sample analogue of $\sqrt{V_{\beta_{CT}}/N}$. Estimating
the standard error of $\widehat{\beta}_{C}$ can follow the same procedure,
as it is a special case with $K=2$.

\subsection{Testing the Treatment Endogeneity\label{subsec:Testing-the-Treatment}}

Testing the significance of the marginal effect difference component
$\Delta_{ME}=\beta_{IV}-\beta_{CT}$ serves as a generalized Durbin--Wu--Hausman
(DWH) test that is robust to the nonlinearity and observed heterogeneity
of the treatment effects. This test further extends the generalized
DWH test proposed in \citet{lochner2015estimating} by allowing for
observed heterogeneity of the effects. Since $\widehat{\beta}_{IV}$
and $\widehat{\beta}_{CT}$ have a common denominator $\frac{1}{N}\sum_{i=1}^{N}\widetilde{X}_{i}\widetilde{Z}_{i}$,
a relevant test statistic is
\begin{equation}
\widehat{T}=\frac{\frac{1}{N}\sum_{i=1}^{N}d_{i}\widetilde{Z}_{i}}{\widehat{S}},\label{eq:test_GDWH}
\end{equation}
where $d_{i}=\widetilde{Y}_{i}-\widetilde{Y}_{2i}$ is the difference
in (residualized) dependent variables between two regressions that
yield $\widehat{\beta}_{IV}$ and $\widehat{\beta}_{CT}$. $\widehat{S}$
is the standard error of the numerator, $\frac{1}{N}\sum_{i=1}^{N}d_{i}\widetilde{Z}_{i}$.
For example, in an i.i.d. heteroscedastic case, 
\begin{equation}
N\widehat{S}^{2}=\frac{1}{N}\sum_{i=1}^{N}\left(d_{i}\widetilde{Z}_{i}-\frac{1}{N}\sum_{j=1}^{N}d_{j}\widetilde{Z}_{j}-v_{1i}\widehat{\widetilde{Z}_{i}}\right)^{2}.\label{eq:test_hc}
\end{equation}
Under a more general distributional assumption, the right hand side
of (\ref{eq:test_hc}) should be replaced by the estimated asymptotic
variance of $\frac{1}{\sqrt{N}}\sum_{i=1}^{N}\left(d_{i}\widetilde{Z}_{i}-\frac{1}{N}\sum_{j=1}^{N}d_{j}\widetilde{Z}_{j}-v_{1i}\widehat{\widetilde{Z}_{i}}\right)$.
Appendix \ref{subsec:Properties-of-GDWH} shows that $\widehat{T}$
converges in distribution to $N(0,1)$ under the null hypothesis $\Delta_{ME}=0$
and diverges under the alternative hypothesis $\Delta_{ME}\ne0$.\footnote{As described in footnote \ref{fn:multIV}, in a setting with multiple
instruments $(Z_{i1},\ldots,Z_{iM})$, the scalar instrument $Z_{i}$
is generated as $Z_{i}=\sum_{m=1}^{M}\pi_{m}Z_{im}$. In this case,
instead of performing the $t$-test with the synthetic $Z_{i}$, one
could separately compute $\frac{1}{N}\sum_{i=1}^{N}d_{i}\widetilde{Z}_{im}$
for each $m=1,\ldots,M$ and perform a chi-squared test to improve
efficiency.} 

Three limitations of this test are worth noting. First, it is a valid
test of endogeneity only when the setup in Section \ref{sec:Econometric-Framework-for}
describes the true model. Most importantly, the marginal effect difference
$\Delta_{ME}$ cannot be attributed to endogeneity bias alone in the
presence of unobserved heterogeneity, as discussed in Section \ref{subsec:Unobserved-Heterogeneity}.\footnote{In a binary treatment context, one might consider the frameworks such
as \citet{donald2014testing} and \citet{mogstad2018using}, which
can test endogeneity bias even with unobserved heterogeneity under
several additional conditions.} Second, this test cannot detect endogeneity bias if the difference
between the IV-identified and OLS-identified marginal effects $\frac{\partial}{\partial x}g(x,w)-\frac{\partial}{\partial x}m(x,w)$
in some regions of $(x,w)$ exactly cancels out the difference in
other regions. While the direction of endogeneity bias is expected
to be unambiguous in many economic contexts, one might consider the
nonparametric test proposed by \citet{blundell2007non} in the contexts
in which the direction of endogeneity may differ across $(x,w)$.
Finally, this test shares the fundamental limitations with the standard
DWH test that the instrument must be valid and that it has little
power to detect endogeneity bias if the instrument is not sufficiently
strong.\footnote{The asymptotic property of the test itself would not be influenced
by the weak instrument problem, since the test statistic in (\ref{eq:test_GDWH})
does not depend on the first stage coefficient, $\frac{1}{N}\sum_{i=1}^{N}\widetilde{X}_{i}\widetilde{Z}_{i}$.
\citet{staiger1997instrumental} show that one version of the DWH
test \citep{Durbin1954} is robust to the weak instrument problem
but two others \citep{Wu1973,hausman1978specification} are not.}

\section{Applications to Return-to-Schooling Estimates\label{sec:Applications}}

This section describes the empirical applications of my framework
to return-to-schooling estimates with three different identification
strategies.\footnote{I use the term ``returns to schooling'' to refer to the causal effect
of schooling on log wages, even though it sometimes denotes the internal
rate of return associated with schooling in its most narrow sense.} First, I use geographic variation in college costs as an instrument
as in \citet{cameron2004estimation} and \citet{carneiro2011estimating}
and estimate the returns to schooling in the National Longitudinal
Survey of Youth 1979 (NLSY79). Next, I exploit a discontinuity in
the minimum school-leaving age in the United Kingdom as in \citet{oreopoulos2006estimating},
using the British General Household Survey (GHS). Finally, I exploit
DID variation in compulsory schooling laws across cohorts and states
in the United States as in \citet*{acemoglu2000large} and estimate
the returns to schooling in the 1960--1980 U.S. Censuses. Appendix
\ref{sec:Data} provides additional details for these analyses.

\subsection{College Cost Instrument with Geographic Variation\label{subsec:College IV}}

This analysis uses the civilian sample from the NLSY79. The original
sample consists of 5,579 males and 5,827 females born in 1957--64.
After dropping persons with missing information about their Armed
Forces Qualification Test (AFQT) score or county of residence at age
14, persons who have not completed 8th grade by age 22, and persons
with no wage information in any year, my sample consists of 4,719
males and 4,986 females. For each person, I use observations between
ages 25 and 54. The outcome $Y$ is the log hourly wage at the current
or most recent job and the treatment $X$ is years of schooling top-coded
at 18 years.

Because minority and economically disadvantaged households are sampled
at higher rates, persons are weighted by the sampling weights throughout
my analysis. Within each person, I use equal weights for the person--year
observations.\footnote{As a result, the weight on a person--year observation is the sampling
weight for the person divided by the number of years for which the
person is observed.} It is important to weight observations appropriately to recover the
population OLS and IV coefficients because my framework does not assume
the linear structural equation and admits a weighted-average interpretation
of the coefficients.

As instruments, I use measures of direct and opportunity costs of
college attendance as in \citet{cameron2004estimation} and \citet{carneiro2011estimating}.\footnote{While I follow their identification strategies in estimating the causal
effects of schooling, I do not follow their empirical specifications
in my analysis. Their original analyses do not exactly fit into my
framework because years of schooling is not the only endogenous variable
in \citet{cameron2004estimation} and the schooling measure is binary
in \citet{carneiro2011estimating}.} In particular, I use the presence of a public four-year college and
tuition rate of the nearest in-state public four-year college in the
county of residence at age 14 to represent the direct cost of attendance.\footnote{In earlier work, \citet{card1995using} and \citet{kane1995labor}
use college proximity as the schooling instrument associated with
the direct costs of college attendance. They do not use opportunity
cost measures in their analyses.} While the tuition rate captures a pecuniary cost of attendance, college
proximity captures both pecuniary (due to reduced costs of room and
board) and nonpecuniary costs of attendance. I use local earnings
and unemployment rate in the county of residence at age 14 in the
year in which the person turns age 17 (1974 for the oldest cohort
and 1981 for the youngest) to represent the opportunity cost of attendance.\footnote{Local earnings are measured at the county level, while the unemployment
rate is at the state level.}

For the vector of covariates $W$, I use the AFQT percentile, age,
an indicator for female, black and Hispanic dummies, indicators for
parental education, the number of siblings, and cohort dummies.\footnote{The specification allows the AFQT scores to have different slopes
across tertiles and the ages to have different slopes across the 25--34,
35--44, and 45--54 age groups.}In addition, I include urban status, Census division dummies, and
the average local earnings and unemployment rate during 1974--81
of the county of residence at age 14 in the covariate vector $W$.
Using the average local labor market conditions as control variables
ensures that variation in the corresponding instruments is driven
by a temporary shock to the local labor market. Otherwise, the instruments
may capture a permanent difference in local labor market conditions,
which can be directly associated with potential earnings. Controlling
for local labor market conditions is also important to ensure the
plausibility of the college proximity instrument, as emphasized by
\citet{cameron2004estimation}. Appendix \ref{subsec:NLSY79} provides
further details of the sample and presents the first-stage regression
results.

Table \ref{Table:decom} reports the decomposition results. The first
three columns of the table report the linear OLS coefficient, the
linear IV coefficient, and the IV--OLS coefficient gap. The next
three columns report the estimates of the covariate weight difference,
the treatment-level weight difference, and the marginal effect difference.
Here, I discuss the first row of the table, which reports the results
from the NLSY79. The second and third rows are discussed in Sections
\ref{subsec:CSL_UK} and \ref{subsec:CSL}. In the first row, the
point estimates of the OLS and IV coefficients are nearly identical,
with the OLS coefficient of 0.065 and the IV coefficient of 0.062.
An empirical researcher may be tempted to conclude from this result
that there is no evidence of ability bias in these data. However,
the decomposition using the IV-weighted OLS coefficients indicates
that the IV coefficient would be well below the OLS coefficient if
they had the same weights on the covariates and treatment levels.
In fact, the covariate weight difference is estimated to be 0.011
and the treatment-level weight difference is estimated to be 0.018.
With the weight difference components accounted for, the marginal
effect difference is --0.032, indicating that the IV-identified returns
to schooling are lower than the OLS-identified returns. This result
is rather consistent with the ability bias story in terms of point
estimates, even though the generalized DWH test fails to reject at
the 5\% level given the large standard error.

I investigate the mechanisms underlying these results by examining
the patterns of the IV and OLS weights. Table \ref{College_weights_W}
presents the total IV and OLS weights on each group of covariates.
The first three columns of the table present the population share,
the total IV weight, and the total OLS weight on each group. Each
set of weights sums to one across the whole sample by construction.
The last column reports the OLS schooling coefficient from a linear
regression performed separately for each group, to illustrate the
difference in OLS-identified schooling effects across groups. While
the OLS weights are close to the population shares, the IV weights
are concentrated on persons with advantaged backgrounds in terms of
AFQT score, parental education, and race/ethnicity. Schooling coefficients
from the separately performed OLS regressions indicate that more-advantaged
groups tend to have higher schooling effects. These data patterns
result in the positive contribution of the covariate weight difference
to the IV--OLS coefficient gap.

The pattern of IV weights is consistent with the empirical observation
by \citet{cameron2004estimation} that persons with advantaged backgrounds
tend to be more sensitive to local college availability in the NLSY79.
However, one may expect that persons with disadvantaged backgrounds
should be weighted more because they are expected to be more sensitive
to college cost instruments given their financial constraints.\footnote{\citet{card1995using} and \citet{kling2001interpreting} find that
persons from less advantaged backgrounds are more sensitive to the
presence of a local college in the sample from the National Longitudinal
Survey of Young Men, which is based on older cohorts than the NLSY79.} Several factors can explain low IV weights on persons with disadvantaged
backgrounds. While \citet{cameron2004estimation} suggest that increased
funding on federal student aid programs in the 1970s is one such factor,
low college attendance and graduation rates among persons from disadvantaged
backgrounds can also be important.\footnote{The share of persons with one or more and four or more years of college
education are 19\% and 4\%, respectively, among the bottom third of
AFQT scores. Among persons in the top third of AFQT scores, 80\% have
one or more years of college education and 56\% have four or more
years of college education.} A low college attendance rate implies that the majority would be
never-takers instead of compliers.\footnote{If the college enrollment decision is explained by a logit or probit
model, the share of compliers is the largest among the group with
a college attendance rate of 50\%.} A low college graduation rate implies that their completed years
of schooling would not be strongly affected, even if the instruments
affect their college attendance decisions.

Panel (a) of Figure \ref{Fig:college_weights_X} illustrates the total
IV and OLS weights on each treatment margin. The IV weights are mostly
placed on college education margins, with the weights on high school
margins close to zero. This weight pattern is consistent with the
expectation that college cost instruments affect years of schooling
through college attendance decisions. Higher IV weights on college
education margins result in the positive contribution of the weight
difference to the IV--OLS coefficient gap because marginal effects
of years in college are much higher than marginal effects of years
in high school in this sample. In fact, the linear OLS coefficient
is 0.031 in the subsample with 12 or fewer years of schooling and
0.071 in the subsample with 12 or more years of schooling.

The overall result from this decomposition exercise that the IV coefficient
is inflated due to the weight difference appears to match the discount
rate bias argument developed by \citet{lang1993ability} and \citet{card1995earnings}.\footnote{Building on the canonical model of \citet{becker1967human}, they
consider a model in which individuals invest in education as long
as the marginal return to an additional year of schooling exceeds
its marginal cost. The model predicts that the marginal return at
the chosen schooling level would be higher for credit-constrained
individuals with higher discount rates. Given this prediction, they
argue that the IV coefficient could exceed the population average
return if credit-constrained individuals are more sensitive to the
instruments and thus weighted more.} However, the estimated weight patterns indicate that the underlying
mechanism is distinct in two ways. First, the IV coefficient places
more weight on advantaged rather than disadvantaged groups in the
population. Higher IV weights on advantaged groups give rise to the
higher IV coefficient because of the higher marginal returns among
advantaged groups. This is the opposite mechanism to the discount
rate bias argument, even though it shifts the IV coefficient in the
same direction. Second, higher marginal returns to college education
give rise to the higher IV coefficient in this sample, given the concentration
of IV weights on college years relative to high school years. The
discount rate bias argument does not consider this possibility, as
it assumes nonincreasing marginal returns to schooling.

\subsection{RD-Based Compulsory Schooling Instrument\label{subsec:CSL_UK}}

As in \citet{oreopoulos2006estimating}, I now exploit the 1947 compulsory
schooling reform in the United Kingdom to construct an RD instrument.
The U.K. government raised the minimum school-leaving age in Great
Britain from 14 to 15 years in 1947. The share of people leaving school
at age 14 or earlier then fell from 56\% in the 1932 birth cohort
turning 14 one year before the reform to 9\% in the 1934 birth cohort
turning 14 one year after the reform. 

The empirical strategy in this analysis closely follows \citet{oreopoulos2006estimating}.\footnote{My regression results slightly differ from the originally published
results in \citet{oreopoulos2006estimating} due to the data correction
\citep{oreopoulos2008estimating} and a top-coding treatment of schooling
described below.} I use the sample of persons younger than 65 years old from the British
General Household Surveys in 1984--98 who turned 14 in 1935--65.
I exclude persons with missing data on earnings or education and persons
leaving school before age 10. The outcome $Y$ is log annual earnings.
The treatment $X$ is years of schooling, which is given by the age
when they left full-time education minus five years, with the top-coding
at 20 years.\footnote{Approximately 2.5\% of persons report having left full-time education
after age 25, and their schooling levels are all treated as 20 years.
\citet{oreopoulos2006estimating} does not make this top-coding treatment.
See Appendix \ref{subsec:GHS} for the analysis without top-coding,
which reaches the same conclusion regarding the relevance of the weight
difference components.} The instrument $Z$ is an indicator for 1933 or later birth cohorts,
who turned 14 in 1947 or later. The covariates $W$ are the fourth-order
polynomials of birth cohort and age.\footnote{\citet{gelman2019high} recommend against the use of high-order polynomials
in RD designs; nevertheless, I follow the original specification in
\citet{oreopoulos2006estimating}. Using two separate quadratic polynomials
for pre- and post-reform cohorts instead of global fourth-order polynomials
slightly pushes up the IV estimate. However, it does not affect the
finding that the weight difference components are important.}

The second row of Table \ref{Table:decom} reports the decomposition
of the IV--OLS coefficient gap in this empirical application. As
the OLS estimate lies above the IV estimate by 0.021, a researcher
who presumes the linear causal model may immediately interpret it
as the result of ability bias. However, adjusting for the weight difference
suggests otherwise. In fact, the decomposition result indicates that
the marginal returns identified by the IV coefficient exceed those
identified by the OLS coefficient by 0.023, after accounting for the
covariate weight difference and the treatment-level weight difference
components. The empirical result is no longer consistent with the
ability bias story as a point estimate, although both standard and
generalized DWH tests fail to reject at the 5\% level due to the imprecise
IV estimate.

Table \ref{Table:UK_wgt_w} presents the IV and OLS weights on the
birth cohorts. The IV weights are concentrated on the birth cohorts
turning age 14 in 1941--50. The linear regression restricted to these
cohorts yields smaller OLS schooling coefficients than those for the
younger cohorts, who receive most of the OLS weights. This explains
the negative contribution of the covariate weight difference to the
IV--OLS coefficient gap. Panel (b) of Figure \ref{Fig:college_weights_X}
shows that the IV weights are almost exclusively placed on the 10th
year of schooling, which is exactly what the 1947 reform mandated.
This weight pattern pushes down the IV coefficient because the 10th
year of schooling has a smaller marginal return than the other schooling
margins. In fact, the linear regression restricted to individuals
with 9 or 10 years of schooling yields an OLS schooling coefficient
of 0.037, which is much smaller than the full-sample OLS coefficient.

This analysis demonstrates that researchers should exercise caution
in extending the intuition of \citet{imbens1994identification} to
an empirical setting with covariates and a multivalued treatment.
Observing a sharp decline in the dropout rate at age 14 after the
reform, \citet{oreopoulos2006estimating} argues that the IV and OLS
coefficients in this setting are expected to identify the treatment
effects for the comparable population, providing an analogy to the
comparison between the LATE and ATE. In principle, however, the LATE
interpretation in \citet{imbens1994identification} draws on a model
with a binary treatment and no covariates. It is still plausible that
the LATE and ATE in this setting match \emph{conditional on} the
schooling level and the birth cohort, judging by the extensive response
to the reform. Nevertheless, the estimated patterns of the weights
suggest that the IV and OLS coefficients identify the effects for
entirely different birth cohorts at completely distinct schooling
margins.

\subsection{Compulsory Schooling Instrument with DID Variation\label{subsec:CSL}}

This analysis exploits DID variation in compulsory schooling laws
across cohorts and regions, following \citet{acemoglu2000large}.\footnote{My specification follows one of their main specifications in estimating
private returns to schooling that uses the 1960--80 Census data with
the child labor laws instrument and no state-of-residence controls
\citep[p.34]{acemoglu2000large}. While they adjust some variables
in the 1960--80 data to incorporate the 1950 Census data in their
alternative specifications, I do not make these adjustments as I focus
only on the 1960--80 Censuses. See Appendix \ref{subsec:Census}
for details.} The analysis sample consists of 40--49-year-old white males born
in the United States from the 1960--80 Censuses. I use log weekly
earnings as the outcome $Y$, and limit my analysis to persons with
positive earnings and working for at least one week in the previous
year. I use years of schooling as the treatment $X$. I include in
a vector of covariates $W$ indicators for the state of birth and
indicators for the year of birth. As instruments, I use the status
of compulsory schooling laws (CSL) at age 14 in the state of birth.
As in \citet{acemoglu2000large}, I construct the CSL instruments
based on the required years of schooling associated with child labor
laws. Appendix \ref{subsec:Census} provides further details of the
sample, first-stage regression results, and decomposition results
using other common compulsory schooling instruments.

The third row of Table \ref{Table:decom} presents the estimated OLS
and IV schooling coefficients and the decomposition of the IV--OLS
coefficient gap. While the IV estimate is slightly above the OLS estimate
with a coefficient gap of 0.017, the decomposition result indicates
that the gap is primarily associated with the weight difference components.
Although the IV--OLS gap is not statistically distinguishable from
zero even without accounting for the weight difference components,
this result reshapes the quantitative implication from the gap. In
particular, \citet{acemoglu2000large} attribute the small positive
IV--OLS gap to modest external returns to schooling. Accounting
for the weight difference components, this result indicates even less
important externality than their original interpretation.

Table \ref{CSL_weights_W} presents the total IV and OLS weights on
each group of covariates. The IV weights attached to some birth cohorts
or birth states are negative. In fact, the IV weights aggregate to
negative values among persons in the 1930--39 birth cohorts and among
persons born in the Midwest or West. Moreover, I find that 54\% of
covariate-specific IV weights are negative and sum to --5.93. 

In a usual IV setting, negative weights imply the presence of both
compliers and defiers. In a DID setting, however, the weights are
not guaranteed to be positive even with the perfect compliance $X=Z$,
as pointed out by \citet{Chaisemartin2020} in the binary treatment
case.\footnote{In fact, the first stage regression indicates the positive relationship
between the CSL requirements and schooling levels, even in the subsample
of the 1930--39 birth cohorts or among persons born in the Midwest
or West.} While the IV-weighted OLS coefficient $\beta_{C}=\int b_{OLS}(w)\overline{\omega}_{Z}(w)dF_{W}(w)$
is a weighted average of the covariate-specific OLS coefficients $b_{OLS}(w)$,
negative weights can push $\beta_{C}$ out of the support of $b_{OLS}(w)$.
In fact, the estimates of $b_{OLS}(w)$ are no greater than $0.076$,
despite $\beta_{C}$ being estimated to be $0.079$. This result suggests
that even a small heterogeneity in treatment effects can give rise
to a large contribution of the weight difference to the IV--OLS coefficient
gap if the IV strategy relies on DID variation.

Panel (c) of Figure \ref{Fig:college_weights_X} reports the total
IV and OLS weights for each schooling level. The IV weights mostly
capture the schooling margins up to the 12th year of schooling, which
is consistent with the context that primary and secondary education
is mandated by the CSL. However, the effect of this treatment-level
weight difference on the IV--OLS coefficient gap is mostly obscured
by the large contribution of the covariate weight difference.

\section{Conclusion}

When OLS and IV estimates differ, empirical researchers typically
consider two explanations. The first takes the linear regression equation
literally and interprets the coefficient gap as endogeneity bias.
The second extends the intuition of the LATE interpretation \citep{imbens1994identification}
to a general regression equation and interprets the coefficient gap
as the weight difference. My paper enables researchers to proceed
a step further and formally quantify the contributions of the weight
difference and endogeneity bias components separately.

I show that the IV--OLS coefficient gap is explained by differences
in the weights on the covariates, the weights on the treatment levels,
and the identified marginal effects. The marginal effect difference
component captures endogeneity bias in the absence of the unobservable-driven
interaction between the treatment effects and treatment responses
to the instrument. I propose a simple two-step regression approach
to perform the decomposition empirically, which can be implemented
in standard statistical packages.

I demonstrate the practical value of my framework through its empirical
applications to return-to-schooling estimates with compulsory schooling
and college cost instruments. The IV--OLS coefficient gaps in these
empirical applications are substantially influenced by the weight
difference components, and accounting for them leads to different
conclusions about the direction or extent of endogeneity bias.

\medskip
\begingroup
\setlength\bibitemsep{0.25em}
\phantomsection

\printbibliography[heading=bibintoc]
\endgroup
\setcounter{secnumdepth}{0}
\phantomsection

\section[Tables and Figures]{}

\begin{table}[H]
\caption{Decomposition of the IV--OLS Gap in Return-to-Schooling Estimates}
\label{Table:decom}

\centering
\begin{threeparttable}
\begin{centering}
\begin{tabular}{cccccccccc}
 &  &  &  &  &  &  &  &  & \tabularnewline
\hline 
\multirow{2}{*}{IV Strategy} & \multirow{2}{*}{Data} &  & \multicolumn{3}{c}{Coefficients} &  & \multicolumn{3}{c}{Decomposition}\tabularnewline
\cline{4-6} \cline{5-6} \cline{6-6} \cline{8-10} \cline{9-10} \cline{10-10} 
 &  &  & OLS & IV & IV--OLS &  & $\Delta_{CW}$ & $\Delta_{TW}$ & $\Delta_{ME}$\tabularnewline
\cline{1-2} \cline{2-2} \cline{4-6} \cline{5-6} \cline{6-6} \cline{8-10} \cline{9-10} \cline{10-10} 
\noalign{\vskip0.25em}
College Cost & \multirow{2}{*}{NLSY79\hspace{0.3em}\tnote{1)}} &  & 0.065 & 0.062 & --0.004 &  & 0.011 & 0.018 & --0.032\tabularnewline
Variation &  &  & (0.003) & (0.087) & (0.087) &  & (0.011) & (0.010) & (0.086)\tabularnewline[0.5em]
Compulsory & \multirow{2}{*}{British GHS\hspace{0.3em}\tnote{2)}} &  & 0.084 & 0.062 & --0.021 &  & --0.016 & --0.029 & 0.023\tabularnewline
Schooling RD &  &  & (0.002) & (0.083) & (0.082) &  & (0.009) & (0.018) & (0.078)\tabularnewline[0.5em]
Compulsory & \multirow{2}{*}{U.S. Census\hspace{0.3em}\tnote{3)}} &  & 0.067 & 0.084 & 0.017 &  & 0.011 & 0.003 & 0.003\tabularnewline
Schooling DID &  &  & (0.0004) & (0.022) & (0.022) &  & (0.004) & (0.003) & (0.021)\tabularnewline[0.25em]
\hline 
\end{tabular}
\par\end{centering}
\vspace{0.25em}
\begin{tablenotes}
\small

\item Notes: Standard errors are in parentheses. The first three
columns report the OLS estimates, the IV estimates, and their gaps.
The next three columns report the estimates of the covariate weight
difference, the treatment-level weight difference, and the marginal
effect difference components. By construction, these three components
sum to the IV--OLS gap. Appendix \ref{subsec:Estimation_Detail}
describes the empirical specification for estimating the decomposition.

\item[1)] Standard errors are robust to heteroskedasticity and correlation
across observations on persons living in the same county at age 14.
The instruments are college cost measures as defined in the main text.

\item[2)] Standard errors are robust to heteroskedasticity and correlation
across observations on the same birth cohort and survey year. The
instrument is an indicator for turning age 14 in 1947 or later.

\item[3)] Standard errors are robust to heteroskedasticity and correlation
across observations on the same state and year of birth. The instruments
are indicators for compulsory schooling requirements implied by child
labor laws (7, 8, and 9 or more years).

\end{tablenotes}
\end{threeparttable}
\end{table}

\begin{table}[H]
\centering
\begin{threeparttable}

\caption{The IV and OLS Weights on the Covariate Groups (NLSY79)}
\label{College_weights_W}
\begin{centering}
\begin{tabular}{cccccc}
 &  &  &  &  & \tabularnewline
\hline 
\multirow{2}{*}{Variable} & \multirow{2}{*}{Group} & Group & OLS & IV & Subsample\tabularnewline
 &  & share & weight & weight & OLS coef.\tabularnewline
\hline 
\multirow{3}{*}{\begin{tabular}{@{}c@{}}AFQT \\ percentile\end{tabular}} & 0--1/3 & 0.32 & 0.25 (0.01) & --0.08 (0.16) & 0.056 (0.006)\tabularnewline
 & 1/3--2/3 & 0.34 & 0.35 (0.01) & 0.19 (0.13) & 0.062 (0.005)\tabularnewline
 & 2/3--1 & 0.34 & 0.40 (0.01) & 0.89 (0.20) & 0.074 (0.005)\tabularnewline
\hline 
Highest & Some HS of less & 0.24 & 0.21 (0.01) & --0.12 (0.15) & 0.055 (0.005)\tabularnewline
parental & HS graduate & 0.42 & 0.41 (0.01) & 0.66 (0.17) & 0.069 (0.005)\tabularnewline
education & Some college & 0.34 & 0.37 (0.01) & 0.46 (0.15) & 0.068 (0.005)\tabularnewline
\hline 
 & Black & 0.14 & 0.14 (0.02) & --0.12 (0.11) & 0.069 (0.005)\tabularnewline
Race/Ethnicity & Hispanic & 0.06 & 0.06 (0.01) & 0.02 (0.03) & 0.064 (0.006)\tabularnewline
 & Other & 0.80 & 0.80 (0.02) & 1.10 (0.11) & 0.064 (0.004)\tabularnewline
\hline 
\multirow{2}{*}{Sex} & Male & 0.50 & 0.49 (0.01) & 0.47 (0.17) & 0.057 (0.004)\tabularnewline
 & Female & 0.50 & 0.51 (0.01) & 0.53 (0.17) & 0.074 (0.005)\tabularnewline
\hline 
Urban residence & Rural & 0.31 & 0.30 (0.03) & 0.53 (0.17) & 0.073 (0.006)\tabularnewline
at age 14 & Urban & 0.69 & 0.70 (0.03) & 0.47 (0.17) & 0.062 (0.004)\tabularnewline
\hline 
 & Northeast & 0.22 & 0.22 (0.04) & 0.18 (0.14) & 0.061 (0.007)\tabularnewline
Region & Midwest & 0.31 & 0.30 (0.04) & 0.51 (0.17) & 0.070 (0.007)\tabularnewline
at age 14 & South & 0.32 & 0.32 (0.04) & 0.09 (0.13) & 0.058 (0.005)\tabularnewline
 & West & 0.15 & 0.16 (0.03) & 0.23 (0.15) & 0.071 (0.005)\tabularnewline
\hline 
\end{tabular}
\par\end{centering}
\begin{tablenotes}
\small

\item Notes: Standard errors are in parentheses and robust to heteroskedasticity
and correlation across observations on persons living in the same
county at age 14. Each set of weights sums to one across the whole
sample. Subsample OLS coefficient of years of schooling is obtained
with the same set of control variables as regressions in Table \ref{Table:decom}.
Appendix \ref{subsec:Estimation_Detail} describes the empirical specification
for estimating the weights.

\end{tablenotes}
\end{threeparttable}
\end{table}

\begin{table}[H]
\caption{The IV and OLS Weights on the Covariate Groups (British GHS)}
\label{Table:UK_wgt_w}

\centering
\begin{threeparttable}
\begin{centering}
\begin{tabular}{ccccc}
 &  &  &  & \tabularnewline
\hline 
\multirow{2}{*}{Year at 14} & Group & OLS & IV & Subsample\tabularnewline
 & share & weights & weights & OLS coef.\tabularnewline
\hline 
1935--40 & \multirow{2}{*}{0.04} & 0.03 & --0.09 & 0.054\tabularnewline
 &  & (0.01) & (0.05) & (0.012)\tabularnewline
1941-45 & \multirow{2}{*}{0.09} & 0.08 & 0.68 & 0.065\tabularnewline
 &  & (0.01) & (0.15) & (0.005)\tabularnewline
1946--50 & \multirow{2}{*}{0.14} & 0.12 & 0.49 & 0.079\tabularnewline
 &  & (0.02) & (0.06) & (0.006)\tabularnewline
1951--55 & \multirow{2}{*}{0.19} & 0.18 & --0.10 & 0.086\tabularnewline
 &  & (0.02) & (0.17) & (0.004)\tabularnewline
1956--60 & \multirow{2}{*}{0.25} & 0.26 & 0.05 & 0.088\tabularnewline
 &  & (0.03) & (0.07) & (0.003)\tabularnewline
1961--65 & \multirow{2}{*}{0.28} & 0.34 & --0.02 & 0.088\tabularnewline
 &  & (0.03) & (0.05) & (0.003)\tabularnewline
\hline 
\end{tabular}
\par\end{centering}
\begin{tablenotes}
\small

\item Notes: Standard errors (in parentheses) are robust to heteroskedasticity
and correlation across observations on the same birth cohort and survey
year. Each set of weights sums to one across the whole sample. Subsample
OLS coefficient of years of schooling is obtained with controlling
for quartic terms of ages and birth cohorts. Appendix \ref{subsec:Estimation_Detail}
describes the empirical specification for estimating the weights.

\end{tablenotes}
\end{threeparttable}
\end{table}

\begin{table}[H]
\caption{The IV and OLS Weights on the Covariate Groups (U.S. Census)}
\label{CSL_weights_W}

\centering
\begin{threeparttable}
\begin{centering}
\begin{tabular}{cccccc}
 &  &  &  &  & \tabularnewline
\hline 
\multirow{2}{*}{Variable} & \multirow{2}{*}{Group} & Group & OLS & IV & Subsample\tabularnewline
 &  & share & weights & weights & OLS coef.\tabularnewline
\hline 
 & 1910--19 & \multirow{2}{*}{0.32} & 0.33 & 1.56 & 0.063\tabularnewline
 &  &  & (0.02) & (0.25) & (0.001)\tabularnewline
Year of & 1920--29 & \multirow{2}{*}{0.35} & 0.36 & 1.25 & 0.070\tabularnewline
birth &  &  & (0.02) & (0.28) & (0.001)\tabularnewline
 & 1930--39 & \multirow{2}{*}{0.33} & 0.31 & --1.81 & 0.067\tabularnewline
 &  &  & (0.02) & (0.38) & (0.001)\tabularnewline
\hline 
 & Northeast & \multirow{2}{*}{0.29} & 0.26 & 0.24 & 0.069\tabularnewline
 &  &  & (0.02) & (0.37) & (0.001)\tabularnewline
 & Midwest & \multirow{2}{*}{0.33} & 0.27 & --1.32 & 0.066\tabularnewline
\multirow{2}{*}{\begin{tabular}{@{}c@{}}Region of \\ birth\end{tabular}} &  &  & (0.02) & (0.34) & (0.001)\tabularnewline
 & South & \multirow{2}{*}{0.30} & 0.39 & 2.95 & 0.068\tabularnewline
 &  &  & (0.02) & (0.41) & (0.001)\tabularnewline
 & West & \multirow{2}{*}{0.09} & 0.08 & --0.87 & 0.064\tabularnewline
 &  &  & (0.01) & (0.15) & (0.001)\tabularnewline
\hline 
\end{tabular}
\par\end{centering}
\begin{tablenotes}
\small

\item Notes: Standard errors (in parentheses) are robust to heteroskedasticity
and correlation across observations on the same state and year of
birth. Each set of weights sums to one across the whole sample. Subsample
OLS coefficient of years of schooling is obtained with controlling
for state of birth and year of birth dummies. Appendix \ref{subsec:Estimation_Detail}
describes the empirical specification for estimating the weights.

\end{tablenotes}
\end{threeparttable}
\end{table}

\begin{figure}[H]
\caption{The IV and OLS Weights on the Treatment Levels}

\label{Fig:college_weights_X}

\vspace{2.0em}
\centering
\begin{threeparttable}

\begin{minipage}[t]{0.49\columnwidth}%
(a) NLSY79 (College Costs IV)

\includegraphics[width=1\columnwidth]{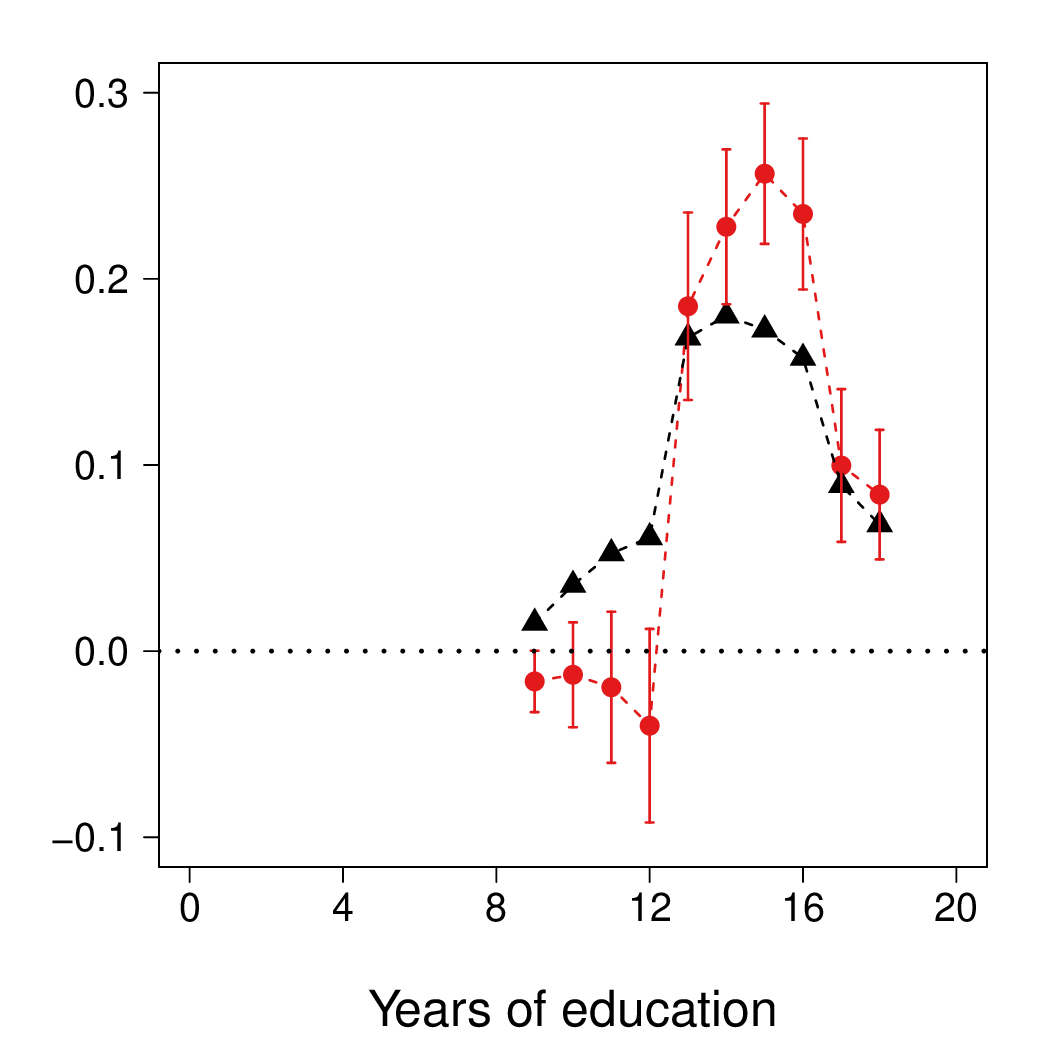}%
\end{minipage}%
\begin{minipage}[t]{0.49\columnwidth}%
(b) British GHS (Compulsory Schooling RD)

\includegraphics[width=1\columnwidth]{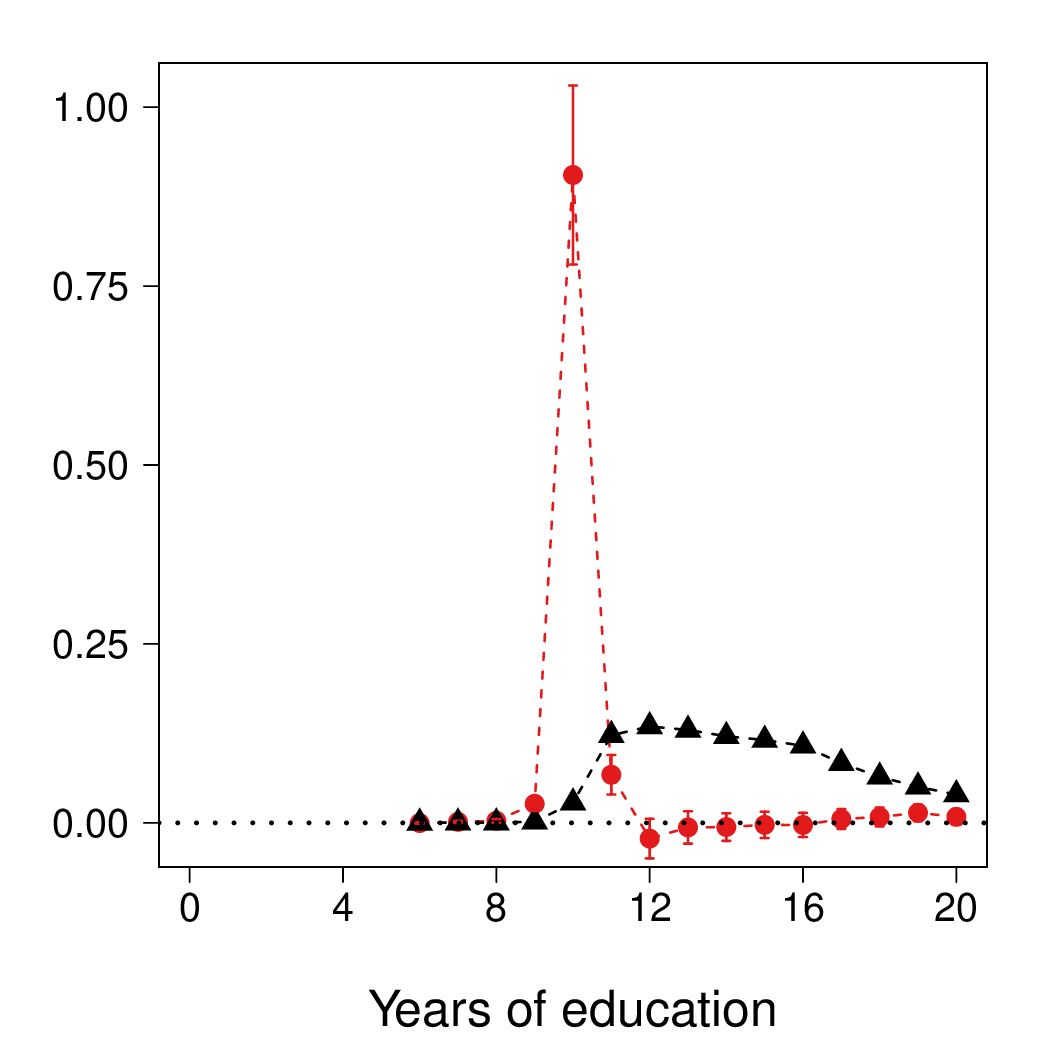}%
\end{minipage}

\vspace{2.0em}

\begin{minipage}[t]{0.49\columnwidth}%
(c) U.S. Census (Compulsory Schooling DID)

\includegraphics[width=1\columnwidth]{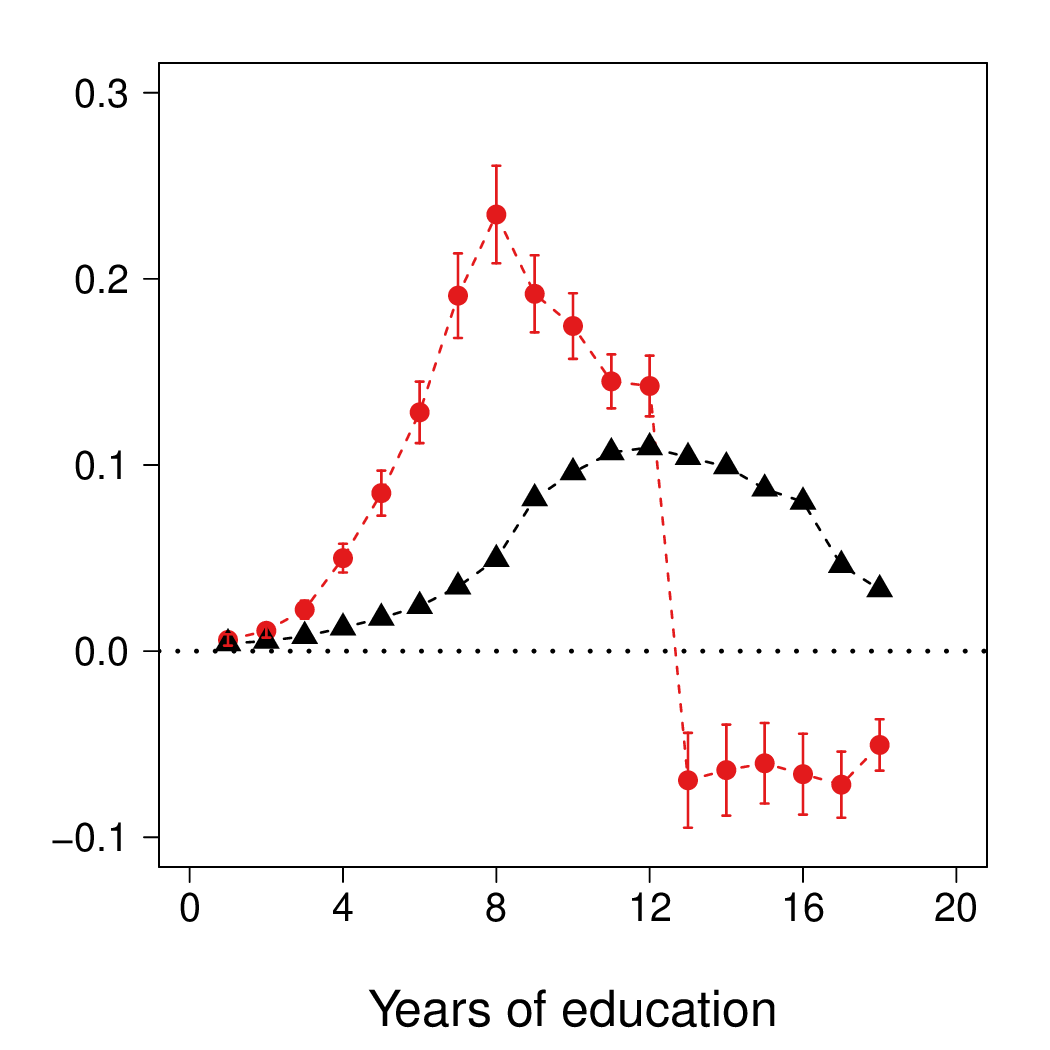}%
\end{minipage}%
\begin{minipage}[t]{0.49\columnwidth}%
\vspace{0.25em}
\begin{center}
\includegraphics[bb=102bp 464bp 402bp 504bp,width=0.95\columnwidth]{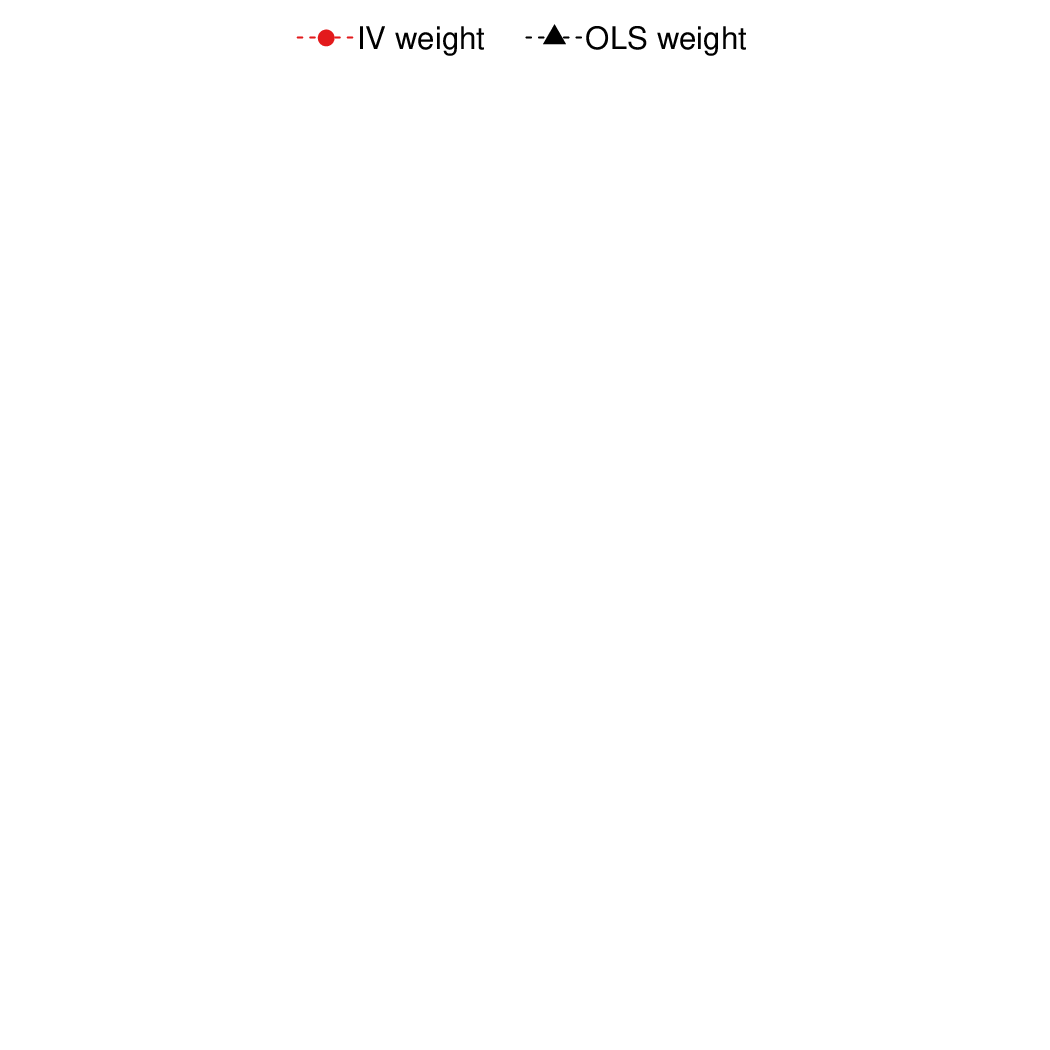}
\par\end{center}%
\end{minipage}

\vspace{1.5em}
\begin{tablenotes}
\small

\item Notes: The IV weights are presented with standard error bars.
The standard error bars for the OLS weights are omitted because they
are graphically negligible. Each set of weights sums to one across
the whole sample. Appendix \ref{subsec:Estimation_Detail} describes
the empirical specification for estimating the weights. The weights
are zero outside the support of years of education by construction,
and thus are not plotted on the graph.

\end{tablenotes}
\end{threeparttable}
\end{figure}

\end{refsection}
\setcounter{secnumdepth}{3}\newpage{}

\appendix

\part[Online Appendix]{}

\vspace{-7em}
\begin{refsection}
\renewcommand*{\thefootnote}{\fnsymbol{footnote}}
\begin{center}
{\large \textbf{Online Appendix: Empirical Decomposition of the IV–OLS Gap with Heterogeneous and Nonlinear Effects}}\\
\vspace{1em}
{\large Shoya Ishimaru}\footnote[1]{Hitotsubashi University, Department of Economics (email: shoya.ishimaru@r.hit-u.ac.jp).}
\end{center}
\renewcommand*{\thefootnote}{\arabic{footnote}}
\setcounter{table}{0}
\setcounter{figure}{0}
\setcounter{equation}{0}
\renewcommand{\thetable}{A.\arabic{table}}
\renewcommand\thefigure{A.\arabic{figure}}
\renewcommand{\theequation}{A.\arabic{equation}}

\section{Proofs\label{sec:Appendix_proof}}

This section provides proofs of key theorems. In the main paper, any
integral expression of $x$ is taken over the support $(\underline{x},\overline{x})$,
but it is kept implicit to simplify the exposition. In this Appendix,
it is kept explicit for rigorousness. 

Some proofs rely on the following lemmas.
\begin{lem}
\label{lem:Int}Suppose $g:(\underline{x},\overline{x})\mapsto\mathbb{R}$
is a differentiable function, where $-\infty\le\underline{x}<\overline{x}\le\infty$.
Then $\int_{\underline{x}}^{\overline{x}}g'(x)\left(\ind_{b\ge x}-\ind_{a>x}\right)dx=g(b)-g(a)$
for any $a,b\in(\underline{x},\overline{x})$.
\end{lem}
\begin{proof}
It immediately follows from the fundamental theorem of calculus, given
the identity that $\ind_{b\ge x}-\ind_{a>x}$ is equal to $1$ if
$a\le x\le b$, to $-1$ if $b<x<a$, and to $0$ otherwise.
\end{proof}
\begin{lem}
\label{lem:Reg}Suppose a random variable $X$ satisfies Assumption
\ref{Ass:con}. Let $R$ be any random variable and $W$ be any random
vector, and define $\widetilde{R}=R-W'E\left[WW'\right]^{-1}E\left[WR\right]$.
Let $h(x,w)$ be a function differentiable in $x$, and suppose that
the total variation $V_{a}^{b}(h,w)=\int_{\min\{a,b\}}^{\max\{a,b\}}|\frac{\partial}{\partial x}h(x,w)|dx$
satisfies $E\left[V_{x_{0}}^{X}(h,W)^{2}\right]<\infty$ for some
$x_{0}\in(\underline{x},\overline{x})$. If $E\left(R|W=w\right)$
is linear in $w$ or $h(x,w)$ is linear in $w$ for each $x\in(\underline{x},\overline{x})$,
then
\[
E\left(h(X,W)\widetilde{R}\right)=\int_{\text{\ensuremath{\underline{x}}}}^{\overline{x}}\int\frac{\partial}{\partial x}h(x,w)E\left(\ind_{X\ge x}\widetilde{R}|W=w\right)dF_{W}(w)dx.
\]
\end{lem}
\begin{proof}
Take any $x_{0}\in(\underline{x},\overline{x})$. Then,
\begin{eqnarray*}
E\left(h(X,W)\widetilde{R}\right) & = & E\left[\left(h(X,W)-h(x_{0},W)\right)\widetilde{R}\right]\\
 & = & E\left[\left(\int_{\text{\ensuremath{\underline{x}}}}^{\overline{x}}\frac{\partial}{\partial x}h(x,W)\left(\ind_{X\ge x}-\ind_{x_{0}>x}\right)dx\right)\widetilde{R}\right]
\end{eqnarray*}
The first equality follows from the linearity of $h(x_{0},w)$ or
$E[R|W=w]$ in $w$. Applying Lemma \ref{lem:Int} by letting $b=X$
and $a=x_{0}$ (for each point in the sample space) yields the second
equality. Note that $(\underline{x},\overline{x})$ can be unbounded
when applying Lemma \ref{lem:Int}, as each realization of $X$ is
finite even if its support $(\underline{x},\overline{x})$ is unbounded.

Continuing the transformation,
\begin{eqnarray*}
E\left(h(X,W)\widetilde{R}\right) & = & \int_{\text{\ensuremath{\underline{x}}}}^{\overline{x}}E\left[\frac{\partial}{\partial x}h(x,W)\left(\ind_{X\ge x}-\ind_{x_{0}>x}\right)\widetilde{R}\right]dx\\
 & = & \int_{\text{\ensuremath{\underline{x}}}}^{\overline{x}}E\left[\frac{\partial}{\partial x}h(x,W)\ind_{X\ge x}\widetilde{R}\right]dx\\
 & = & \int_{\text{\ensuremath{\underline{x}}}}^{\overline{x}}\int\frac{\partial}{\partial x}h(x,w)E\left(\ind_{X\ge x}\widetilde{R}|W=w\right)dF_{W}(w)dx
\end{eqnarray*}
The first equality uses Fubini's theorem. The second equality follows
from the linearity of $h(x,w)$ or $E[R|W=w]$ in $w$. The third
equality follows from the law of iterated expectations. Note that
Fubini's theorem can be applied here because
\begin{eqnarray*}
E\left[\int_{\text{\ensuremath{\underline{x}}}}^{\overline{x}}\left|\frac{\partial}{\partial x}h(x,W)\left(\ind_{X\ge x}-\ind_{x_{0}>x}\right)\widetilde{R}\right|dx\right] & = & E\left[|\widetilde{R}|\cdot\int_{\text{\ensuremath{\underline{x}}}}^{\overline{x}}\left|\frac{\partial}{\partial x}h(x,W)\right|\left|\ind_{X\ge x}-\ind_{x_{0}>x}\right|dx\right]\\
 & = & E\left[|\widetilde{R}|\cdot V_{x_{0}}^{X}(h,W)\right]\\
 & \le & \sqrt{E\left[\widetilde{R}^{2}\right]E\left[V_{x_{0}}^{X}(h,w)^{2}\right]}\\
 & < & \infty,
\end{eqnarray*}
where the second equality follows from the identity that $\left|\ind_{X\ge x}-\ind_{x_{0}>x}\right|$
is equal to $1$ if $x$ is between $x_{0}$ and $X$ and to 0 otherwise.
This completes the proof.
\end{proof}

\subsection{Proof of Theorem \ref{Thm:Interpret}}
\begin{proof}
The IV coefficient is given by
\begin{align}
\beta_{IV} & =\frac{E(\widetilde{Y}\widetilde{Z})}{E(\widetilde{X}\widetilde{Z})}\nonumber \\
 & =\frac{E\left[g(X,W)\widetilde{Z}\right]}{E(\widetilde{X}\widetilde{Z})}\nonumber \\
 & =\int_{\text{\ensuremath{\underline{x}}}}^{\overline{x}}\int\frac{\partial}{\partial x}g(x,w)\frac{E(\ind_{X\ge x}\widetilde{Z}|W=w)}{E(\widetilde{X}\widetilde{Z})}dF_{W}(w)dx\nonumber \\
 & =\int_{\text{\ensuremath{\underline{x}}}}^{\overline{x}}\int\frac{\partial}{\partial x}g(x,w)\omega_{Z}(x,w)dF_{W}(w)dx.\label{eq:IV_proof}
\end{align}
The second equality uses Assumptions \ref{Ass:sep} and \ref{Ass:iv}--(i).
The third equality uses Lemma \ref{lem:Reg} by letting $R=Z$ and
$h=g$, where Assumptions \ref{Ass:deriv}--(i) and \ref{Ass:linear}--(i)
are required to use the lemma. The last equality uses Assumption \ref{Ass:linear}--(i).

Since $E\left[Y\widetilde{X}\right]=E\left[E[Y|X,W]\widetilde{X}\right]=E\left[m(X,W)\widetilde{X}\right]$,
letting $R=X$ and $h=m$ in Lemma \ref{lem:Reg} gives the weighted-average
expression for $\beta_{OLS}$ based on the same argument as above.

Finally, letting $Y=X$ gives $\frac{\partial}{\partial x}g(x,w)=1$
and $\beta_{IV}=1$ in (\ref{eq:IV_proof}), which jointly yield $1=\int_{\text{\ensuremath{\underline{x}}}}^{\overline{x}}\int\omega_{Z}(x,w)dF_{W}(w)dx$.
Letting $Y=X$ also gives $\frac{\partial}{\partial x}m(x,w)=1$ and
$\beta_{OLS}=1$, which jointly imply $1=\int_{\text{\ensuremath{\underline{x}}}}^{\overline{x}}\int\omega_{X}(x,w)dF_{W}(w)dx$.
\end{proof}

\subsection{Proof of Theorem \ref{Thm:avgw}}
\begin{proof}
$E\left[Y\widetilde{Z}\right]=E\left[b_{IV}(W)X\widetilde{Z}\right]$
follows from the equalities below.
\begin{eqnarray*}
E\left[\left(Y-b_{IV}(W)X\right)\widetilde{Z}\right] & = & E\left[\left(Y-\frac{Cov(Y,Z|W)}{Cov(X,Z|W)}X\right)\widetilde{Z}\right]\\
 & = & E\left[E\left(Y\widetilde{Z}|W\right)-\frac{Cov(Y,Z|W)}{Cov(X,Z|W)}E\left(X\widetilde{Z}|W\right)\right]\\
 & = & E\left[Cov\left(Y,Z|W\right)-Cov(Y,Z|W)\right]\\
 & = & 0.
\end{eqnarray*}
The first equality uses the definition of $b_{IV}(w)$, the second
equality uses the law of iterated expectations, and the third equality
uses Assumption \ref{Ass:linear}--(i). Given $E\left[Y\widetilde{Z}\right]=E\left[b_{IV}(W)X\widetilde{Z}\right]$,
letting $h(x,w)=b_{IV}(w)x$ in Lemma \ref{lem:Reg} yields $\beta_{IV}=\int b_{IV}(w)\overline{\omega}_{Z}(w)dF_{W}(w).$
The same argument yields $\beta_{OLS}=\int b_{OLS}(w)\overline{\omega}_{X}(w)dF_{W}(w)$
by letting $Z=X$.
\end{proof}

\subsection{Proof of Theorem \ref{Thm:avgx_w}}
\begin{proof}
Using Theorem \ref{Thm:Interpret} conditional on $W=w$ immediately
gives these results. See also \citet{yitzhaki1996using} and \citet{schechtman2004gini}. 
\end{proof}

\subsection{Proof of Theorem \ref{Thm:DID_RD}}
\begin{proof}
Since $E\left[Y\widetilde{Z}\right]=E\left[g(X,W)\widetilde{Z}\right]$
due to Assumptions \ref{Ass:sep} and \ref{Ass:iv}--(i), letting
$R=Z$ and $h=g$ in Lemma \ref{lem:Reg} gives the following weighted-average
expression for $\beta_{IV}$, where Assumptions \ref{Ass:deriv}--(i)
and \ref{Ass:DID_RD} are required to use the lemma.
\begin{equation}
\beta_{IV}=\int_{\text{\ensuremath{\underline{x}}}}^{\overline{x}}\int\frac{\partial}{\partial x}g(x,w)\frac{E(\ind_{X\ge x}\widetilde{Z}|W=w)}{E(\widetilde{X}\widetilde{Z})}dF_{W}(w)dx.\label{eq:IV_proof_did}
\end{equation}

Since $\frac{\partial}{\partial x}g(x,w)$ is linear in $w$ for each
$x\in(\underline{x},\overline{x})$ due to Assumption \ref{Ass:DID_RD},
$\frac{E(\ind_{X\ge x}\widetilde{Z}|W=w)}{E(\widetilde{X}\widetilde{Z})}$
can be replaced with its linear projection $\omega_{Z}^{*}(x,w)=L_{w}(\ind_{X\ge x}\widetilde{Z})/E(\widetilde{X}\widetilde{Z})$,
as well as any function $r(x,w)$ with $L_{w}\left(r(x,W)\right)=\omega_{Z}^{*}(x,w)$.
Finally, letting $Y=X$ in (\ref{eq:IV_proof_did}) gives $\frac{\partial}{\partial x}g(x,w)=1$
and $\beta_{IV}=1$, which jointly yield $\int_{\text{\ensuremath{\underline{x}}}}^{\overline{x}}\int\omega_{Z}^{*}(x,w)dF_{W}(w)dx=1$.
\end{proof}

\subsection{Proof of Theorem \ref{Thm:Unobs}}
\begin{proof}
Take any $x_{0}\in(\underline{x},\overline{x})$. Then,
\begin{eqnarray*}
E\left(Y\widetilde{Z}\right) & = & E\left[\left(Y(X)-Y(x_{0})\right)\widetilde{Z}\right]\\
 & = & E\left[\left(\int_{\text{\ensuremath{\underline{x}}}}^{\overline{x}}Y'(x)\left(\ind_{X\ge x}-\ind_{x_{0}>x}\right)dx\right)\widetilde{Z}\right]\\
 & = & \int_{\text{\ensuremath{\underline{x}}}}^{\overline{x}}E\left[Y'(x)\left(\ind_{X\ge x}-\ind_{x_{0}>x}\right)\widetilde{Z}\right]dx\\
 & = & \int_{\text{\ensuremath{\underline{x}}}}^{\overline{x}}E\left[Y'(x)\ind_{X\ge x}\widetilde{Z}\right]dx.
\end{eqnarray*}
The first and last equalities follow from Assumptions \ref{Ass:pot}--(i)
and \ref{Ass:linear}--(i). Applying Lemma \ref{lem:Int} by letting
$b=X$ and $a=x_{0}$ (for each point in the sample space) yields
the second equality. The third equality uses Fubini's theorem, which
can be applied because
\begin{eqnarray*}
E\left[\int_{\text{\ensuremath{\underline{x}}}}^{\overline{x}}\left|Y'(x)\left(\ind_{X\ge x}-\ind_{x_{0}>x}\right)\widetilde{Z}\right|dx\right] & = & E\left[|\widetilde{Z}|\cdot\int_{\min\{x_{0},X\}}^{\max\{x_{0},X\}}\left|Y'(x)\right|dx\right]\\
 & = & E\left[V_{x_{0}}^{X}\left(Y(\cdot)\right)|\widetilde{Z}|\right]\\
 & \le & \sqrt{E\left[V_{x_{0}}^{X}\left(Y(\cdot)\right)^{2}\right]E\left[\widetilde{Z}^{2}\right]}\\
 & < & \infty.
\end{eqnarray*}

Continuing the transformation,
\begin{eqnarray*}
E\left(Y\widetilde{Z}\right) & = & \int_{\text{\ensuremath{\underline{x}}}}^{\overline{x}}E\left[E\left[Y'(x)E[\ind_{X\ge x}\widetilde{Z}|Y'(x),W]|W\right]\right]dx\\
 & = & \int_{\text{\ensuremath{\underline{x}}}}^{\overline{x}}E\left[E\left[Y'(x)\lambda(Y'(x)|x,W)|W\right]\omega_{Z}(x,W)\right]dx\\
 & = & \int_{\text{\ensuremath{\underline{x}}}}^{\overline{x}}\int\tau_{IV}(x,w)\omega_{Z}(x,w)dF_{W}(w)dx.
\end{eqnarray*}
The first equality uses the law of iterated expectations. The second
equality follows from the definition of the weight functions $\lambda$
and $\omega_{Z}$. The last equality uses the definition of $\tau_{IV}(x,w)$.
This completes the proof.
\end{proof}

\section{Additional Extensions\label{sec:Additional-Extensions}}

This section provides additional econometric results that extend the
analyses in the main paper.

\subsection{When Linearity Is Not a Good Approximation\label{subsec:Nonlinear}}

Except for DID and RD designs, Assumption \ref{Ass:linear} can be
made as close an approximation as possible with sufficient data. However,
it can be restrictive in empirical applications with a limited sample
size and many covariates, in which it is difficult to make a covariate
vector sufficiently flexible. Without the assumption, both $\beta_{IV}$
and $\beta_{OLS}$ may have omitted variable bias associated with
unaccounted nonlinear effects of covariates.

Removing Assumption \ref{Ass:linear} from Theorem \ref{Thm:Interpret}
yields the following result.
\begin{thm}
\label{Thm:Interpret_NL}Let $\omega_{R}(x,w)=\frac{Cov(\ind_{X\ge x},R|W=w)}{E\left[Cov(\ind_{X\ge x},R|W)\right]}$
for $R=Z,X$.
\begin{enumerate}
\item[(i)] With Assumptions \ref{Ass:sep}, \ref{Ass:con}, \ref{Ass:deriv}--(i),
and \ref{Ass:iv}, the difference between the IV coefficient $\beta_{IV}$
and the weighted average 
\begin{align*}
\beta_{IV}^{*} & =\int_{\text{\ensuremath{\underline{x}}}}^{\overline{x}}\int\frac{\partial}{\partial x}g(x,w)\omega_{Z}(x,w)dF_{W}(w)dx
\end{align*}
is given by
\begin{align*}
\beta_{IV} & -\beta_{IV}^{*}=\frac{E\left[\left(E(Y-\beta_{IV}^{*}X|W)-L_{W}(Y-\beta_{IV}^{*}X)\right)\left(E(Z|W)-L_{W}(Z)\right)\right]}{E\left[\widetilde{X}\widetilde{Z}\right]}.
\end{align*}
\item[(ii)] With Assumptions \ref{Ass:con}, \ref{Ass:deriv}--(ii), and \ref{Ass:ols},
the difference between the OLS coefficient $\beta_{OLS}$ and the
weighted average
\begin{align*}
\beta_{OLS}^{*} & =\int_{\text{\ensuremath{\underline{x}}}}^{\overline{x}}\int\frac{\partial}{\partial x}m(x,w)\omega_{X}(x,w)dF_{W}(w)dx.
\end{align*}
is given by
\begin{align*}
\beta_{OLS} & -\beta_{OLS}^{*}=\frac{E\left[\left(E(Y-\beta_{OLS}^{*}X|W)-L_{W}(Y-\beta_{OLS}^{*}X)\right)\left(E(X|W)-L_{W}(X)\right)\right]}{E\left[\widetilde{X}^{2}\right]}.
\end{align*}
\end{enumerate}
\end{thm}
Theorem \ref{Thm:Interpret_NL} implies that the weighted-average
interpretation of the IV and OLS coefficients remains valid but it
accompanies approximation errors associated with omitted nonlinear
effects of covariates $W$.\footnote{It is conceptually possible to isolate the approximation error by
estimating a model that allows for a greater flexibility in the covariate
vector $W$. In this situation, however, one could use the framework
in Section \ref{sec:Econometric-Framework-for} by estimating a more
flexible model in the first place.} Nonetheless, it is still possible to estimate the weight difference
components, $\Delta_{CW}$ and $\Delta_{TW}$, using parametric approximations
of weights $\omega_{Z}(x,w)$ and $\omega_{X}(x,w)$ and effects $b_{OLS}(w)$
and $\frac{\partial}{\partial x}m(x,w)$. Thus, the decomposition
continues to provide relevant information about how much the weight
difference matters for the IV--OLS coefficient gap. However, it may
also require caution when interpreting the remaining gap because the
approximation errors, as well as endogeneity bias, can influence the
gap.

The proof of Theorem \ref{Thm:Interpret_NL} is presented below.
\begin{proof}
By construction, $\beta_{IV}^{*}$ is given by
\begin{align}
\beta_{IV}^{*} & =\int_{\text{\ensuremath{\underline{x}}}}^{\overline{x}}\int\frac{\partial}{\partial x}g(x,w)\omega_{Z}(x,w)dF_{W}(w)dx\nonumber \\
 & =\int b_{IV}(w)\overline{\omega}_{Z}(w)dF_{W}(w)\nonumber \\
 & =\frac{E\left[Cov(Y,Z|W)\right]}{E\left[Cov(X,Z|W)\right]}\nonumber \\
 & =\frac{E\left[\left(Y-E(Y|W)\right)\left(Z-E(Z|W)\right)\right]}{E\left[\left(X-E(X|W)\right)\left(Z-E(Z|W)\right)\right]}.\label{eq:IVstar}
\end{align}
Then,
\begin{align*}
E\left[\widetilde{Y}\widetilde{Z}\right]= & E\left[\left(Y-E(Y|W)\right)\left(Z-E(Z|W)\right)+\left(E(Y|W)-L_{W}(Y)\right)\left(E(Z|W)-L_{W}(Z)\right)\right]\\
= & \beta_{IV}^{*}E\left[\left(X-E(X|W)\right)\left(Z-E(Z|W)\right)\right]+E\left[\left(E(Y|W)-L_{W}(Y)\right)\left(E(Z|W)-L_{W}(Z)\right)\right]\\
= & \beta_{IV}^{*}E\left[\widetilde{X}\widetilde{Z}-\left(E(X|W)-L_{W}(X)\right)\left(E(Z|W)-L_{W}(Z)\right)\right]\\
 & +E\left[\left(E(Y|W)-L_{W}(Y)\right)\left(E(Z|W)-L_{W}(Z)\right)\right]\\
= & \beta_{IV}^{*}E\left[\widetilde{X}\widetilde{Z}\right]+E\left[\left(E(Y-\beta_{IV}^{*}X|W)-L_{W}(Y-\beta_{IV}^{*}X)\right)\left(E(Z|W)-L_{W}(Z)\right)\right].
\end{align*}
Note that the second equality uses (\ref{eq:IVstar}) and the other
ones follow from the law of iterated expectations and simple algebra.
Therefore, $\beta_{IV}$ can be expressed as follows.
\begin{eqnarray*}
\beta_{IV} & = & E(\widetilde{Y}\widetilde{Z})/E(\widetilde{X}\widetilde{Z})\\
 & = & \beta_{IV}^{*}+\frac{E\left[\left(E(Y-\beta_{IV}^{*}X|W)-L_{W}(Y-\beta_{IV}^{*}X)\right)\left(E(Z|W)-L_{W}(Z)\right)\right]}{E(\widetilde{X}\widetilde{Z})}
\end{eqnarray*}
The same argument yields the expression for $\beta_{OLS}$ by letting
$Z=X$.
\end{proof}

\subsection{Decomposition with Invalid Instrument\label{subsec:Invalid_IV}}

Allowing the instrument to be correlated with the unobservable $U$
by removing Assumption \ref{Ass:iv}--(i), the interpretation of
the IV coefficient changes as follows.
\begin{thm}
With Assumptions \ref{Ass:sep}, \ref{Ass:con}, \ref{Ass:deriv},
\ref{Ass:iv}--(ii), and \ref{Ass:linear}--(i), the linear IV coefficient
is
\[
\beta_{IV}=\int_{\text{\ensuremath{\underline{x}}}}^{\overline{x}}\int\frac{\partial}{\partial x}g(x,w)\omega_{Z}(x,w)dF_{W}(w)dx+\frac{E[U\widetilde{Z}]}{E[X\widetilde{Z}]}.
\]
\end{thm}
\begin{proof}
The proof is virtually identical to that of Theorem \ref{Thm:Interpret}.
\end{proof}
Because of the correlation between the instrument and the unobservable
$U$, the IV coefficient is biased from the weighted average of the
causal effects. The interpretation of the decomposition (\ref{eq:decom1}--\ref{eq:decom3})
does not change except for (\ref{eq:decom3}), which is replaced by
\[
\Delta_{ME}=\int_{\text{\ensuremath{\underline{x}}}}^{\overline{x}}\int\left(\frac{\partial}{\partial x}g(x,w)-\frac{\partial}{\partial x}m(x,w)\right)\omega_{Z}(x,w)dF_{W}(w)dx+\frac{E[U\widetilde{Z}]}{E[X\widetilde{Z}]}.
\]
The first term still captures the dependence of the treatment $X$
on the unobservable $U$, as implied by (\ref{eq:m-g}). The newly
added second term captures the correlation between the instrument
$Z$ and the unobservable $U$ after controlling for covariates $W$.\footnote{This correlation can arise from both the failure of exclusion restriction
and the failure of random assignment, although this analysis does
not distinguish between them. For example, $U$ could be decomposed
as $U=h(Z)+V$, where the failure of exclusion restriction implies
$h\ne0$ and the failure of random assignment implies $E(V\widetilde{Z})\ne0$.}

\subsection{Decomposition with Unobserved Heterogeneity and DID/RD-Based Identification\label{subsec:DID_RD_u}}

While Section \ref{sec:more_general} allows for either a DID/RD-based
identification or unobserved heterogeneity, this extension allows
for both at the same time. Starting from the setting in Section \ref{subsec:Unobserved-Heterogeneity},
I remove Assumption \ref{Ass:linear}, which does not hold for DID
and RD cases. Then, I make the following assumption on the potential
outcome $Y(x)$.

\renewcommand{\theassumptionx}{LP}
\begin{assumptionx}

\label{Ass:pot-L}\textup{(Linearity of Conditional Potential Outcomes)}
$E\left[Y(x)|W=w\right]$ is linear in $w$ for each $x\in(\underline{x},\overline{x})$.

\end{assumptionx}

In a DID setting discussed in Section \ref{subsec:DID_RD}, Assumption
\ref{Ass:pot-L} can be written as $E[Y(x)|W]=\sum_{g=1}^{G}\gamma_{g}(x)d_{g}+\sum_{t=1}^{T-1}\delta_{t}(x)D_{t}$,
which implies parallel trends of the average potential outcome for
each treatment level $x$. In an RD setting, Assumption \ref{Ass:pot-L}
can be written as $E[Y(x)|C]=\sum_{k=1}^{K}\gamma_{k}(x)p_{k}(C)$.
This implies that the mean potential outcome conditional on a running
variable $C$ is well approximated by a linear combination of the
basis functions $\left(p_{1}(C),\ldots,p_{K}(C)\right)$, which also
implies continuity at the cutoff point $C=c$. Note that Assumption
\ref{Ass:pot}--(i) trivially holds in these settings since $Cov\left(Y(x),Z|W\right)=0$
immediately follows from $Var(Z|W)=0$.
\begin{thm}
\label{thm:DID_RD_u}With Assumptions \ref{Ass:pot}, \ref{Ass:con},
\ref{Ass:iv}--(ii), and \ref{Ass:pot-L}, the linear IV regression
coefficient is
\begin{align}
\beta_{IV} & =\int_{\text{\ensuremath{\underline{x}}}}^{\overline{x}}\int\tau_{IV}(x,w)\omega_{Z}(x,w)dF_{W}(w)dx,\label{eq:DID_RD_u}
\end{align}
\textup{where }the IV-identified marginal effect $\tau_{IV}(x,w)$
at each $(x,w)$ is given by
\[
\tau_{IV}(x,w)=E\left[Y'(x)\lambda\left(Y'(x)|x,w\right)|W=w\right]
\]
 with $\lambda(t|x,w)=\frac{E\left[\ind_{X\ge x}\widetilde{Z}|Y'(x)=t,W=w\right]}{E\left[\ind_{X\ge x}\widetilde{Z}|W=w\right]}$
and $E\left[\lambda(Y'(x)|x,w)|W=w\right]=1$. The weight function
$\omega_{Z}(x,w)$ is given by

\[
\omega_{Z}(x,w)=\frac{E\left[\ind_{X\ge x}\widetilde{Z}|W=w\right]}{E\left[X\widetilde{Z}\right]}.
\]
\end{thm}
Two important implications of Theorem \ref{thm:DID_RD_u} are worth
noting. First, the marginal effect difference component $\Delta_{ME}$
of the decomposition in (\ref{eq:decom1}--\ref{eq:decom3}) should
be attributed to not only endogeneity bias but also the unobservable-driven
weight difference, as discussed in Section \ref{subsec:Unobserved-Heterogeneity}. 

Second, unlike in Section \ref{subsec:DID_RD}, the weight function
$\omega_{Z}(x,w)=\frac{E\left[\ind_{X\ge x}\widetilde{Z}|W=w\right]}{E\left[X\widetilde{Z}\right]}$
cannot be reduced to its linearly projected counterpart $\omega_{Z}^{*}(x,w)=\frac{L_{w}\left(\ind_{X\ge x}\widetilde{Z}\right)}{E\left(X\widetilde{Z}\right)}$
in general. Even though the AME $\tau(x,w)=E\left[Y'(x)|W=w\right]$
is linear in $w$ due to Assumption \ref{Ass:pot-L}, the IV-identified
marginal effect $\tau_{IV}(x,w)=E\left[Y'(x)\lambda\left(Y'(x)|x,w\right)|W=w\right]$
does not necessarily satisfy the linearity condition due to the weight
function $\lambda\left(Y'(x)|x,w\right)$ that depends on $w$.

An interesting special case of Theorem \ref{thm:DID_RD_u} is a sharp
DID or RD with $X=Z$. In this case, the equation (\ref{eq:DID_RD_u})
simplifies to
\[
\beta_{IV}=\int_{\text{\ensuremath{\underline{x}}}}^{\overline{x}}\int\tau(x,w)\omega_{Z}^{*}(x,w)dF_{W}(w)dx.
\]
Even with unobserved heterogeneity, perfect compliance due to $X=Z$
eliminates the unobservable-driven weight difference. This makes the
weighted-average expression essentially identical to Theorem \ref{Thm:DID_RD},
which rules out unobserved heterogeneity.

The proof of Theorem \ref{thm:DID_RD_u} is presented below. This
proof is similar to that of Theorem \ref{Thm:Unobs}.
\begin{proof}
Take any $x_{0}\in(\underline{x},\overline{x})$. Then,
\begin{align*}
E\left(Y\widetilde{Z}\right) & =E\left[\left(Y(X)-Y(x_{0})\right)\widetilde{Z}\right]\\
 & =E\left[\left(\int_{\text{\ensuremath{\underline{x}}}}^{\overline{x}}Y'(x)\left(\ind_{X\ge x}-\ind_{x_{0}>x}\right)dx\right)\widetilde{Z}\right]\\
 & =\int_{\text{\ensuremath{\underline{x}}}}^{\overline{x}}E\left[Y'(x)\left(\ind_{X\ge x}-\ind_{x_{0}>x}\right)\widetilde{Z}\right]dx\\
 & =\int_{\text{\ensuremath{\underline{x}}}}^{\overline{x}}E\left[Y'(x)\ind_{X\ge x}\widetilde{Z}\right]dx.
\end{align*}
The first and last equalities follow from Assumptions \ref{Ass:pot}--(i)
and \ref{Ass:pot-L}. Applying Lemma \ref{lem:Int} by letting $b=X$
and $a=x_{0}$ (for each point in the sample space) yields the second
equality. The third equality uses Fubini's theorem.

Continuing the transformation,
\begin{align*}
E\left(Y\widetilde{Z}\right) & =\int_{\text{\ensuremath{\underline{x}}}}^{\overline{x}}E\left[E\left[Y'(x)E[\ind_{X\ge x}\widetilde{Z}|Y'(x),W]|W\right]\right]dx\\
 & =\int_{\text{\ensuremath{\underline{x}}}}^{\overline{x}}E\left[E\left[Y'(x)\lambda(Y'(x)|x,W)|W\right]\omega_{Z}(x,W)\right]dx\\
 & =\int_{\text{\ensuremath{\underline{x}}}}^{\overline{x}}\int\tau_{IV}(x,w)\omega_{Z}(x,w)dF_{W}(w)dx.
\end{align*}
The first equality uses the law of iterated expectations. The second
equality follows from the definition of the weight functions $\lambda$
and $\omega_{Z}$. The last equality uses the definition of $\tau_{IV}(x,w)$.
\end{proof}

\section{Other Econometric Details}

This section provides technical details and additional econometric
results omitted from the main paper.

\subsection{Derivation of the Marginal OLS Weights on Treatment Levels\label{subsec:Derivation-of-the}}

The derivation of (\ref{eq:OLSweightx}) uses the following lemma.
\begin{lem}
\label{lem:sandwich}Let $X$ be any random variable with mean $\mu_{X}$.
Then, for any real number $x$,
\[
E\left[\ind_{X\ge x}(X-\mu_{X})\right]=\int_{\text{\ensuremath{\underline{x}}}}^{\overline{x}}\int_{\text{\ensuremath{\underline{x}}}}^{\overline{x}}(x_{1}-x_{2})\ind_{x_{2}<x\le x_{1}}dF_{X}(x_{1})dF_{X}(x_{2}).
\]
\end{lem}
\begin{proof}
As in \citet{yitzhaki1996using}, it is a well-known property of the
U-statistics that $Cov\left(Y,X\right)=\frac{1}{2}E\left[(Y_{1}-Y_{2})(X_{1}-X_{2})\right]$,
where $(X_{i},Y_{i})\overset{i.i.d.}{\sim}F_{X,Y}$ for $i=1,2$.
Letting $Y=\ind_{X\ge x}$ in this result yields
\begin{align*}
E\left[\ind_{X\ge x}(X-\mu_{X})\right] & =\frac{1}{2}E\left[\left(\ind_{X_{1}\ge x}-\ind_{X_{2}\ge x}\right)\left(X_{1}-X_{2}\right)\right]\\
 & =E\left[\ind_{X_{1}>X_{2}}\left(\ind_{X_{1}\ge x}-\ind_{X_{2}\ge x}\right)\left(X_{1}-X_{2}\right)\right]\\
 & =\int_{\text{\ensuremath{\underline{x}}}}^{\overline{x}}\int_{\text{\ensuremath{\underline{x}}}}^{\overline{x}}(x_{1}-x_{2})\ind_{x_{2}<x\le x_{1}}dF_{X}(x_{1})dF_{X}(x_{2}).
\end{align*}
This completes the proof.
\end{proof}
Using Lemma \ref{lem:sandwich}, $\overline{\omega}_{X}(x)$ is given
by
\begin{align*}
\overline{\omega}_{X}(x) & =\frac{E\left[\ind_{X\ge x}\widetilde{X}\right]}{E\left[\widetilde{X}^{2}\right]}\\
 & =\frac{E\left[E\left[\ind_{X\ge x}(X-E[X|W])|W\right]\right]}{E\left[\widetilde{X}^{2}\right]}\\
 & =E\left[\widetilde{X}^{2}\right]^{-1}\int\int_{\text{\ensuremath{\underline{x}}}}^{\overline{x}}\int_{\text{\ensuremath{\underline{x}}}}^{\overline{x}}(x_{1}-x_{2})\ind_{x_{2}<x\le x_{1}}dF_{X}(x_{1}|w)dF_{X}(x_{2}|w)dF_{W}(w).
\end{align*}
Note that the second equality uses Assumption \ref{Ass:linear}--(i)
and the law of iterated expectations. The last equality uses Lemma
\ref{lem:sandwich} conditional on $W=w$. This confirms the equation
(\ref{eq:OLSweightx}).

\subsection[Comparing the Weight Function with de Chaisemartin and D'Haultfoeuille
(2020)]{Comparing the Weight Function with \citet{Chaisemartin2020}\label{subsec:Comparing-the-Weight}}

Important special cases of Theorem \ref{Thm:DID_RD} are DID settings
studied by \citet{Chaisemartin2020}, in which $X$ is a binary treatment
($x=0,1$) and a covariate vector consists of time and group dummies
as $W=(d_{1},\ldots,d_{G},D_{1},\ldots,D_{T-1})$. They consider a
sharp DID setting with $X=Z$ and a fuzzy DID setting in which $X\ne Z$.
Using the notation in the main paper, the weight on the covariate
value $W=w$ they suggest can be expressed as $\frac{E\left(X\tilde{Z}|W=w\right)}{E\left(X\tilde{Z}\right)}$.
This weight function differs from $\overline{\omega}_{Z}^{*}(w)=\frac{L_{w}\left(X\widetilde{Z}\right)}{E\left(X\tilde{Z}\right)}$
suggested in Section \ref{subsec:DID_RD}. 

The discrepancy of the weight functions arises because \citet{Chaisemartin2020}
assume a parallel trend only for a potential outcome without treatment
in deriving the weights. The parallel trend assumption is usually
justified by the absence of group--period specific economic shocks
coinciding with the instrument $Z$. Thus, assuming parallel trends
for both the treated and untreated outcomes is no more economically
restrictive than assuming so only for the untreated outcome in many
cases. In fact, \citet{Chaisemartin2020} themselves assume a parallel
trend for a potential outcome with treatment when they propose their
alternative estimator. Note that the treatment effects are still allowed
to be heterogeneous across groups and periods in an additive way. 

Parallel trends for both treated and untreated outcomes imply the
linearity of the treatment effect $g(1,w)-g(0,w)$. Therefore, their
weight function $\frac{E\left(X\tilde{Z}|W=w\right)}{E\left(X\tilde{Z}\right)}$
can be linearly projected onto $W$ without affecting the weighted
average. The linear projection in this case means expressing the weight
function as additive in group dummies $d_{g}$ and time dummies $D_{t}$.
The projected weight function exactly matches the weight function
$\overline{\omega}_{Z}^{*}(w)$ in Section \ref{subsec:DID_RD}.

There are two important exceptions in which it can be reasonable to
use the original weight function in \citet{Chaisemartin2020} without
linearly projecting it. The first exception arises when treatment
effects are dynamic. In this situation, the treatment effect can depend
on the number of periods since the treatment assignment in a nonlinear
way. As a result, the treatment effect (and thus the weights) can
be non-additive in group and period effects, as in \citet{Borusyak2018}.
The second exception can arise in a fuzzy DID setting due to a subtle
effect of unobserved heterogeneity as discussed in Section \ref{subsec:DID_RD_u}.
Due to the unobservable-driven weight difference, the IV-identified
marginal effect $\tau_{IV}(x,w)$ can be nonlinear in $w$ even if
the AME $\tau(x,w)$ is linear in $w$.

\subsection{Relation to the LATE and MTE Interpretations\label{subsec:LATE_MTE}}

Section \ref{subsec:Unobserved-Heterogeneity} explores the weighted-average
interpretation of the IV coefficient without assuming that the treatment
is monotonic in the instrument. This section considers the weighted-average
interpretation under the monotonicity condition, as in \citet{angrist1995two}
and \citet{heckman2006understanding}. Let $X(z)$ be the potential
treatment associated with the instrument level $z$. The observed
treatment is $X=X(Z)$. I make the following set of assumptions, modifying
Assumption \ref{Ass:pot}.

\renewcommand{\theassumptionx}{P$'$}
\begin{assumptionx}

\label{Ass:Pot2} The process of the potential outcome and treatment
$\left(Y(\cdot),X(\cdot)\right)$ satisfies:
\begin{enumerate}
\item[(i)] \textup{(Conditional Independence)} $\left(Y(\cdot),X(\cdot)\right)$
and $Z$ are independent given $W$;
\item[(ii)] \textup{(Differentiability)} $Y'(x)$ exists and the total variation
$V_{x_{0}}^{X}\left(Y(\cdot)\right)=\int_{\min\{x_{0},X\}}^{\max\{x_{0},X\}}\left|Y'(x)\right|dx$
satisfies $E\left[V_{x_{0}}^{X}\left(Y(\cdot)\right)^{2}\right]$
for some $x_{0}\in(\underline{x},\overline{x})$;
\item[(iii)] \textup{(Monotonicity)} $X(z')\ge X(z)$ for any $z'\ge z$.
\end{enumerate}
\end{assumptionx}

Given the monotonicity condition, there exists a stochastic process
$V(\cdot)$ (valued on the extended real line) that satisfies $X\ge x\Leftrightarrow Z\ge V(x)$
for any $x\in\mathbb{R}$. $V(x)$ is an inverse of $X(z)$ defined
such that the treatment $X$ exceeds the level $x$ whenever the instrument
$Z$ exceeds the cutoff $V(x)$. While such $V(x)$ is not unique
if $X(z)$ is not strictly increasing or if $Z$ is discrete, how
$V(x)$ is defined does not influence the following results.
\begin{thm}
\label{thm:Unobs2}With Assumptions \ref{Ass:Pot2}, \ref{Ass:con},
and \ref{Ass:linear}--(i), the IV coefficient $\beta_{IV}$ is given
by
\begin{align*}
\beta_{IV} & =\int_{\text{\ensuremath{\underline{x}}}}^{\overline{x}}\int\tau_{IV}(x,w)\omega_{Z}(x,w)dF_{W}(w)dx,
\end{align*}
\textup{where }the IV-identified marginal effect $\tau_{IV}(x,w)$
at each $(x,w)$ is expressed as

\begin{equation}
\tau_{IV}(x,w)=\int E\left[Y'(x)|V(x)=v,W=w\right]\kappa(v|x,w)dF_{V(x)|W}(v|w).\label{eq:MTE}
\end{equation}
The weight function $\kappa(v|x,w)$ is
\[
\kappa(v|x,w)=\frac{\int\int(z_{1}-z_{2})\ind_{z_{1}\ge v>z_{2}}dF_{Z|W}(z_{1}|w)dF_{Z|W}(z_{2}|w)}{Cov\left(\ind_{X\ge x},Z|W\right)}
\]
and integrates to one: $\int\kappa(v|x,w)dF_{V(x)|W}(v|w)=1.$
\end{thm}
Theorem \ref{thm:Unobs2} is merely a reinterpretation of Theorem
\ref{Thm:Unobs} under the stronger monotonicity condition. The equation
(\ref{eq:MTE}) represents the IV-identified effect $\tau_{IV}(x,w)$
using the unobserved threshold $V(x)$. This representation generalizes
the weighted-average expression based on the marginal treatment effect
(MTE) in \citet{heckman2006understanding} to the continuous treatment
case. 

The equation (\ref{eq:MTE}) can also be expressed using the pairwise
LATE interpretation in \citet{imbens1994identification}, \citet{angrist1995two},
and \citet{angrist2000interpretation}.\footnote{Although the term ``LATE'' is sometimes exclusively reserved for
a binary treatment case (in fact, \citet{angrist1995two} and \citet{angrist2000interpretation}
do not use this term), I use it here for the lack of a better expression.} Rewriting the equation yields
\[
\tau_{IV}(x,w)=\int\int E\left[Y'(x)|X(z_{1})\ge x>X(z_{2}),W=w\right]\phi(z_{1},z_{2}|x,w)dF_{Z|W}(z_{1}|w)dF_{Z}(z_{2}|w),
\]
where the weight on each $(z_{1},z_{2})$ is given by
\[
\phi(z_{1},z_{2}|x,w)=\frac{(z_{1}-z_{2})Pr\left[X(z_{1})\ge x>X(z_{2})|W=w\right]}{Cov\left(\ind_{X\ge x},Z|W=w\right)}.
\]
This expression implies that the IV-identified effect $\tau_{IV}(x,w)$
is the weighted average of pairwise LATE associated with two alternative
instrument assignments, $Z_{1},Z_{2}\overset{i.i.d.}{\sim}F_{Z|W}$.
The weight on each pairwise LATE is proportional to the difference
between two assignments and how likely the difference induces individuals
to cross the treatment level $x$.\footnote{The pairwise LATE interpretation in \citet{imbens1994identification},
\citet{angrist1995two}, and \citet{angrist2000interpretation} use
adjacent pairs to construct the weighted average, which makes their
weighted-average expressions more difficult to interpret.} If the instrument is binary, $\tau_{IV}(x,w)$ simplifies to the
$(x,w)$-specific LATE: $\tau_{IV}(x,w)=E\left[Y'(x)|X(1)\ge x>X(0),W=w\right]$. 

I now present the proof of Theorem \ref{thm:Unobs2}.
\begin{proof}
This proof is given by slightly modifying the proof of Theorem \ref{Thm:Unobs}.
\begin{eqnarray*}
E\left(Y\widetilde{Z}\right) & = & \int_{\text{\ensuremath{\underline{x}}}}^{\overline{x}}E\left[Y'(x)\ind_{X\ge x}\widetilde{Z}\right]dx\\
 & = & \int_{\text{\ensuremath{\underline{x}}}}^{\overline{x}}E\left[Y'(x)\ind_{Z\ge V(x)}\widetilde{Z}\right]dx\\
 & = & \int_{\text{\ensuremath{\underline{x}}}}^{\overline{x}}\int\int E\left[Y'(x)\ind_{Z\ge v}\widetilde{Z}|V(x)=v,W=w\right]dF_{V(x)|W}(v|w)dF_{W}(w)dx\\
 & = & \int_{\text{\ensuremath{\underline{x}}}}^{\overline{x}}\int\int E\left[Y'(x)|V(x)=v,W=w\right]E\left[\ind_{Z\ge v}\widetilde{Z}|W=w\right]dF_{V(x)|W}(v|w)dF_{W}(w)dx.
\end{eqnarray*}
The first equality is shown in the proof of Theorem \ref{Thm:Unobs}.
The second equality follows from the existence of $V(x)$ due to Assumption
\ref{Ass:Pot2}--(iii). The third equality uses the law of iterated
expectations. The fourth equality follows from $V(x)\perp Z|W$ due
to the definition of $V(x)$ and Assumption \ref{Ass:Pot2}--(i).

Using Lemma \ref{lem:sandwich} and Assumption \ref{Ass:linear}--(i),
the weight function is given by
\begin{eqnarray*}
\kappa(v|x,w) & = & \frac{E\left[\ind_{Z\ge v}\widetilde{Z}|W=w\right]}{E\left[\ind_{X\ge x}\tilde{Z}|W=w\right]}\\
 & = & \frac{\int\int(z_{1}-z_{2})\ind_{z_{1}\ge v>z_{2}}dF_{Z|W}(z_{1}|w)dF_{Z|W}(z_{2}|w)}{Cov\left(\ind_{X\ge x},Z|W\right)}.
\end{eqnarray*}

Finally, integrating the weight function $\kappa(v|x,w)$ over $v$
yields
\begin{eqnarray*}
\int\kappa(v|x,w)dF_{V(x)|W}(v|w) & = & \frac{\int E\left[\ind_{Z\ge v}\widetilde{Z}|W=w\right]dF_{V(x)|W}(v|w)}{E\left[\ind_{X\ge x}\tilde{Z}|W=w\right]}\\
 & = & \frac{\int E\left[\ind_{Z\ge v}\widetilde{Z}|W=w,V(x)=v\right]dF_{V(x)|W}(v|w)}{E\left[\ind_{X\ge x}\tilde{Z}|W=w\right]}\\
 & = & \frac{E\left[\ind_{Z\ge V(x)}\widetilde{Z}|W=w\right]}{E\left[\ind_{X\ge x}\tilde{Z}|W=w\right]}\\
 & = & 1,
\end{eqnarray*}
where the second equality follows from $V(x)\perp Z|W$.
\end{proof}

\subsection{Statistical Properties of the Decomposition Estimators\label{subsec:Statistical-Properties-of}}

The following discussion focuses on the statistical properties of
$\widehat{\beta}_{CT}$, since the estimation of $\widehat{\beta}_{C}$
can be considered as its special case. To simplify the exposition,
I use $\widehat{\beta}$ to denote $\widehat{\beta}_{CT}$. I focus
on the parametric case, given the equivalence result in \citet{ackerberg2012practical}. 

Let $(Y_{i},X_{i},Z_{i},W_{i})_{i=1}^{N}$ be an i.i.d. random sample
that satisfies the set of assumptions in Section \ref{sec:Econometric-Framework-for}.
The first step solves the minimization problem
\begin{equation}
\underset{\theta}{\min}\frac{1}{N}\sum_{i=1}^{N}\left(Y_{i}-B_{i}'\theta\right)^{2},\label{eq:step1_LS}
\end{equation}
where $B_{i}'=\left(q_{k1}(W_{i})P_{ik},\ldots,q_{kL^{(k)}}(W_{i})P_{ik}\right)_{k=1}^{K}$
is a vector that stacks the basis functions and $\theta'=\left(\theta_{k1},\ldots,\theta_{kL^{(k)}}\right)_{k=1}^{K}$
is a vector that stacks the parameters. This yields the vector moment
condition 
\begin{equation}
0=\frac{1}{N}\sum_{i=1}^{N}B_{i}\left(Y_{i}-B_{i}'\widehat{\theta}\right).\label{eq:mom1}
\end{equation}
Then, the second step solves for the scalar moment condition $0=\frac{1}{N}\sum_{i=1}^{N}h_{i}(\widehat{\beta},\widehat{\theta},\widehat{\gamma})$
with
\begin{equation}
h_{i}(\widehat{\beta},\widehat{\theta},\widehat{\gamma})=\left(\sum_{k=1}^{K}\sum_{\ell=1}^{L^{(k)}}\widehat{\theta}_{k\ell}q_{k\ell}(W_{i})\left(P_{ik}-W_{i}'\widehat{\gamma}_{P_{k}}\right)-\widehat{\beta}\left(X_{i}-W_{i}'\widehat{\gamma}_{X}\right)\right)\left(Z_{i}-W_{i}'\widehat{\gamma}_{Z}\right),\label{eq:mom2}
\end{equation}
where $\widehat{\gamma}_{R}=\left(\sum_{i=1}^{N}W_{i}W_{i}'\right)^{-1}\left(\sum_{i=1}^{N}W_{i}R_{i}\right)$
represents a regression coefficient of $R_{i}=X_{i},Z_{i},P_{ik}$
on $W_{i}$.

The moment conditions (\ref{eq:mom1}--\ref{eq:mom2}) jointly yield
a stacked method-of-moments problem. Therefore, $\widehat{\beta}$
can be shown to be $\sqrt{N}$-consistent and asymptotically normal
using the standard econometric results. In particular, under the regularity
conditions in \citet[Section 6]{newey1994large}, the asymptotic expansion
yields
\begin{align*}
 & E\left[\frac{\partial}{\partial\beta}h_{i}(\beta,\theta,\gamma)\right]\sqrt{N}\left(\widehat{\beta}-\beta\right)\\
= & \frac{1}{\sqrt{N}}\sum_{i=1}^{N}h_{i}(\beta,\theta,\gamma)+E\left[\frac{\partial}{\partial\theta'}h_{i}(\beta,\theta,\gamma)\right]E\left[B_{i}B_{i}'\right]^{-1}\frac{1}{\sqrt{N}}\sum_{i=1}^{N}B_{i}\left(Y_{i}-B_{i}'\theta\right)\\
 & +E\left[\frac{\partial}{\partial\gamma'}h_{i}(\beta,\theta,\gamma)\right]\sqrt{N}\left(\widehat{\gamma}-\gamma\right)+o_{P}(1).
\end{align*}

Note that 
\[
E\left[\frac{\partial}{\partial\gamma_{X}}h_{i}(\beta,\theta,\gamma)\right]=\beta E\left[W_{i}\widetilde{Z}_{i}\right]=0.
\]
In addition, due to Assumption \ref{Ass:linear}--(i),
\begin{align*}
E\left[\frac{\partial}{\partial\theta_{k\ell}}h_{i}(\beta,\theta,\gamma)\right] & =E\left[q_{k\ell}(W_{i})\widetilde{P}_{ik}\widetilde{Z}_{i}\right]=E\left[q_{k\ell}(W_{i})P_{ik}\widetilde{Z}_{i}\right],\\
E\left[\frac{\partial}{\partial\gamma_{P_{k}}}h_{i}(\beta,\theta,\gamma)\right] & =-E\left[W_{i}\widetilde{Z}_{i}\sum_{\ell=1}^{L^{(k)}}\theta_{k\ell}q_{k\ell}(W_{i})\right]=0.
\end{align*}
Finally,
\[
E\left[\frac{\partial}{\partial\gamma_{Z}}h_{i}(\beta,\theta,\gamma)\right]=-E\left[W_{i}\sum_{k=1}^{K}\alpha_{k}(W_{i})\widetilde{P}_{ik}\right].
\]
Combining these results,
\begin{align}
 & E\left[\widetilde{X}_{i}\widetilde{Z}_{i}\right]\sqrt{N}\left(\widehat{\beta}-\beta\right)\nonumber \\
= & \frac{1}{\sqrt{N}}\sum_{i=1}^{N}\left(Y_{2i}-\beta\widetilde{X}_{i}\right)\widetilde{Z}_{i}+\frac{1}{\sqrt{N}}\sum_{i=1}^{N}(Y_{i}-B_{i}'\theta)B_{i}'E\left[B_{i}B_{i}'\right]^{-1}E\left[B_{i}\widetilde{Z}_{i}\right]\nonumber \\
 & -\frac{1}{\sqrt{N}}\sum_{i=1}^{N}\widetilde{Z}_{i}W_{i}'E\left[\widetilde{W}_{i}\widetilde{W}_{i}'\right]^{-1}E\left[W_{i}Y_{2i}\right]\\
= & \frac{1}{\sqrt{N}}\sum_{i=1}^{N}\left(\widetilde{Y}_{2i}-\beta\widetilde{X}_{i}\right)\widetilde{Z}_{i}+\frac{1}{\sqrt{N}}\sum_{i=1}^{N}(Y_{i}-B_{i}'\theta)B_{i}'E\left[B_{i}B_{i}'\right]^{-1}E\left[B_{i}\widetilde{Z}_{i}\right]+o_{P}(1),\label{eq:Asy_Expansion}
\end{align}
where $Y_{2i}=\sum_{k=1}^{K}\alpha_{k}(W_{i})\widetilde{P}_{ik}$.
This completes the derivation of (\ref{eq:Std_Error}).

With a DID- or RD-type instrument, (\ref{eq:mom2}) is modified as
\[
h_{i}(\widehat{\beta},\widehat{\theta},\widehat{\gamma})=\left(\sum_{k=1}^{K}\left(\sum_{\ell=1}^{L^{(k)}}\widehat{\theta}_{k\ell}W_{i}'\widehat{\gamma}_{q_{k\ell}}\right)P_{ik}-\widehat{\beta}\left(X_{i}-W_{i}'\widehat{\gamma}_{X}\right)\right)\left(Z_{i}-W_{i}'\widehat{\gamma}_{Z}\right),
\]
where $\widehat{\gamma}_{q_{k\ell}}=\left(\sum_{i=1}^{N}W_{i}W_{i}'\right)^{-1}\left(\sum_{i=1}^{N}W_{i}q_{k\ell}(W_{i})\right)$.
For each $k=1,\ldots,K$,
\begin{align*}
 & \sum_{\ell=1}^{L^{(k)}}E\left[\frac{\partial}{\partial\gamma_{q_{k\ell}}}h_{i}(\beta,\theta,\gamma)\right]\sqrt{N}\left(\widehat{\gamma}_{q_{k\ell}}-\gamma_{q_{k\ell}}\right)\\
= & \sum_{\ell=1}^{L^{(k)}}\theta_{k\ell}E\left[P_{ik}W_{i}\widetilde{Z}_{i}\right]E\left[W_{i}W_{i}'\right]^{-1}\frac{1}{\sqrt{N}}\sum_{i=1}^{N}W_{i}\left(q_{k\ell}(W_{i})-W_{i}'\gamma_{q_{k\ell}}\right)\\
= & \frac{1}{\sqrt{N}}\sum_{i=1}^{N}L_{W_{i}}\left(P_{ik}\widetilde{Z}_{i}\right)\left(\alpha_{k}(W_{i})-L_{W_{i}}\left(\alpha_{k}(W_{i})\right)\right).
\end{align*}

Then, the asymptotic expansion (\ref{eq:Asy_Expansion}) should be
modified as
\begin{align*}
 & E\left[\widetilde{X}_{i}\widetilde{Z}_{i}\right]\sqrt{N}\left(\widehat{\beta}-\beta\right)\\
= & \frac{1}{\sqrt{N}}\sum_{i=1}^{N}\left(\widetilde{Y}_{2i}-\beta\widetilde{X}_{i}\right)\widetilde{Z}_{i}+\frac{1}{\sqrt{N}}\sum_{i=1}^{N}(Y_{i}-B_{i}'\theta)B_{i}'E\left[B_{i}B_{i}'\right]^{-1}E\left[B_{i}\widetilde{Z}_{i}\right]\\
 & +\frac{1}{\sqrt{N}}\sum_{i=1}^{N}\sum_{k=1}^{K}L_{W_{i}}\left(P_{ik}\widetilde{Z}_{i}\right)\left(\alpha_{k}(W_{i})-L_{W_{i}}\left(\alpha_{k}(W_{i})\right)\right)+o_{P}(1),
\end{align*}
where $Y_{2i}=\sum_{k=1}^{K}L_{W_{i}}\left(\alpha_{k}(W_{i})\right)P_{ik}$
and $B_{i}'=\left(L_{W_{i}}\left(q_{k1}(W_{i})\right)P_{ik},\ldots,L_{W_{i}}\left(q_{kL^{(k)}}(W_{i})\right)P_{ik}\right)_{k=1}^{K}$.
Note that if $\alpha_{k}(W_{i})$ is restricted to be linear, this
expression is identical to (\ref{eq:Asy_Expansion}) except that $Y_{2i}$
is differently defined.

\subsection{Statistical Properties of the Generalized DWH Test Statistics\label{subsec:Properties-of-GDWH}}

Let $(Y_{i},X_{i},Z_{i},W_{i})_{i=1}^{N}$ be an i.i.d. random sample
that satisfies the set of assumptions in Section \ref{sec:Econometric-Framework-for}.
The numerator of the test statistic (\ref{eq:test_GDWH}) defined
in Section \ref{subsec:Testing-the-Treatment} can be expressed as
\[
\widehat{\tau}=\frac{1}{N}\sum_{i=1}^{N}h_{i}(\widehat{\theta},\widehat{\gamma}),
\]
where

\begin{equation}
h_{i}(\widehat{\theta},\widehat{\gamma})=\left\{ \left(Y_{i}-W_{i}'\widehat{\gamma}_{Y}\right)-\sum_{k=1}^{K}(P_{ik}-W_{i}'\widehat{\gamma}_{P_{k}})\sum_{\ell=1}^{L^{(k)}}\widehat{\theta}_{k\ell}q_{k\ell}(W_{i})\right\} (Z_{i}-W_{i}'\widehat{\gamma}_{Z}).\label{eq:momGDWH}
\end{equation}
The population value of $\widehat{\tau}$ is given by 
\[
\tau=E\left[h_{i}(\theta,\gamma)\right]=E\left[\left(\widetilde{Y}_{i}-Y_{2i}\right)\widetilde{Z}_{i}\right]=E\left[\left(\widetilde{Y}_{i}-\widetilde{Y}_{2i}\right)\widetilde{Z}_{i}\right],
\]
where $Y_{2i}=\sum_{k=1}^{K}\alpha_{k}(W_{i})\widetilde{P}_{ik}$.

To perform the asymptotic expansion, consider derivatives of $E\left[h_{i}(\theta,\gamma)\right]$.
Due to Assumption \ref{Ass:linear}--(i),
\begin{align*}
E\left[\frac{\partial}{\partial\gamma_{P_{k}}}h_{i}(\theta,\gamma)\right] & =E\left[W_{i}\widetilde{Z}_{i}\sum_{\ell=1}^{L^{(k)}}\theta_{k\ell}q_{k\ell}(W_{i})\right]=0\\
E\left[\frac{\partial}{\partial\theta_{k\ell}}h_{i}(\theta,\gamma)\right] & =E\left[q_{k\ell}(W_{i})\widetilde{P}_{ik}\widetilde{Z}_{i}\right]=E\left[q_{k\ell}(W_{i})P_{ik}\widetilde{Z}_{i}\right].
\end{align*}
In addition,
\begin{align*}
E\left[\frac{\partial}{\partial\gamma_{Z}}h_{i}(\theta,\gamma)\right] & =E\left[W_{i}Y_{2i}\right],\\
E\left[\frac{\partial}{\partial\gamma_{Y}}h_{i}(\theta,\gamma)\right] & =0.
\end{align*}
Combining these, the standard asymptotic formula in \citet[Section 6]{newey1994large}
yields
\begin{align}
\sqrt{N}\left(\widehat{\tau}-\tau\right) & =\frac{1}{\sqrt{N}}\sum_{i=1}^{N}\left\{ \left(\widetilde{Y}_{i}-Y_{2i}\right)\widetilde{Z}_{i}-E\left[\left(\widetilde{Y}_{i}-\widetilde{Y}_{2i}\right)\widetilde{Z}_{i}\right]\right\} \nonumber \\
 & -\frac{1}{\sqrt{N}}\sum_{i=1}^{N}(Y_{i}-B_{i}'\theta)B_{i}'E\left[B_{i}B_{i}'\right]^{-1}E\left[B_{i}\widetilde{Z}_{i}\right]\nonumber \\
 & +\frac{1}{\sqrt{N}}\sum_{i=1}^{N}\widetilde{Z}_{i}W_{i}'E\left[W_{i}W_{i}\right]^{-1}E\left[W_{i}Y_{2i}\right]+o_{P}(1)\nonumber \\
= & \frac{1}{\sqrt{N}}\sum_{i=1}^{N}\left(\left(\widetilde{Y}_{i}-\widetilde{Y}_{2i}\right)\widetilde{Z}_{i}-E\left[\left(\widetilde{Y}_{i}-\widetilde{Y}_{2i}\right)\widetilde{Z}_{i}\right]-v_{1i}\widehat{\widetilde{Z}}_{i}\right)+o_{P}(1)\label{eq:Asy_GDWH}
\end{align}
where $v_{1i}=Y_{i}-B_{i}'\theta$, and $\widehat{\widetilde{Z}}_{i}=B_{i}'E\left[B_{i}B_{i}'\right]^{-1}E\left[B_{i}\widetilde{Z}_{i}\right]$.
Then, the central limit theorem implies $\sqrt{N}\left(\widehat{\tau}-\tau\right)\overset{d}{\to}N\left(0,\Sigma\right)$,
where
\[
\Sigma=E\left[\left(\left(\widetilde{Y}_{i}-\widetilde{Y}_{2i}\right)\widetilde{Z}_{i}-E\left[\left(\widetilde{Y}_{i}-\widetilde{Y}_{2i}\right)\widetilde{Z}_{i}\right]-v_{1i}\widehat{\widetilde{Z}}_{i}\right)^{2}\right].
\]
In addition,
\[
N\widehat{S}^{2}=\frac{1}{N}\sum_{i=1}^{N}\left(\left(\widetilde{Y}_{i}-\widetilde{Y}_{2i}\right)\widetilde{Z}_{i}-\frac{1}{N}\sum_{j=1}^{N}\left(\widetilde{Y}_{j}-\widetilde{Y}_{2j}\right)\widetilde{Z}_{j}-v_{1i}\widehat{\widetilde{Z}}_{i}\right)^{2}\overset{p}{\to}\Sigma.
\]

Therefore, the test statistic $\widehat{T}=\widehat{\tau}/\widehat{S}$
converges in distribution to $N(0,1)$ if $\tau=0$ and diverges if
$\tau\ne0$. Since $\tau=E\left[\widetilde{X}_{i}\widetilde{Z}_{i}\right]\cdot\Delta_{ME}$
and $E\left[\widetilde{X}_{i}\widetilde{Z}_{i}\right]\ne0$, $\widehat{T}$
converges in distribution to $N(0,1)$ if $\Delta_{ME}=0$ and diverges
if $\Delta_{ME}\ne0$.\footnote{Since a feasible estimator for $\widehat{S}$ would be constructed
using $Y_{2i}=\sum_{k=1}^{K}\widehat{\alpha}_{k}(W_{i})\widetilde{P}_{ik}$
instead of $Y_{2i}=\sum_{k=1}^{K}\alpha_{k}(W_{i})\widetilde{P}_{ik}$,
an additional regularity condition such as $\sup_{w}|\widehat{\alpha}_{k}(w)-\alpha_{k}(w)|\overset{p}{\to}0$
would be required in practice for the convergence $N\widehat{S}^{2}\overset{p}{\to}\Sigma$.}

With a DID- or RD-type instrument, (\ref{eq:momGDWH}) is modified
as
\[
h_{i}(\widehat{\theta},\widehat{\gamma})=\left\{ (Y_{i}-W_{i}'\widehat{\gamma}_{Y})-\sum_{k=1}^{K}P_{ik}\sum_{\ell=1}^{L^{(k)}}\widehat{\theta}_{k\ell}W_{i}'\widehat{\gamma}_{q_{k\ell}}\right\} (Z_{i}-W_{i}'\widehat{\gamma}_{Z}).
\]
For each $k=1,\ldots,K$,
\begin{align*}
 & \sum_{\ell=1}^{L^{(k)}}E\left[\frac{\partial}{\partial\gamma_{q_{k\ell}}}h_{i}(\theta,\gamma)\right]\sqrt{N}\left(\widehat{\gamma}_{q_{k\ell}}-\gamma_{q_{k\ell}}\right)\\
=- & \sum_{\ell=1}^{L^{(k)}}\theta_{k\ell}E\left[P_{ik}W_{i}\widetilde{Z}_{i}\right]E\left[W_{i}W_{i}'\right]^{-1}\frac{1}{\sqrt{N}}\sum_{i=1}^{N}W_{i}\left(q_{k\ell}(W_{i})-W_{i}'\gamma_{q_{k\ell}}\right)\\
=- & \frac{1}{\sqrt{N}}\sum_{i=1}^{N}L_{W_{i}}\left(P_{ik}\widetilde{Z}_{i}\right)\left(\alpha_{k}(W_{i})-L_{W_{i}}\left(\alpha_{k}(W_{i})\right)\right).
\end{align*}
Then the asymptotic expansion in (\ref{eq:Asy_GDWH}) is modified
as

\begin{align*}
\sqrt{N}\left(\widehat{\tau}-\tau\right) & =\frac{1}{\sqrt{N}}\sum_{i=1}^{N}\left(\left(\widetilde{Y}_{i}-\widetilde{Y}_{2i}\right)\widetilde{Z}_{i}-E\left[\left(\widetilde{Y}_{i}-\widetilde{Y}_{2i}\right)\widetilde{Z}_{i}\right]-v_{1i}\widehat{\widetilde{Z}}_{i}\right)\\
 & -\frac{1}{\sqrt{N}}\sum_{i=1}^{N}\sum_{k=1}^{K}L_{W_{i}}\left(P_{ik}\widetilde{Z}_{i}\right)\left(\alpha_{k}(W_{i})-L_{W_{i}}\left(\alpha_{k}(W_{i})\right)\right)+o_{P}(1),
\end{align*}
where $Y_{2i}=\sum_{k=1}^{K}\alpha_{k}(W_{i})P_{ik}$, $B_{i}'=\left(L_{W_{i}}\left(q_{k1}(W_{i})\right)P_{ik},\ldots,L_{W_{i}}\left(q_{kL^{(k)}}(W_{i})\right)P_{ik}\right)_{k=1}^{K}$,
$v_{1i}=Y_{i}-B_{i}'\theta$, and $\widehat{\widetilde{Z}}_{i}=B_{i}'E\left[B_{i}B_{i}'\right]^{-1}E\left[B_{i}\widetilde{Z}_{i}\right]$.
If $\alpha_{k}(W_{i})$ is restricted to be linear, then this expression
is identical to (\ref{eq:Asy_GDWH}) except for the definition of
$Y_{2i}$.

\subsection{Monte Carlo Analysis\label{subsec:Monte-Carlo-Analysis}}

This section provides a Monte Carlo analysis of the decomposition
estimators and the generalized DWH test proposed proposed in the main
paper. I consider a setting that closely follows a Monte Carlo analysis
in \citet{lochner2015estimating}, which originally draws on the schooling
model in \citet{card1995using}. An individual $i$ chooses a schooling
level $x\in\{0,1,\ldots,20\}$ to maximize $Y_{i}(x)-C_{i}(x)$, where
$Y_{i}(x)$ represents potential log earnings associated with the
schooling level $x$ and $C_{i}(x)$ represents the cost to attain
$x$ years of education. Each individual has an observed characteristics
$W_{i}$ and a cost shifter $Z_{i}$. $W_{i}$ and $Z_{i}$ are mutually
independent and each of them takes $0$ or $1$ with equal probability. 

The potential earnings are given by
\[
Y_{i}(x)=b_{i}x+\kappa\ind_{x\ge12}+\varepsilon_{i}.
\]
The parameter $\kappa$ represents the ``sheepskin effect'' of high
school graduation. While \citet{lochner2015estimating} set the coefficient
$b_{i}=0.04$ for all individuals, I allow it to be heterogeneous.
The schooling costs are given by
\[
C_{i}(x)=\eta_{i}x+d_{i}Z_{i}(x-12)\ind_{x>12}+\frac{\gamma}{2}x^{2}+\kappa\ind_{x\ge12}+(b_{i}-E[b_{i}])x.
\]
Following \citet{lochner2015estimating}, I include $\kappa\ind_{x\ge12}$
in the cost function to ensure that the nonlinearity of the earnings
function does not influence the schooling decision. In a similar spirit,
I include $(b_{i}-E[b_{i}])x$ to ensure that the schooling decision
does not vary across different degrees of heterogeneity in $b_{i}$.
This specification allows the exploration of different scenarios without
changing the distribution of schooling and the IV and OLS weights.
The coefficient $d_{i}$ represents the sensitivity of the schooling
cost to the cost shifter $Z_{i}$. While \citet{lochner2015estimating}
set $d_{i}=0.01$ for everyone and assume that the cost shifter $Z_{i}$
influences all schooling margins, I allow $d_{i}$ to be heterogeneous
and assume that the cost shifter affects college education margins
$(x>12$) only. $(\varepsilon_{i},\eta_{i})$ follows a bivariate
normal distribution independently from $(W_{i},Z_{i})$, with zero
means and a variance matrix
\[
\left(\begin{array}{cc}
\sigma_{\varepsilon}^{2} & \rho_{\varepsilon\eta}\sigma_{\varepsilon}\sigma_{\eta}\\
\rho_{\varepsilon\eta} & \sigma_{\eta}^{2}
\end{array}\right).
\]
Since $\rho_{\varepsilon\eta}$ governs the correlation between the
potential earnings and the schooling cost, $\rho_{\varepsilon\eta}\ne0$
gives rise to endogeneity bias of the OLS coefficient. Conditional
on $W_{i}$, $(b_{i},d_{i})$ is independent from $(Z_{i},\varepsilon_{i},\eta_{i})$
and follows a bivariate normal distribution 
\[
\left[\begin{array}{c}
b_{i}\\
\ln d_{i}
\end{array}\right]|W_{i}\sim Normal\left(\left[\begin{array}{c}
\mu_{b}+\delta_{b}(W_{i}-\frac{1}{2})\\
\ln\mu_{d}+\delta_{d}(W_{i}-\frac{1}{2})
\end{array}\right],\left[\begin{array}{cc}
\sigma_{b}^{2} & \rho_{bd}\sigma_{b}\sigma_{d}\\
\rho_{bd}\sigma_{b}\sigma_{d} & \sigma_{d}^{2}
\end{array}\right]\right),
\]
where $d_{i}$ is log-transformed to ensure a positive response to
the cost shifter. 

Following \citet{lochner2015estimating}, I set $\mu_{b}=0.04$, $\mu_{d}=0.01$,
$\gamma=0.003$, $\sigma_{\varepsilon}=0.5$, and $\sigma_{\eta}=0.01$.\footnote{Since the cost function specification slightly differs from \citet{lochner2015estimating},
I set the parameter $\sigma_{\eta}$ higher than $\sigma_{\eta}=0.007$
in \citet{lochner2015estimating} to retain a reasonable schooling
distribution. Years of schooling in the simulated sample has the mean
$12.6$ and the standard deviation $2.9$, which are comparable to
the values in the 1980 U.S. Census data.} In addition, I set $\delta_{d}=1$ and $\sigma_{d}=0.5$ to introduce
observed and unobserved heterogeneity in the sensitivity coefficient
$d_{i}$. Given the cost function specification, the IV weights are
expected to be higher for a group with $W_{i}=1$ and college education
margins ($x>12$). For other parameters, I consider five cases below.
\begin{description}
\item [{Case\,1:}] The earnings function is linear: $\kappa=0$. The schooling
coefficient $b_{i}$ is homogeneous in both observed and unobserved
dimensions: $(\delta_{b},\sigma_{b})=(0,0)$.
\item [{Case\,2:}] The earnings function is nonlinear: $\kappa=0.1$.
The schooling coefficient $b_{i}$ is homogeneous in both observed
and unobserved dimensions: $(\delta_{b},\sigma_{b})=(0,0)$.
\item [{Case\,3:}] The earnings function is nonlinear: $\kappa=0.1$.
The schooling coefficient $b_{i}$ is heterogeneous only in an observed
dimension: $(\delta_{b},\sigma_{b})=(-0.04,0)$.
\item [{Case\,4:}] The earnings function is nonlinear: $\kappa=0.1$.
The schooling coefficient $b_{i}$ is heterogeneous in both observed
and unobserved dimensions: $(\delta_{b},\sigma_{b})=(-0.04,0.02)$.
However, these is no unobservable-driven correlation between $b_{i}$
and the sensitivity $d_{i}$ to the instrument: $\rho_{bd}=0$. 
\item [{Case\,5:}] The earnings function is nonlinear: $\kappa=0.1$.
The schooling coefficient $b_{i}$ is heterogeneous in both observed
and unobserved dimensions: $(\delta_{b},\sigma_{b})=(-0.04,0.02)$.
These is unobservable-driven correlation between $b_{i}$ and $d_{i}$:
$\rho_{bd}=-0.25$. 
\end{description}
I explore three different levels of endogeneity, $\rho_{\varepsilon\eta}\in\{0,0.1,0.2\}$,
for each of the five cases.\footnote{$\rho_{\varepsilon\eta}>0$ gives rise to downward endogeneity bias
of the OLS coefficient, which is opposite to the expected direction
of endogeneity bias in the return-to-schooling context. Nevertheless,
I consider a positive $\rho_{\varepsilon\eta}$ following the setting
in \citet{lochner2015estimating}.} 

$\kappa>0$ in Cases 2--5 implies that the marginal return to the
12th year of education is higher than the marginal return to college
education. Given that the instrument $Z_{i}$ influences college education
costs, it is expected that the treatment-level weight difference pushes
down the IV coefficient. $\delta_{b}<0$ in Cases 3--5 implies that
individuals with $W_{i}=1$, who are more sensitive to the instrument,
have smaller returns to schooling. Therefore, it is expected that
the covariate weight difference pushes down the IV coefficient. $\rho_{bd}<0$
in Case 5 implies that there is an unobservable factor that increases
sensitivity to the instrument and decreases returns to schooling.
This is expected to further push down the IV coefficient by making
the IV-identified marginal effect smaller than the AME.

For each possible combination of parameters, I perform 1,000 Monte
Carlo simulations with the sample size $N=$ 5,000. In each simulation,
I estimate the OLS and IV coefficients and the decomposition estimators.
I perform three endogeneity tests below.
\begin{enumerate}
\item A standard DWH test, which examines the significance of the raw IV--OLS
gap $\widehat{\beta}_{IV}-\widehat{\beta}_{OLS}$. I use the heteroskedasticity-robust
standard errors unlike in the classic version of the test.
\item A generalized DWH test proposed by \citet{lochner2015estimating}.
\item A generalized DWH test proposed in the main paper.
\end{enumerate}
For expositional convenience, I refer to the first one as the standard
DWH test, the second one as the DWH--LM test, and the third one as
the DWH--I test.

Table \ref{Table:MC} and \ref{Table:MC2} report Monte Carlo simulation
results. For each model, the tables report the means and the standard
deviations of the point estimates of the OLS coefficient, the IV coefficient,
the IV--OLS gap, and three components of the decomposition. In addition,
the table reports the fraction of simulation draws in which each of
the three endogeneity tests rejects the null at a 5\% significance
level. The fraction of rejection of an asymptotically unbiased test
is expected to be close to 5\% with $\rho_{\varepsilon\eta}=0$ and
larger than 5\% with $\rho_{\varepsilon\eta}\ne0$.

Since Case 1 boils down to a classic linear model, all three versions
of the DWH test are expected to be asymptotically unbiased. The first
panel of Table \ref{Table:MC} confirms this expectation. The standard
DWH test has a slightly smaller fraction of Type II errors than the
other two tests. Even though the true earnings function is linear,
endogeneity of schooling $X_{i}$ makes the conditional mean function
$m(x,w)$ nonlinear. As a result, the DWH--LM test and the DWH--I
test attribute a part of the endogeneity bias to the treatment-level
weight difference.\footnote{While the spurious nonlinearity of the conditional mean function works
in favor of the standard DWH test and against the other two tests
in this particular case, it can work in the opposite way depending
on the IV weight patterns.} On the other hand, I find no sizable difference in performance across
the DWH--LM test and the DWH--I test.

Case 2 corresponds to a model considered in \citet{lochner2015estimating}.
The true relationship between earnings and schooling is nonlinear
but homogeneous. The second panel of Table \ref{Table:MC} confirms
that the standard DWH test is no longer valid due to nonlinearity.
On the other hand, the other two tests perform well, with similar
fractions of Type I and Type II errors.

Case 3 corresponds to a separable model considered in Section 2 of
the main paper. The true relationship between earnings and schooling
is nonlinear and observably heterogeneous. While the DWH--I test
is expected to work, the other two tests are expected to fail due
to heterogeneity. In this setting, the marginal return to $x$th year
of education is $0.02W_{i}+0.06(1-W_{i})+0.1\ind_{x=12}$ and the
IV weights are concentrated on a group with $W_{i}=1$ and college
education margins ($x>12$). As a result, both the covariate weight
difference and the treatment-level weight difference negatively contributes
to the IV--OLS gap. As confirmed by the third panel of Table \ref{Table:MC},
only the DWH--I test performs well as it successfully accounts for
both types of the weight difference contributions. 

Case 4 and Case 5 demonstrate that the decomposition framework in
the main paper provides relevant information even when the separability
assumption does not hold. Both of these cases fit into a nonseparable
model considered in Section \ref{subsec:Unobserved-Heterogeneity}
of the main paper. The true relationship between earnings and schooling
is nonlinear and heterogeneous in both observable and unobservable
dimensions. Unobservable-driven correlation between the treatment
effect and the treatment sensitivity to the instrument is absent in
Case 4 but present in Case 5. The DWH--I test is expected to still
work in Case 4, since there is no discrepancy between the IV-identified
marginal effect and the AME in this case. This expectation is confirmed
by the first panel of Table \ref{Table:MC2}. On the other hand, the
DWH--I test is no longer valid in Case 5, as confirmed by the second
panel of Table \ref{Table:MC2}. For example, even with $\rho_{\varepsilon\eta}=0$,
the estimated marginal effect difference component $\widehat{\Delta}_{ME}$
is biased downward from $0$. This is because $\widehat{\Delta}_{ME}$
captures the negative gap between the IV-identified marginal effect
and the AME, which arises from the negative correlation between $b_{i}$
and $d_{i}$. Nevertheless, $\widehat{\Delta}_{ME}$ is closer to
zero than the raw IV--OLS gap, as it still accounts for the covariate
weight difference and the treatment-level weight difference components.

\section{Empirical Details and Additional Results\label{sec:Data}}

This section provides additional information and results for empirical
applications in the main paper.

\subsection{College Cost Instrument with Geographic Variation\label{subsec:NLSY79}}

Table \ref{Table:NLSY_Selection} reports the sample selection criteria
for the analysis sample in Section \ref{subsec:College IV} in detail.
Most individuals have hourly wages reported at least once, which makes
sample selection bias associated with labor force participation less
of a concern. 

Table \ref{Table:NLSY_Descriptive} presents the descriptive statistics
of the analysis sample. As the wage measure, I use the log hourly
wage at the current or most recent job, deflated using CPI--U (with
the base year of 1982--84), bottom-coded at \$1, and top-coded at
\$100. Years of education is based on the highest grade completed
as of May 1 of the survey year (HGCREV), top-coded at 18. I use the
age-adjusted Armed Forces Qualification Test (AFQT) percentile (AFQT-3)
as the AFQT score measure. In the regressions, I use piecewise linear
functions of AFQT and age with two kink points to ensure the flexible
functional form. In particular, AFQT, $\max\left\{ \text{AFQT}-\frac{1}{3},0\right\} $,
$\max\left\{ \text{AFQT}-\frac{2}{3},0\right\} $, age, $\max\left\{ \text{age}-35,0\right\} $,
and $\max\left\{ \text{age}-45,0\right\} $ are included in a vector
of covariates. Parental education dummies in the covariate vector
follow the categories defined in Table \ref{Table:NLSY_Descriptive}.
Local earnings are based on the annual total earnings divided by the
total employment, taken from the county-level economic accounts by
the Bureau of Economic Analysis. Local unemployment rates are based
on the state-level annual unemployment rates from the Bureau of Labor
Statistics (BLS).\footnote{Because the BLS website only provides the data from 1976, I take the
1974--75 data from the 1977 Statistical Abstract of the United States.} Information about public four-year colleges is based on the Higher
Education General Information Survey (HEGIS) in 1977. Following \citet{carneiro2011estimating},
I use the HEGIS data compiled by \citet{kling2001interpreting} to
construct the college variables. The urban status of each county is
based on whether the majority of the county's population lives in
an urbanized area.\footnote{Urbanized areas are defined by the U.S. Census Bureau based on the
1990 Census.} 

Table \ref{Table: NLSY_FS} reports the first-stage estimates. Local
college availability and local tuition rate coefficients have the
expected signs: smaller pecuniary or psychological costs of college
attendance result in longer years of schooling. Local earnings and
unemployment coefficients also have the expected signs: worse local
labor market conditions reduce opportunity costs of college attendance
and thus increase years of schooling.

Table \ref{Table: College_decom_alt} explores the IV--OLS coefficient
gap and its decomposition in alternative specifications. It is common
across most specifications that the weight difference components positively
contribute to the IV--OLS gap. However, it is notable that the weight
difference contributions are estimated to be negative for the female
sample and the sample of blacks and Hispanics.

Table \ref{Table:coll_wa} explores the mechanisms underlying the
difference in the decomposition results across different sample specifications.
Panel (A) of the table shows the OLS weights on AFQT and parental
education groups, high school margins of schooling, and college margins
of schooling, for different samples. Panel (B) of the table presents
the IV weights analogously. Panel (C) presents the OLS schooling coefficient
given by a regression restricted to each covariate group or each schooling
margin. I find no substantial difference in weight patterns across
the male and female samples, although point estimates of the IV weights
among males are somewhat more concentrated on advantaged groups and
college education margins. The male sample has much larger disparity
in OLS-identified returns to schooling across personal backgrounds
and across schooling levels. Among blacks and Hispanics, the IV weights
are not particularly concentrated on advantaged groups. In addition,
their OLS-identified returns to schooling differs less across high
school and college margins, compared with non-black, non-Hispanic
individuals.

\subsection{RD-Based Compulsory Schooling Instrument\label{subsec:GHS}}

In Section \ref{subsec:CSL_UK}, I use the British General Household
Survey (GHS) data provided by \citet{oreopoulos2008estimating}, which
makes several corrections to the data used in \citet{oreopoulos2006estimating}.\footnote{In addition to the corrections made by \citet{oreopoulos2008estimating},
I correct a small number of observations with erratic reported ages.
In particular, I compare the reported age with the difference between
survey and birth years, and replace the former with the latter if
two age measures differ by more than one year.} Table \ref{Table:GHS_Descriptive} presents the descriptive statistics
from British and Northern Irish GHS data. While my main analysis focuses
on the British GHS in due to its much larger sample size, \citet{oreopoulos2006estimating}
estimates the returns to schooling in Northern Ireland in addition
to those in Britain.

Table \ref{Table:NI_decom} explores several alternative specifications.
The first row of the table reproduces the main result in Table \ref{Table:decom}.
The second row presents the decomposition result with no top-coding
of years schooling, following the original specification in \citet{oreopoulos2006estimating}.
The third and fourth rows show the decomposition results using the
Northern Irish GHS with and without top-coding of the schooling levels.
Because Northern Ireland raised the minimum school-leaving age in
1957, the instrument is an indicator for 1943 or later birth cohorts,
who turned age 14 in 1957 or later. All the other variables are defined
in the same manner as in the British GHS. Across all four specifications,
the weight difference components are estimated to negatively contribute
to the IV--OLS coefficient gap.

Table \ref{Table:NI_wgt_w} presents the estimated weights on covariate
groups in the Northern Irish GHS. The IV weights are concentrated
on individuals turning age 14 around the reform year, which is 10
years later than the British sample. Because the cohorts influenced
by the reform have similar levels of OLS-identified returns to schooling
to the other cohorts, the weight difference across cohorts are expected
to contribute little to the IV--OLS coefficient gap. This expectation
is consistent with a small impact of the covariate weight difference
in Table \ref{Table:NI_decom}.

\begin{figure}[H]
\caption{The IV and OLS Weights on the Treatment Levels (Northern Irish GHS)}

\label{Fig:UK_wgt_x_NI}

\vspace{1.0em}
\centering
\begin{threeparttable}

\includegraphics[width=0.7\columnwidth]{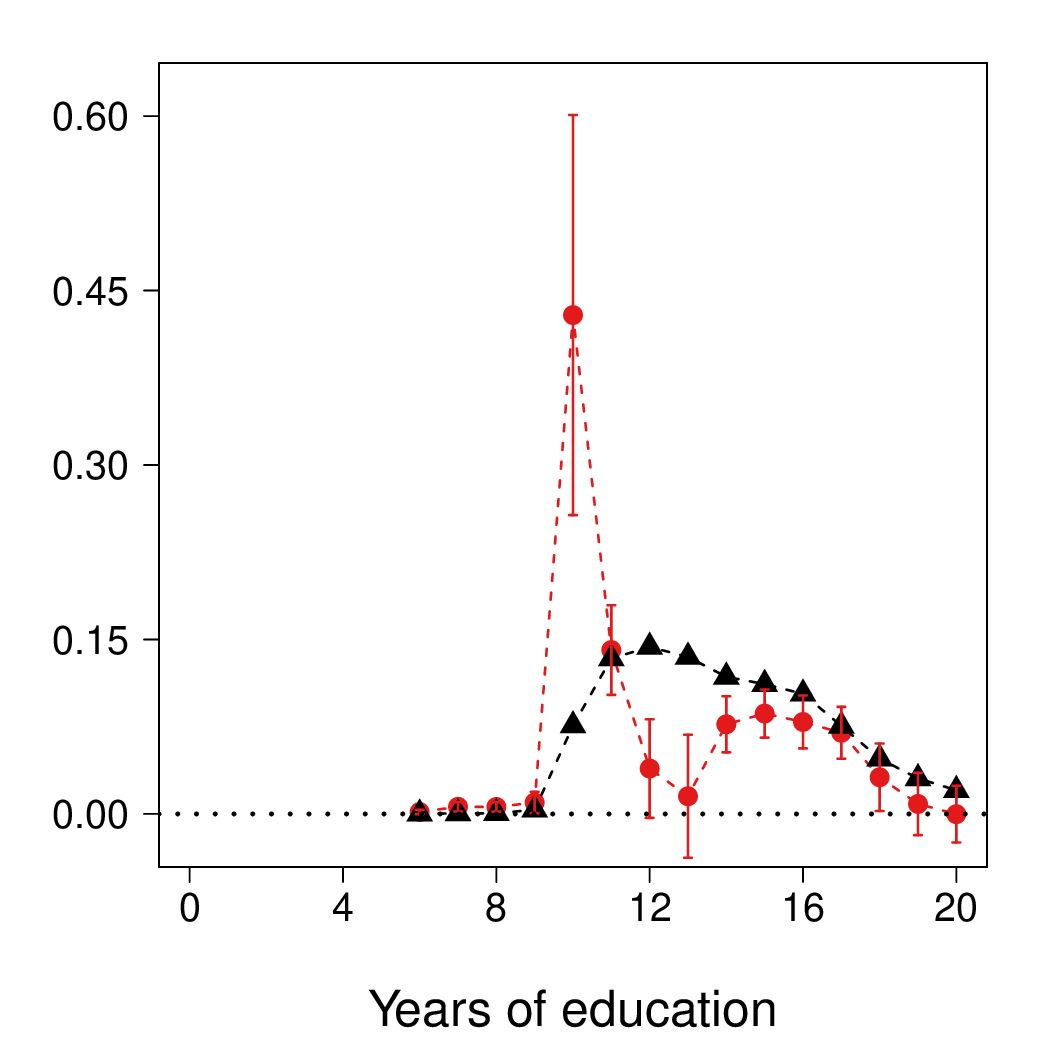}
\begin{centering}
\includegraphics[bb=0bp 464bp 504bp 504bp,width=0.85\columnwidth]{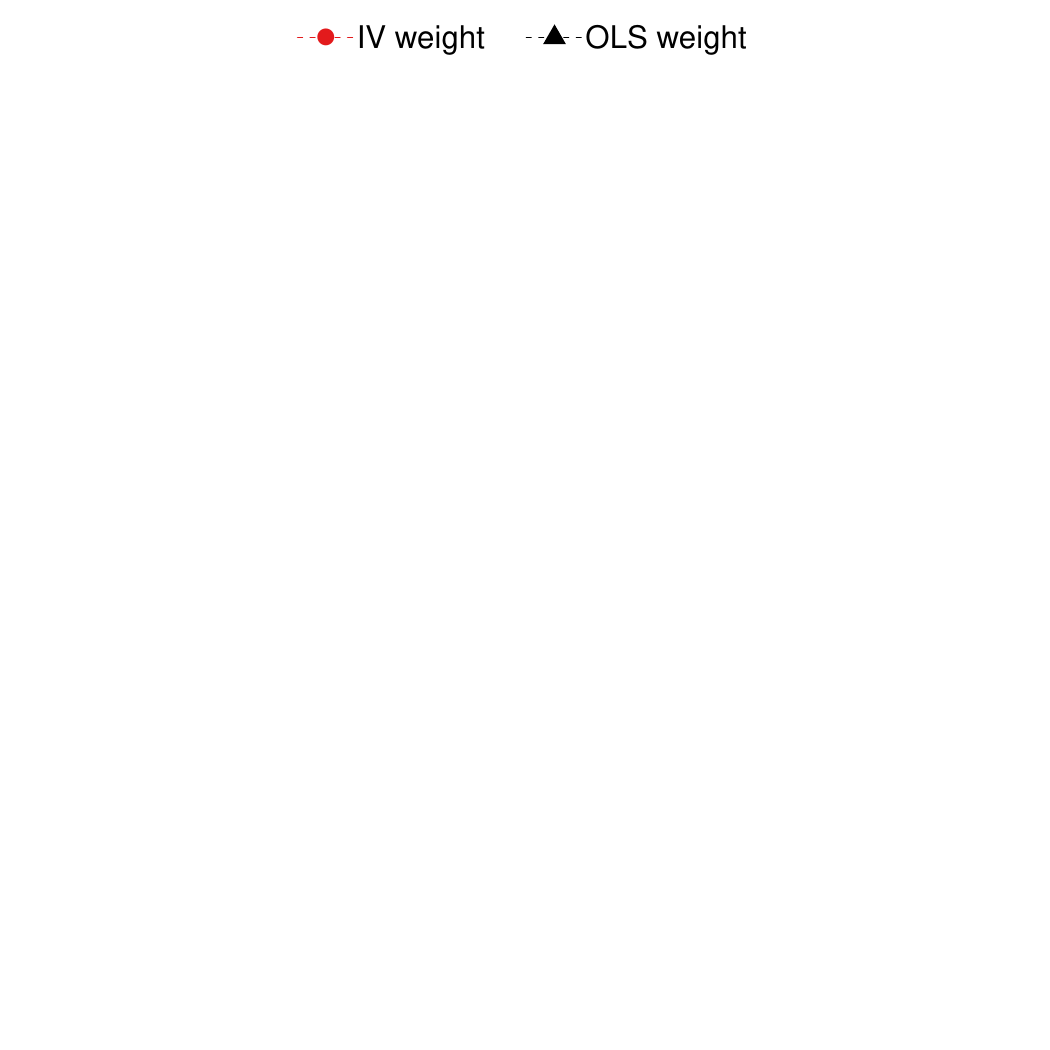}
\par\end{centering}
\begin{tablenotes}
\footnotesize

\item Notes: The IV weights are presented with standard error bars.
The standard error bars for the OLS weights are omitted because they
are graphically negligible. Each set of weights sums to one across
the whole sample. Section \ref{subsec:Estimation_Detail} describes
the empirical specification for estimating the weights.

\end{tablenotes}
\end{threeparttable}
\end{figure}

Figure \ref{Fig:UK_wgt_x_NI} illustrates the patterns of OLS and
IV weights on schooling levels. Unlike in the British sample, the
IV weights are not entirely concentrated on the 10th year of schooling.
While the IV weights are not precisely estimated due to a small sample
size, it is possible that cohorts turning 14 around the reform year
were exposed to other concurrent institutional or economic shocks
to schooling levels.

\subsection{Compulsory Schooling Instrument with DID Variation\label{subsec:Census}}

I use the 1960--80 U.S. Census data compiled by IPUMS \citep{ruggles2020ipums}
in Section \ref{subsec:CSL}. Following \citet{acemoglu2000large}
, I use the 1\% sample from the 1960 Census, a pooled 2\% sample from
the 1970 Census by combining the 1\% State Form 1 and Form 2 samples,
and the 5\% sample from the 1980 Census. Due to the inconsistency
of the sampling rates across Census years, I weight observations from
the 1960 Census by $5$ and observations from the 1970 Census by $2.5$,
relative to observations from the 1980 Census. I use the compulsory
schooling laws (CSL) data provided by \citet{acemoglu2000large}.
Due to missing CSL status for Alaska and Hawaii in their data, I drop
individuals born in these two states from the analysis sample.

\citet{acemoglu2000large} use the 1960--80 U.S. Census data in their
main specification, which my main analysis follows. As they use the
1950 and 1990 Census data in some other specifications, they modify
several variables in the 1960--80 Census data to maintain consistency
across all Census years. I do not implement these modifications because
they are unnecessary for analyzing the 1960--80 Census data alone.
There are three main differences between my specification and theirs
for this reason. First, I top-code years of schooling at 18 because
the 1960--70 Censuses report years of schooling with a cap at 18
(the cap is 20 years in the 1980 Census). On the other hand, \citet{acemoglu2000large}
use 17 years as a common cap to the 1950 Census. Second, \citet{acemoglu2000large}
impute year of birth, which is unavailable in the 1950 Census data,
from age as of the Census day and quarter of birth even in other Census
years. I do not follow this approach because year of birth is always
available in the 1960--80 Census data. Finally, in all Census years,
they determine the CSL status from age as of Census day and state
of birth. I determine the CSL status from year of birth and state
of birth to ensure that individuals in the same cohort and the same
state are assigned with the same CSL status.

Table \ref{Table:Census_Descriptive} presents descriptive statistics
for each Census year. I compute weekly earnings by dividing wage and
salary income by weeks worked in the previous year. Because weeks
worked are given in intervals in the 1960--70 Censuses, I use group
means in the 1980 Census to represent the intervals. I bottom-code
weekly earnings at \$20 for the 1960 sample, at \$26 for the 1970
sample, and at \$58 for the 1980 sample, which respectively correspond
to 20 times the federal minimum wage in 1959, 1969, and 1979. Income
is reported with a top-coding at \$25,000, \$50,000, and \$75,000
in each decennial Census. Thus, I impose a top-coding on weekly earnings
at \$25,000/52, \$50,000/52, and \$75,000/52 for each Census year.
Table \ref{Table: CSL_FS} provides the first-stage coefficients of
the CSL instruments. Patterns of the estimated coefficients are consistent
with the expectation that more stringent CSL requirements result in
longer years of schooling.

Table \ref{Table:QOB_decom} presents the decomposition results with
alternative compulsory schooling instruments. The first row of the
table reproduces the main result in Table \ref{Table:decom} that
uses required years of schooling implied by child labor laws. The
second row presents the result with required years of schooling implied
by compulsory attendance laws.\footnote{The definition of compulsory attendance laws follows \citet{acemoglu2000large}.
In computing the required number of years, I use the corrected formula
suggested by \citet{goldin2011} and \citet{Stephens2014}, which
replaces the max function in the original formula in \citet{acemoglu2000large}
with the min function.} The third row shows the result with required schooling years jointly
implied by child labor and compulsory attendance laws, as compiled
by \citet{Stephens2014}. The fourth row shows the result using quarter-of-birth
dummies interacted with dummies for the Census years as the instruments.
The weight difference components positively contribute to the IV--OLS
gap in the decomposition results with compulsory attendance laws or
required schooling years, similarly to the result with child labor
laws. On the other hand, with quarter-of-birth instruments, the IV
and OLS estimates are nearly identical and the weight difference components
are close to zero. 

Table \ref{Table:QOBL_weights_W} shows the estimated OLS and IV weights
on covariate groups. The patterns of the compulsory attendance and
required schooling IV weights are similar to that of the child labor
IV weights presented in Table \ref{CSL_weights_W} in the main paper.
On the other hand, the quarter-of-birth IV weights do not substantially
differ from the OLS weights.

\begin{figure}[H]
\caption{The IV and OLS Weights on the Treatment Levels with Alternative Instruments
(U.S. Census)}

\label{Fig:QOB_weights_X}

\vspace{2.0em}
\centering
\begin{threeparttable}

\begin{minipage}[t]{0.48\columnwidth}%
(a) Compulsory Attendance Laws

\includegraphics[width=1\columnwidth]{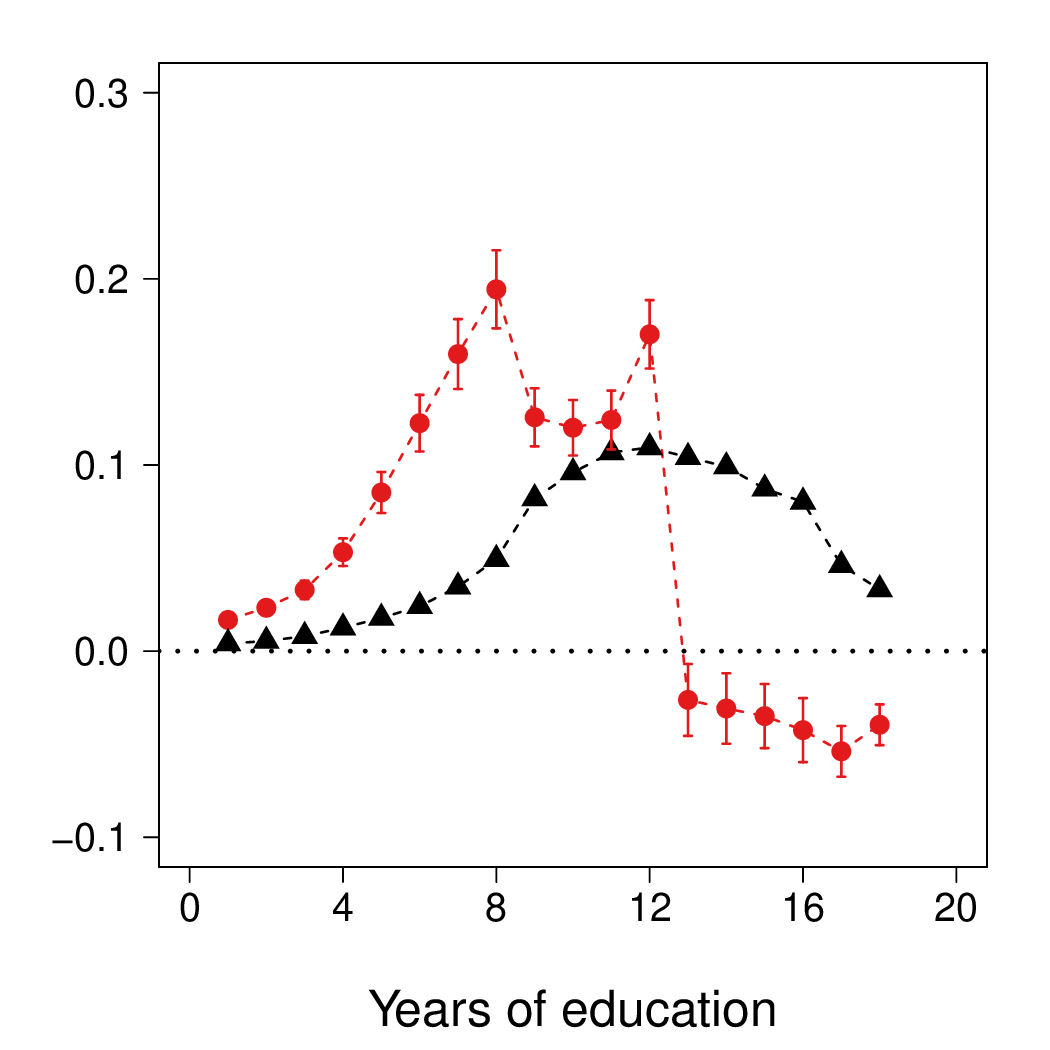}%
\end{minipage}%
\begin{minipage}[t]{0.48\columnwidth}%
(b) Required Schooling

\includegraphics[width=1\columnwidth]{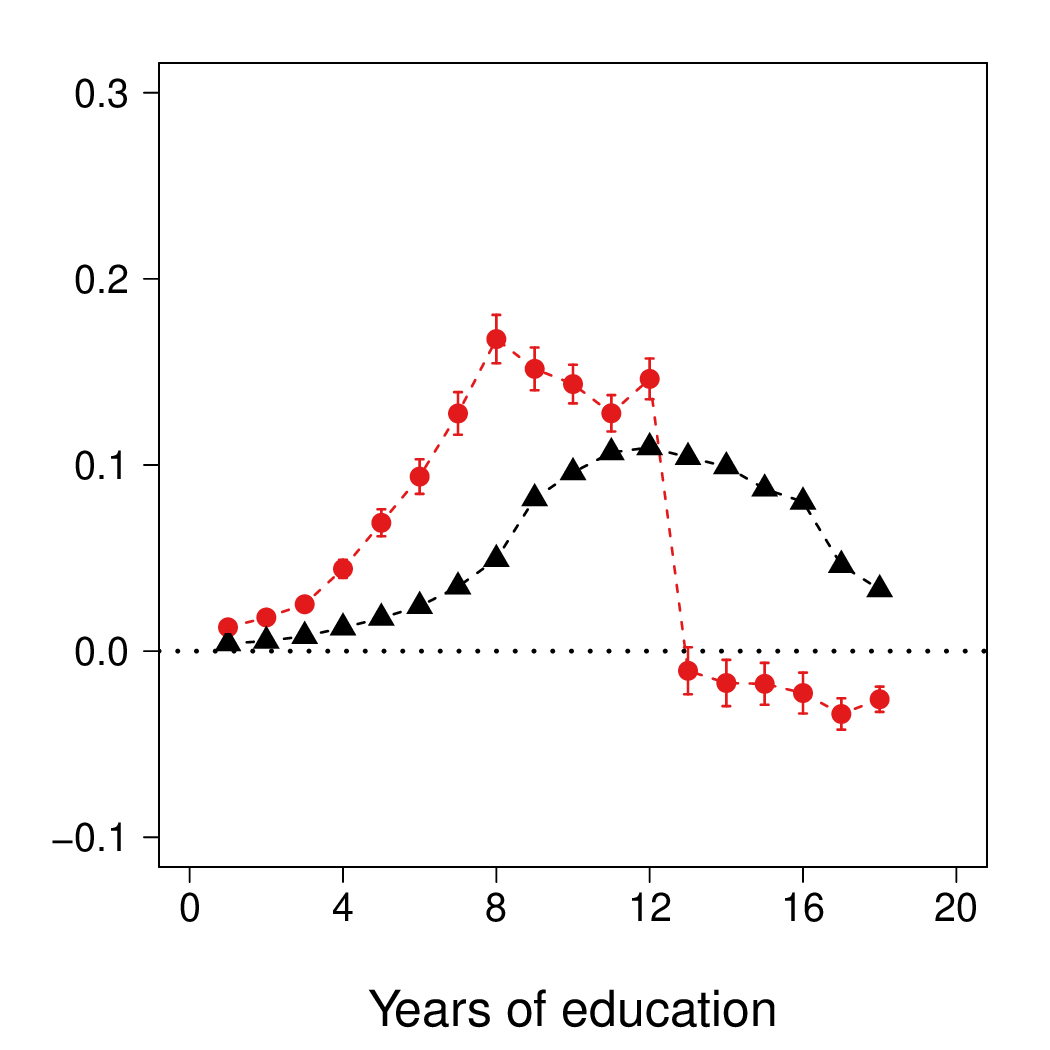}%
\end{minipage}

\vspace{2.0em}

\begin{minipage}[t]{0.48\columnwidth}%
(c) Quarter of Birth

\includegraphics[width=1\columnwidth]{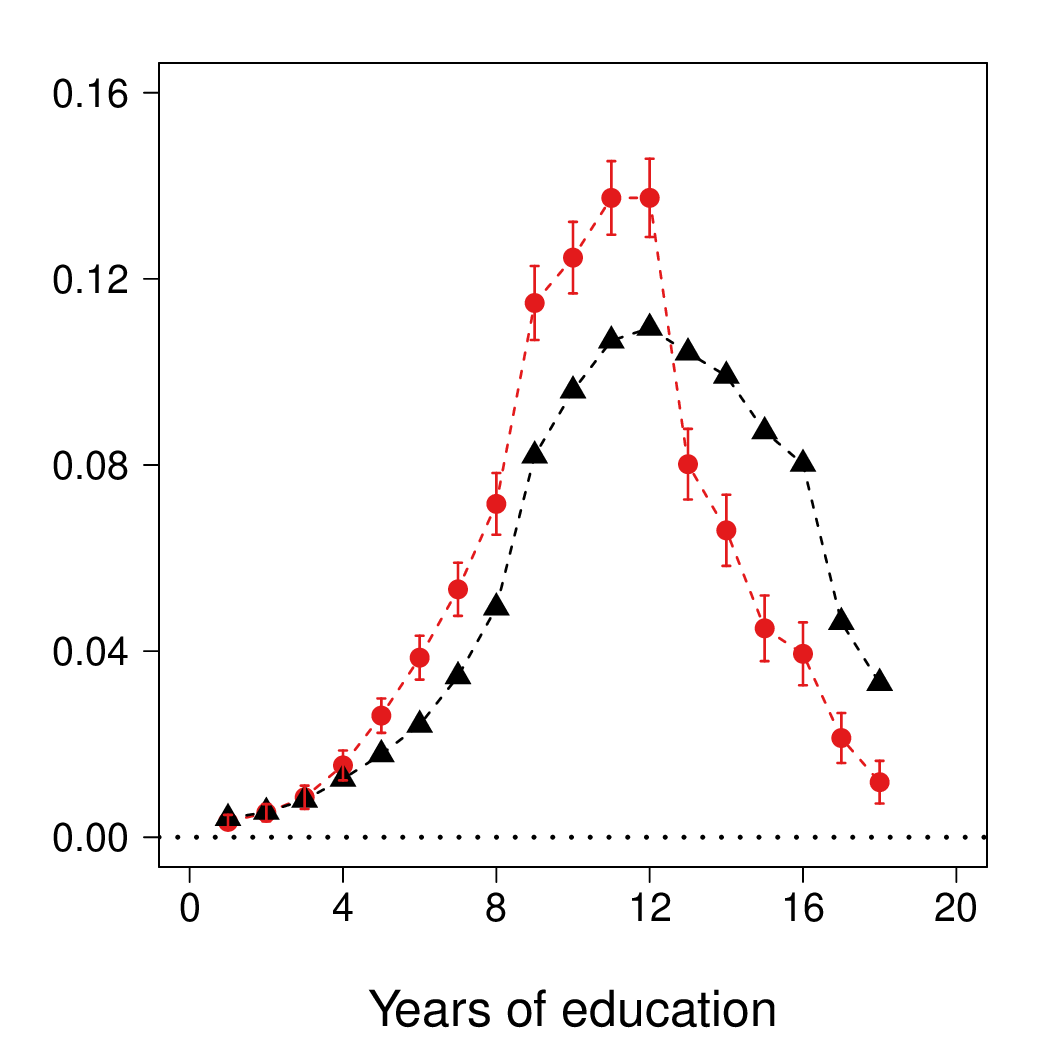}%
\end{minipage}%
\begin{minipage}[t]{0.49\columnwidth}%
\vspace{0.25em}
\begin{center}
\includegraphics[bb=102bp 464bp 402bp 504bp,width=0.95\columnwidth]{legend_x}
\par\end{center}%
\end{minipage}

\vspace{2.0em}
\begin{tablenotes}
\footnotesize

\item Notes: The IV weights are presented with standard error bars.
The standard error bars for the OLS weights are omitted because they
are graphically negligible. Each set of weights sums to one across
the whole sample. Section \ref{subsec:Estimation_Detail} describes
the empirical specification for estimating the weights.

\end{tablenotes}
\end{threeparttable}
\end{figure}

The estimated OLS and IV weights on treatment levels are illustrated
in Figure \ref{Fig:QOB_weights_X}. The patterns of the IV weights
associated with compulsory attendance laws or required schooling years
are close to those of the child labor IV weights illustrated in Panel
(c) of Figure \ref{Fig:college_weights_X} in the main paper. On the
other hand, the quarter-of-birth IV weights are somewhat closer to
the OLS weights, although the difference is still visible. In fact,
55\% of the OLS weights are placed on the 1st--12th years of schooling,
while 74\% of the quarter-of-birth IV weights are placed on the same
schooling margins. 

Higher quarter-of-birth IV weights on primary and secondary schooling
margins are consistent with the expectation that the season of birth
should influence schooling because of the age-based CSL. However,
it should be noted that the non-negligible share of IV weights is
placed on college education years. While it is possible that students
who are forced to complete an additional year of schooling may proceed
with further schooling, the weight patterns associated with the other
two compulsory schooling instruments in Section \ref{sec:Applications}
do not indicate such dynamic effects. Therefore, this evidence is
rather consistent with the argument of \citet{bound2000compulsory}
that the influence of the season of birth on schooling is too strong
to be fully explained by the CSL. It is likely that the season of
birth influences schooling through other factors, such as the persistence
of a relative age effect during childhood.\footnote{In fact, \citet{kawaguchi2011actual} finds that early-born children
in a given cohort tend to complete more years of schooling in Japan,
where the CSL are cohort-based instead of age-based.}

\subsection{Empirical Specification of the Decomposition Estimators\label{subsec:Estimation_Detail}}

I use the two-step method proposed in Section \ref{subsec:IV-weighted-OLS-Estimators}
to obtain the IV-weighted OLS estimates. I use the parametric specifications
\global\long\def\ind{\mathbbm{1}}%
\[
m(x,w)=w'\gamma_{1}+\left(w'\gamma_{2}\right)x,
\]
for estimating $\beta_{C}$ and
\[
m(x,w)=w'\gamma_{1}+\left(w'\gamma_{2}\right)x+\left(w'\gamma_{3}\right)\max\left\{ x-\overline{x},0\right\} +\sum_{x}\delta_{x}\ind_{X\ge x},
\]
for estimating $\beta_{CT}$. $\overline{x}$ is a constant that represents
the last year of secondary school, where $\overline{x}=12$ in Sections
\ref{subsec:College IV} and \ref{subsec:CSL} and $\overline{x}=11$
in Section \ref{subsec:CSL_UK}. Interacting $\max\left\{ x-\overline{x},0\right\} $
with $w$ allows return to schooling at post-secondary schooling margins
to have different patterns of heterogeneity from primary and secondary
schooling margins. Including $\ind_{X\ge x}$ in the equation allows
arbitrary nonlinearity of $m(x,w)$ in $x$. Since the covariate vector
$w$ in Section \ref{subsec:CSL} consists of a large number of variables
(i.e., year of birth and state of birth dummies), I restrict $w'\gamma_{1}$,
$w'\gamma_{2}$, and $w'\gamma_{3}$ to be additive in 5-year cohort
group dummies and Census division dummies to avoid overstating heterogeneity.

I estimate the OLS and IV covariate weights and treatment-level weights
from the data counterparts of how they are defined in Section \ref{subsec:Interpretation}
and \ref{subsec:DID_RD}. I estimate the OLS and IV weight on treatment
level $x$, $\overline{\omega}_{X}(x)$ and $\overline{\omega}_{Z}(x)$
(or $\overline{\omega}_{Z}^{*}(x)$ equivalently), from
\[
\frac{\sum_{i=1}^{N}\widetilde{\ind}_{X_{i}\ge x}\widetilde{X}_{i}}{\sum_{i=1}^{N}\widetilde{X}_{i}^{2}},
\]
\[
\frac{\sum_{i=1}^{N}\widetilde{\ind}_{X_{i}\ge x}\widetilde{Z}_{i}}{\sum_{i=1}^{N}\widetilde{X}_{i}\widetilde{Z}_{i}}.
\]
I estimate the total OLS weight on covariate group $G$, $\int_{w\in G}\overline{\omega}_{X}(w)dF_{W}(w)$,
from
\[
\frac{\sum_{i=1}^{N}\ind_{W_{i}\in G}\widetilde{X}_{i}^{2}}{\sum_{i=1}^{N}\widetilde{X}_{i}^{2}}.
\]
In Section \ref{subsec:College IV}, I estimate the total IV weight
on covariate group $G$, $\int_{w\in G}\overline{\omega}_{Z}(w)dF_{W}(w)$,
from
\[
\frac{\sum_{i=1}^{N}\ind_{W_{i}\in G}\widetilde{X}_{i}\widetilde{Z}_{i}}{\sum_{i=1}^{N}\widetilde{X}_{i}\widetilde{Z}_{i}},
\]
In Sections \ref{subsec:CSL_UK} and \ref{subsec:CSL}, which use
RD-based and DID-based instruments, I estimate the IV weight on covariate
group $G$, $\int_{w\in G}\overline{\omega}_{Z}^{*}(w)dF_{W}(w)$,
from
\[
\frac{\sum_{i=1}^{N}\ind_{W_{i}\in G}\widehat{L}_{W_{i}}\left(X_{i}\widetilde{Z}_{i}\right)}{\sum_{i=1}^{N}\widetilde{X}_{i}\widetilde{Z}_{i}},
\]
where $\widehat{L}_{W_{i}}\left(X_{i}\widetilde{Z}_{i}\right)=W_{i}'\left(\sum_{j=1}^{N}W_{j}W_{j}'\right)^{-1}\left(\sum_{j=1}^{N}W_{j}X_{j}\widetilde{Z}_{j}\right)$.

\medskip
\begingroup
\setlength\bibitemsep{0.25em}
\phantomsection

\printbibliography[heading=bibintoc,title={Appendix References}]

\endgroup
\setcounter{secnumdepth}{0}
\phantomsection

\section[Appendix Tables and Figures]{}

\vspace{-5em}
\begin{sidewaystable}[H]
\caption{Monte Carlo Simulation (Cases 1--3)}
\label{Table:MC}

\centering
\begin{threeparttable}
\setlength{\tabcolsep}{0.5em}
\renewcommand{\arraystretch}{1.2}
\begin{centering}
\begin{tabular}{cccccccccccc}
 &  &  &  &  &  &  &  &  &  &  & \tabularnewline
\hline 
\multicolumn{1}{c}{} &  & \multicolumn{6}{c}{Estimators} &  & \multicolumn{3}{c}{Fraction of rejection of DWH test}\tabularnewline
\cline{3-8} \cline{4-8} \cline{5-8} \cline{6-8} \cline{7-8} \cline{8-8} \cline{10-12} \cline{11-12} \cline{12-12} 
{\small{}$\rho_{\varepsilon\eta}$} &  & {\small{}$\widehat{\beta}_{OLS}$} & {\small{}$\widehat{\beta}_{IV}$} & {\small{}$\widehat{\beta}_{IV}-\widehat{\beta}_{OLS}$} & {\small{}$\widehat{\Delta}_{CW}$} & {\small{}$\widehat{\Delta}_{TW}$} & {\small{}$\widehat{\Delta}_{ME}$} &  & {\small{}Standard DWH} & {\small{}DWH--LM} & {\small{}DWH--I}\tabularnewline
\hline 
\multicolumn{12}{c}{Case 1: $\kappa=0$, $\delta_{b}=0$, $\sigma_{b}=0$, and $\rho_{bd}=0$}\tabularnewline
\hline 
0 &  & 0.040 & 0.040 & 0.000 & 0.000 & 0.000 & 0.000 &  & 0.048 & 0.046 & 0.050\tabularnewline
 &  & (0.002) & (0.010) & (0.010) & (0.001) & (0.004) & (0.008) &  &  &  & \tabularnewline
0.1 &  & 0.024 & 0.040 & 0.015 & 0.000 & 0.004 & 0.011 &  & 0.379 & 0.264 & 0.249\tabularnewline
 &  & (0.002) & (0.010) & (0.010) & (0.001) & (0.004) & (0.009) &  &  &  & \tabularnewline
0.2 &  & 0.009 & 0.040 & 0.031 & 0.000 & 0.009 & 0.022 &  & 0.903 & 0.760 & 0.726\tabularnewline
 &  & (0.002) & (0.010) & (0.010) & (0.001) & (0.004) & (0.008) &  &  &  & \tabularnewline
\hline 
\multicolumn{12}{c}{Case 2: $\kappa=0.1$, $\delta_{b}=0$, $\sigma_{b}=0$, and $\rho_{bd}=0$}\tabularnewline
\hline 
0 &  & 0.051 & 0.040 & --0.011 & 0.000 & --0.011 & 0.000 &  & 0.214 & 0.061 & 0.058\tabularnewline
 &  & (0.002) & (0.010) & (0.010) & (0.001) & (0.004) & (0.001) &  &  &  & \tabularnewline
0.1 &  & 0.035 & 0.039 & 0.004 & 0.000 & --0.007 & 0.010 &  & 0.070 & 0.241 & 0.238\tabularnewline
 &  & (0.002) & (0.010) & (0.010) & (0.001) & (0.004) & (0.009) &  &  &  & \tabularnewline
0.2 &  & 0.020 & 0.040 & 0.020 & 0.000 & --0.002 & 0.022 &  & 0.587 & 0.766 & 0.742\tabularnewline
 &  & (0.002) & (0.010) & (0.009) & (0.001) & (0.004) & (0.008) &  &  &  & \tabularnewline
\hline 
\multicolumn{12}{c}{Case 3: $\kappa=0.1$, $\delta_{b}=-0.04$, $\sigma_{b}=0$, and $\rho_{bd}=0$}\tabularnewline
\hline 
0 &  & 0.052 & 0.035 & --0.017 & --0.006 & --0.011 & 0.000 &  & 0.425 & 0.110 & 0.050\tabularnewline
 &  & (0.003) & (0.010) & (0.009) & (0.001) & (0.004) & (0.009) &  &  &  & \tabularnewline
0.1 &  & 0.037 & 0.036 & --0.001 & --0.006 & --0.006 & 0.011 &  & 0.054 & 0.098 & 0.268\tabularnewline
 &  & (0.003) & (0.010) & (0.010) & (0.001) & (0.004) & (0.009) &  &  &  & \tabularnewline
0.2 &  & 0.021 & 0.035 & 0.014 & --0.006 & --0.002 & 0.022 &  & 0.330 & 0.488 & 0.744\tabularnewline
 &  & (0.003) & (0.010) & (0.010) & (0.001) & (0.004) & (0.009) &  &  &  & \tabularnewline
\hline 
\end{tabular}
\par\end{centering}
\begin{tablenotes}
\footnotesize

\item Notes: Columns for estimators report the average of the estimates,
with the standard deviation in parentheses. Columns for fraction of
rejection report the fraction of simulations finding the statistically
significant (at a 5\% level) endogeneity bias. The standard DWH test
examines the significance of $\widehat{\beta}_{IV}-\widehat{\beta}_{OLS}$
using robust standard errors. DWH--LM refers to a generalized DWH
test proposed by \citet{lochner2015estimating}. DWH--I refers to
a generalized DWH test proposed in the main paper, which tests the
significance of $\widehat{\Delta}_{ME}$.

\end{tablenotes}
\end{threeparttable}
\end{sidewaystable}

\begin{sidewaystable}[H]
\caption{Monte Carlo Simulation (Cases 4--5)}
\label{Table:MC2}

\centering
\begin{threeparttable}
\setlength{\tabcolsep}{0.5em}
\renewcommand{\arraystretch}{1.2}
\begin{centering}
\begin{tabular}{cccccccccccc}
 &  &  &  &  &  &  &  &  &  &  & \tabularnewline
\hline 
\multicolumn{1}{c}{} &  & \multicolumn{6}{c}{Estimators} &  & \multicolumn{3}{c}{Fraction of rejection of DWH test}\tabularnewline
\cline{3-8} \cline{4-8} \cline{5-8} \cline{6-8} \cline{7-8} \cline{8-8} \cline{10-12} \cline{11-12} \cline{12-12} 
{\small{}$\rho_{\varepsilon\eta}$} &  & {\small{}$\widehat{\beta}_{OLS}$} & {\small{}$\widehat{\beta}_{IV}$} & {\small{}$\widehat{\beta}_{IV}-\widehat{\beta}_{OLS}$} & {\small{}$\widehat{\Delta}_{CW}$} & {\small{}$\widehat{\Delta}_{TW}$} & {\small{}$\widehat{\Delta}_{ME}$} &  & {\small{}Standard DWH} & {\small{}DWH--LM} & {\small{}DWH--I}\tabularnewline
\hline 
\multicolumn{12}{c}{Case 4: $\kappa=0.1$, $\delta_{b}=-0.04$, $\sigma_{b}=0.02$, and
$\rho_{bd}=0$}\tabularnewline
\hline 
0 &  & 0.052 & 0.034 & --0.018 & --0.006 & --0.011 & 0.000 &  & 0.373 & 0.107 & 0.058\tabularnewline
 &  & (0.003) & (0.011) & (0.011) & (0.001) & (0.004) & (0.010) &  &  &  & \tabularnewline
0.1 &  & 0.037 & 0.035 & --0.002 & --0.006 & --0.007 & 0.011 &  & 0.049 & 0.067 & 0.180\tabularnewline
 &  & (0.003) & (0.011) & (0.011) & (0.001) & (0.005) & (0.010) &  &  &  & \tabularnewline
0.2 &  & 0.021 & 0.035 & 0.014 & --0.006 & --0.002 & 0.022 &  & 0.275 & 0.404 & 0.611\tabularnewline
 &  & (0.003) & (0.011) & (0.011) & (0.001) & (0.004) & (0.010) &  &  &  & \tabularnewline
\hline 
\multicolumn{12}{c}{Case 5: $\kappa=0.1$, $\delta_{b}=-0.04$, $\sigma_{b}=0.02$, and
$\rho_{bd}=-0.25$}\tabularnewline
\hline 
0 &  & 0.054 & 0.034 & --0.020 & --0.006 & --0.008 & --0.006 &  & 0.452 & 0.240 & 0.104\tabularnewline
 &  & (0.003) & (0.011) & (0.011) & (0.001) & (0.004) & (0.010) &  &  &  & \tabularnewline
0.1 &  & 0.038 & 0.034 & --0.004 & --0.006 & --0.003 & 0.005 &  & 0.077 & 0.052 & 0.073\tabularnewline
 &  & (0.003) & (0.011) & (0.011) & (0.001) & (0.005) & (0.010) &  &  &  & \tabularnewline
0.2 &  & 0.023 & 0.034 & 0.012 & --0.006 & 0.002 & 0.016 &  & 0.197 & 0.191 & 0.389\tabularnewline
 &  & (0.003) & (0.011) & (0.011) & (0.001) & (0.005) & (0.010) &  &  &  & \tabularnewline
\hline 
\end{tabular}
\par\end{centering}
\begin{tablenotes}
\footnotesize

\item Notes: Columns for estimators report the average of the estimates,
with the standard deviation in parentheses. Columns for fraction of
rejection report the fraction of simulations finding the statistically
significant (at a 5\% level) endogeneity bias. The standard DWH test
examines the significance of $\widehat{\beta}_{IV}-\widehat{\beta}_{OLS}$
using robust standard errors. DWH--LM refers to a generalized DWH
test proposed by \citet{lochner2015estimating}. DWH--I refers to
a generalized DWH test proposed in the main paper, which tests the
significance of $\widehat{\Delta}_{ME}$.

\end{tablenotes}
\end{threeparttable}
\end{sidewaystable}

\begin{table}[H]
\caption{NLSY79 Sample Selection}
\label{Table:NLSY_Selection}

\centering
\begin{threeparttable}
\vspace{0.5em}
\begin{centering}
\begin{tabular}{llccccc}
\hline 
 &  & \multicolumn{2}{c}{Males} &  & \multicolumn{2}{c}{Females}\tabularnewline
\cline{3-4} \cline{4-4} \cline{6-7} \cline{7-7} 
 &  & Persons & Person--Years &  & Persons & Person--Years\tabularnewline
\cline{1-4} \cline{2-4} \cline{3-4} \cline{4-4} \cline{6-7} \cline{7-7} 
\multicolumn{2}{l}{Civilian Sample} & 5,579 &  &  & 5,827 & \tabularnewline
\hline 
\multicolumn{7}{l}{Restrictions to Persons}\tabularnewline
 & AFQT available & --343 &  &  & --307 & \tabularnewline
 & Age-14 location available & --273 &  &  & --265 & \tabularnewline
 & 8th grade until age 22 & --60 &  &  & --72 & \tabularnewline
\multicolumn{2}{l}{Subtotal} & 4,903 & 107,137 &  & 5,183 & 115,684\tabularnewline
\hline 
\multicolumn{7}{l}{Restrictions to Person-Year Observations}\tabularnewline
 & Between age 25 and 54 & --93 & --35,108 &  & --62 & --36,655\tabularnewline
 & Not currently enrolled & --15 & --2,948 &  & --4 & --4,351\tabularnewline
 & Wage observed & --76 & --8,569 &  & --131 & --15,589\tabularnewline
\multicolumn{2}{l}{Subtotal} & 4,719 & 60,512 &  & 4,986 & 59,089\tabularnewline
\hline 
\end{tabular}
\par\end{centering}
\begin{tablenotes}
\footnotesize

\item Notes: Negative values refer to the number of persons or observations
dropped from the sample. Positive values indicate the number of persons
or observations remaining in the sample.

\end{tablenotes}
\end{threeparttable}
\end{table}

\begin{table}[H]
\caption{NLSY79 Descriptive Statistics}
\label{Table:NLSY_Descriptive}

\centering
\begin{threeparttable}
\vspace{0.5em}
\begin{centering}
\begin{tabular}{llccc}
\hline 
 &  &  & Mean & (S.D.)\tabularnewline
\hline 
\multicolumn{5}{l}{Variables by Person--Years}\tabularnewline
 & Hourly wage (\$) &  & 9.56 & (10.31)\tabularnewline
 & Years of schooling &  & 13.41 & (2.30)\tabularnewline
 & Age &  & 35.56 & (8.57)\tabularnewline
\hline 
\multicolumn{5}{l}{Personal Background Variables}\tabularnewline
 & AFQT percentile &  & 0.506 & (0.286)\tabularnewline
 & Number of siblings &  & 3.07 & (1.71)\tabularnewline
 & Female &  & 0.504 & \tabularnewline
 & Black &  & 0.141 & \tabularnewline
 & Hispanic &  & 0.055 & \tabularnewline
\hline 
\multicolumn{5}{l}{Geographic Background Variables (age-14 county)}\tabularnewline
 & Local earnings in 1974--81 (\$1,000) &  & 19.44 & (3.55)\tabularnewline
 & Unemployment rate in 1974--81 (\%) &  & 7.05 & (1.27)\tabularnewline
 & Local earnings at age 17 (\$1,000) &  & 19.48 & (3.74)\tabularnewline
 & Unemployment rate at age 17 (\%) &  & 7.04 & (1.85)\tabularnewline
 & Tuition rate of public 4-year (\$1,000) &  & 1.09 & (0.40)\tabularnewline
 & Urban status		 &  & 0.689 & \tabularnewline
 & Presence of public 4-year &  & 0.666 & \tabularnewline
\hline 
\multicolumn{2}{l}{Parental Education} &  & Father & Mother\tabularnewline
 & Missing &  & 0.096 & 0.048\tabularnewline
 & Less than high school &  & 0.148 & 0.102\tabularnewline
 & Some high school &  & 0.149 & 0.190\tabularnewline
 & High school graduate (no college) &  & 0.332 & 0.452\tabularnewline
 & Some college &  & 0.103 & 0.111\tabularnewline
 & College graduate &  & 0.173 & 0.097\tabularnewline
\hline 
\end{tabular}
\par\end{centering}
\begin{tablenotes}
\footnotesize

\item Notes: Persons are weighted based on the sampling weights.
Person--year observations are equally weighted within each person.
Standard deviations are in parentheses for nonbinary variables. The
number of siblings is top-coded at 6. Hourly wage is bottom-coded
at \$1 and top-coded at \$100. Hourly wage, local earnings, and tuition
rate are deflated using CPI--U with the base year of 1982--84.

\end{tablenotes}
\end{threeparttable}
\end{table}

\begin{table}[H]
\caption{First-Stage Coefficients of the College Cost Instruments}
\label{Table: NLSY_FS}

\centering
\begin{threeparttable}
\vspace{0.5em}
\setlength{\tabcolsep}{0.5em}
\begin{centering}
\begin{tabular}{llccc}
\hline 
 &  &  & Coef. & S.E.\tabularnewline
\hline 
\multicolumn{5}{l}{Geographic Costs of Attendance}\tabularnewline
 & Public 4-year present &  & 0.080 & (0.059)\tabularnewline
 & Tuition rate of public 4-year (\$1,000) &  & --0.148 & (0.087)\tabularnewline
\multicolumn{5}{l}{Opportunity Costs of Attendance}\tabularnewline
 & Log local earnings at age 17 &  & --0.886 & (0.475)\tabularnewline
 & Unemployment rate at age 17 (\%) &  & 0.021 & (0.022)\tabularnewline
\hline 
\multicolumn{2}{l}{Effective F-stat.} &  & \multicolumn{2}{c}{2.64}\tabularnewline
\hline 
\end{tabular}
\par\end{centering}
\begin{tablenotes}
\footnotesize

\item Notes: Standard errors (in parentheses) are robust to heteroskedasticity
and correlation across observations on persons living in the same
county at age 14. The effective F-statistics is based on \citet{olea2013robust}.

\end{tablenotes}
\end{threeparttable}
\end{table}

\begin{table}[H]
\caption{Alternative Specifications for Decomposing the IV--OLS Coefficient
Gap (NLSY79)}
\label{Table: College_decom_alt}

\centering
\begin{threeparttable}
\setlength{\tabcolsep}{0.4em}
\begin{centering}
\begin{tabular}{lllccccccccc}
 &  &  &  &  &  &  &  &  &  &  & \tabularnewline
\hline 
 &  &  & \multicolumn{3}{c}{Coefficients} &  & \multicolumn{3}{c}{Decomposition} &  & No. of\tabularnewline
\cline{4-6} \cline{5-6} \cline{6-6} \cline{8-10} \cline{9-10} \cline{10-10} 
 &  &  & OLS & IV & IV--OLS &  & $\Delta_{CW}$ & $\Delta_{TW}$ & $\Delta_{ME}$ &  & Obs.\tabularnewline
\cline{1-2} \cline{2-2} \cline{4-6} \cline{5-6} \cline{6-6} \cline{8-10} \cline{9-10} \cline{10-10} \cline{12-12} 
1. & Base specification &  & 0.065 & 0.062 & --0.004 &  & 0.011 & 0.018 & --0.032 &  & \multirow{2}{*}{119,601}\tabularnewline
 &  &  & (0.003) & (0.087) & (0.087) &  & (0.011) & (0.010) & (0.086) &  & \tabularnewline
2. & Cross-sectional &  & 0.065 & 0.064 & --0.001 &  & 0.009 & 0.011 & --0.021 &  & \multirow{2}{*}{73,167}\tabularnewline
 & sample\hspace{0.15em}\tnote{a} &  & (0.004) & (0.077) & (0.077) &  & (0.012) & (0.009) & (0.076) &  & \tabularnewline
3. & Add controls for &  & 0.064 & 0.043 & --0.021 &  & --0.005 & 0.037 & --0.053 &  & \multirow{2}{*}{55,908}\tabularnewline
 & family income\hspace{0.15em}\tnote{b} &  & (0.004) & (0.111) & (0.111) &  & (0.016) & (0.023) & (0.110) &  & \tabularnewline
4. & Females sample &  & 0.074 & 0.074 & 0.000 &  & --0.004 & --0.005 & 0.008 &  & \multirow{2}{*}{59,089}\tabularnewline
 &  &  & (0.005) & (0.110) & (0.110) &  & (0.012) & (0.011) & (0.106) &  & \tabularnewline
5. & Males sample &  & 0.057 & 0.034 & --0.023 &  & 0.038 & 0.037 & --0.098 &  & \multirow{2}{*}{60,512}\tabularnewline
 &  &  & (0.004) & (0.106) & (0.106) &  & (0.029) & (0.029) & (0.121) &  & \tabularnewline
6. & Black or Hispanic &  & 0.067 & 0.077 & 0.010 &  & --0.005 & --0.007 & 0.022 &  & \multirow{2}{*}{53,808}\tabularnewline
 &  &  & (0.004) & (0.108) & (0.108) &  & (0.014) & (0.014) & (0.106) &  & \tabularnewline
7. & Non-black &  & 0.064 & 0.058 & --0.007 &  & 0.008 & 0.015 & --0.029 &  & \multirow{2}{*}{65,793}\tabularnewline
 & and non-Hispanic &  & (0.004) & (0.073) & (0.073) &  & (0.011) & (0.009) & (0.073) &  & \tabularnewline
8. & Direct cost &  & 0.065 & 0.008 & --0.057 &  & 0.012 & 0.013 & --0.083 &  & \multirow{2}{*}{119,601}\tabularnewline
 & instruments only\hspace{0.15em}\tnote{c} &  & (0.003) & (0.133) & (0.133) &  & (0.013) & (0.013) & (0.130) &  & \tabularnewline
9. & Opportunity cost &  & 0.065 & 0.131 & 0.066 &  & 0.009 & 0.016 & 0.033 &  & \multirow{2}{*}{119,601}\tabularnewline
 & instruments only\hspace{0.15em}\tnote{d} &  & (0.003) & (0.113) & (0.113) &  & (0.020) & (0.033) & (0.109) &  & \tabularnewline
\hline 
\end{tabular}
\par\end{centering}
\begin{tablenotes}
\footnotesize\item Notes: Standard errors (in parentheses) are robust to heteroskedasticity
and correlation across observations on persons living in the same
county at age 14. Section \ref{subsec:Estimation_Detail} describes
the empirical specification for estimating the decomposition.

\item[a] Excludes oversamples of minority or economically disadvantaged
individuals.

\item[b] Adds family income percentile to covariates, allowing different
slopes across tertiles. Excludes persons with no observation of family
income until age 17. 1957--60 birth cohorts are entirely excluded
as a result.

\item[c] Uses college proximity and tuition rate as instruments.

\item[d] Uses local earnings and unemployment rate at age 17 as instruments.

\end{tablenotes}
\end{threeparttable}
\end{table}

\begin{sidewaystable}[H]
\centering
\begin{threeparttable}
\setlength{\tabcolsep}{0.25em}
\renewcommand{\arraystretch}{1.2}

\caption{The IV and OLS Weights on the Covariate Groups and the Treatment Levels
(NLSY79)}
\label{Table:coll_wa}
\begin{centering}
\begin{tabular}{ccccccccccccccccc}
 &  &  &  &  &  &  &  &  &  &  &  &  &  &  &  & \tabularnewline
\hline 
\multirow{2}{*}{Variable} & \multirow{2}{*}{Group} &  & \multicolumn{4}{c}{(A) OLS Weight} &  & \multicolumn{4}{c}{(B) IV Weight} &  & \multicolumn{4}{c}{(C) Subsample OLS Coefficient}\tabularnewline
\cline{4-7} \cline{5-7} \cline{6-7} \cline{7-7} \cline{9-12} \cline{10-12} \cline{11-12} \cline{12-12} \cline{14-17} \cline{15-17} \cline{16-17} \cline{17-17} 
 &  &  & {\footnotesize{}Male} & {\footnotesize{}Female} & \scriptsize \begin{tabular}{@{}c@{}}Non-black, \\ non-Hispanic\end{tabular} & \scriptsize \begin{tabular}{@{}c@{}}Black, \\ Hispanic\end{tabular} &  & {\footnotesize{}Male} & {\footnotesize{}Female} & \scriptsize \begin{tabular}{@{}c@{}}Non-black, \\ non-Hispanic\end{tabular} & \scriptsize \begin{tabular}{@{}c@{}}Black, \\ Hispanic\end{tabular} &  & {\footnotesize{}Male} & {\footnotesize{}Female} & \scriptsize \begin{tabular}{@{}c@{}}Non-black, \\ non-Hispanic\end{tabular} & \scriptsize \begin{tabular}{@{}c@{}}Black, \\ Hispanic\end{tabular}\tabularnewline
\cline{1-2} \cline{2-2} \cline{4-7} \cline{5-7} \cline{6-7} \cline{7-7} \cline{9-12} \cline{10-12} \cline{11-12} \cline{12-12} \cline{14-17} \cline{15-17} \cline{16-17} \cline{17-17} 
 & 0--1/3 &  & 0.25 & 0.25 & 0.17 & 0.59 &  & --0.23 & 0.10 & 0.01 & 0.32 &  & 0.044 & 0.066 & 0.048 & 0.063\tabularnewline
 &  &  & (0.02) & (0.01) & (0.01) & (0.02) &  & (0.29) & (0.17) & (0.12) & (0.27) &  & (0.009) & (0.006) & (0.009) & (0.005)\tabularnewline
AFQT & 1/3--2/3 &  & 0.32 & 0.37 & 0.36 & 0.29 &  & 0.20 & 0.18 & 0.24 & 0.59 &  & 0.045 & 0.075 & 0.060 & 0.063\tabularnewline
percentile &  &  & (0.02) & (0.01) & (0.01) & (0.02) &  & (0.19) & (0.19) & (0.11) & (0.25) &  & (0.007) & (0.007) & (0.006) & (0.009)\tabularnewline
 & 2/3--1 &  & 0.43 & 0.38 & 0.47 & 0.12 &  & 1.03 & 0.72 & 0.75 & 0.09 &  & 0.071 & 0.076 & 0.072 & 0.093\tabularnewline
 &  &  & (0.02) & (0.01) & (0.01) & (0.01) &  & (0.37) & (0.22) & (0.15) & (0.12) &  & (0.008) & (0.007) & (0.006) & (0.011)\tabularnewline
\cline{1-2} \cline{2-2} \cline{4-7} \cline{5-7} \cline{6-7} \cline{7-7} \cline{9-12} \cline{10-12} \cline{11-12} \cline{12-12} \cline{14-17} \cline{15-17} \cline{16-17} \cline{17-17} 
 & Some HS of less &  & 0.20 & 0.22 & 0.15 & 0.48 &  & --0.03 & --0.15 & --0.02 & 0.76 &  & 0.034 & 0.076 & 0.044 & 0.065\tabularnewline
 &  &  & (0.02) & (0.01) & (0.01) & (0.02) &  & (0.20) & (0.19) & (0.11) & (0.24) &  & (0.008) & (0.007) & (0.008) & (0.005)\tabularnewline
Parental & HS graduate &  & 0.41 & 0.42 & 0.44 & 0.32 &  & 0.24 & 0.95 & 0.62 & 0.17 &  & 0.061 & 0.077 & 0.069 & 0.062\tabularnewline
education\hspace{0.15em}\tnote{a} &  &  & (0.02) & (0.02) & (0.02) & (0.02) &  & (0.26) & (0.24) & (0.14) & (0.20) &  & (0.007) & (0.007) & (0.006) & (0.006)\tabularnewline
 & Some college &  & 0.39 & 0.35 & 0.41 & 0.20 &  & 0.79 & 0.20 & 0.40 & 0.07 &  & 0.063 & 0.072 & 0.067 & 0.080\tabularnewline
 &  &  & (0.02) & (0.02) & (0.02) & (0.02) &  & (0.32) & (0.18) & (0.13) & (0.17) &  & (0.007) & (0.008) & (0.006) & (0.011)\tabularnewline
\cline{1-2} \cline{2-2} \cline{4-7} \cline{5-7} \cline{6-7} \cline{7-7} \cline{9-12} \cline{10-12} \cline{11-12} \cline{12-12} \cline{14-17} \cline{15-17} \cline{16-17} \cline{17-17} 
 & 9--12th years\hspace{0.15em}\tnote{b} &  & 0.19 & 0.14 & 0.14 & 0.26 &  & --0.21 & 0.07 & --0.00 & 0.16 &  & 0.019 & 0.052 & 0.024 & 0.049\tabularnewline
Years of &  &  & (0.01) & (0.01) & (0.01) & (0.01) &  & (0.24) & (0.11) & (0.09) & (0.18) &  & (0.009) & (0.009) & (0.009) & (0.006)\tabularnewline
Schooling & 13--18th years\hspace{0.15em}\tnote{c} &  & 0.81 & 0.86 & 0.86 & 0.74 &  & 1.21 & 0.93 & 1.00 & 0.84 &  & 0.066 & 0.076 & 0.070 & 0.072\tabularnewline
 &  &  & (0.01) & (0.01) & (0.01) & (0.01) &  & (0.24) & (0.11) & (0.09) & (0.18) &  & (0.005) & (0.005) & (0.004) & (0.005)\tabularnewline
\hline 
\end{tabular}
\par\end{centering}
\begin{tablenotes}
\footnotesize
\setlength{\itemsep}{0.15em}

\item Notes: Standard errors are in parentheses and robust to heteroskedasticity
and correlation across observations on persons living in the same
county at age 14. Each set of weights sums to one across the whole
sample. Subsample OLS coefficient of years of schooling is obtained
with the same set of control variables as regressions in Table \ref{Table: College_decom_alt}.
Section \ref{subsec:Estimation_Detail} describes the empirical specification
for estimating the weights.

\item[a] The highest level of education among paternal and maternal
figures.

\item[b] The OLS and IV weights on 9th, 10th, 11th, and 12th years
of schooling are aggregated. The subsample OLS coefficient is taken
from the OLS regression restricted to persons with 12 or less years
of schooling.

\item[c] The OLS and IV weights on 13th, 14th, 15th, 16th, 17th,
and 18th years of schooling are aggregated. The subsample OLS coefficient
is taken from the OLS regression restricted to persons with 12 or
more years of schooling.

\end{tablenotes}
\end{threeparttable}
\end{sidewaystable}

\begin{table}[H]
\caption{Descriptive Statistics (U.K. General Household Surveys 1984--98)}
\label{Table:GHS_Descriptive}

\centering
\begin{threeparttable}
\vspace{0.5em}
\begin{centering}
\begin{tabular}{llcccc}
\hline 
 &  &  & \multirow{2}{*}{Britain} &  & \multicolumn{1}{c}{Northern}\tabularnewline
 &  &  &  &  & Ireland\tabularnewline
\hline 
\multicolumn{2}{l}{Annual earnings (£1,000)} &  & 11.01 &  & 8.89\tabularnewline
 &  &  & (12.10) &  & (6.57)\tabularnewline
\multicolumn{2}{l}{Years of schooling} &  & 11.38 &  & 11.22\tabularnewline
 &  &  & (2.70) &  & (2.61)\tabularnewline
\multicolumn{2}{l}{Age} &  & 48.47 &  & 48.70\tabularnewline
 &  &  & (7.40) &  & (7.36)\tabularnewline
\hline 
\multicolumn{2}{l}{Sample size} &  & 55,088 &  & 8,398\tabularnewline
\hline 
\end{tabular}
\par\end{centering}
\begin{tablenotes}
\footnotesize

\item Notes: Standard deviations are in parentheses.

\end{tablenotes}
\end{threeparttable}
\end{table}
\begin{table}[H]
\caption{Alternative Specifications for Decomposing the IV--OLS Coefficient
Gap (British and Northern Irish GHS)}
\label{Table:NI_decom}

\centering
\begin{threeparttable}
\begin{centering}
\begin{tabular}{lllccccccc}
 &  &  &  &  &  &  &  &  & \tabularnewline
\hline 
 &  &  & \multicolumn{3}{c}{Coefficients} &  & \multicolumn{3}{c}{Decomposition}\tabularnewline
\cline{4-6} \cline{5-6} \cline{6-6} \cline{8-10} \cline{9-10} \cline{10-10} 
 &  &  & OLS & IV & IV--OLS &  & $\Delta_{CW}$ & $\Delta_{TW}$ & $\Delta_{ME}$\tabularnewline
\cline{1-2} \cline{2-2} \cline{4-6} \cline{5-6} \cline{6-6} \cline{8-10} \cline{9-10} \cline{10-10} 
1. & British sample &  & 0.084 & 0.062 & --0.021 &  & --0.016 & --0.029 & 0.023\tabularnewline
 & (base specification) &  & (0.002) & (0.083) & (0.082) &  & (0.009) & (0.018) & (0.078)\tabularnewline
2. & British sample &  & 0.064 & 0.062 & --0.002 &  & --0.005 & --0.023 & 0.026\tabularnewline
 & without top-coding &  & (0.001) & (0.083) & (0.083) &  & (0.008) & (0.019) & (0.078)\tabularnewline
3. & Northern Ireland sample &  & 0.114 & 0.224 & 0.109 &  & --0.003 & --0.023 & 0.135\tabularnewline
 &  &  & (0.003) & (0.144) & (0.144) &  & (0.005) & (0.019) & (0.146)\tabularnewline
4. & Northern Ireland sample &  & 0.105 & 0.209 & 0.104 &  & --0.002 & --0.013 & 0.119\tabularnewline
 & without top-coding &  & (0.004) & (0.142) & (0.142) &  & (0.006) & (0.019) & (0.139)\tabularnewline
\hline 
\end{tabular}
\par\end{centering}
\begin{tablenotes}
\footnotesize

\item Notes: Standard errors (in parentheses) are robust to heteroskedasticity
and correlation across observations on the same birth cohort and survey
year. For the British sample, the instrument is an indicator for turning
age 14 in 1947 or later. For the Northern Ireland sample, the instrument
is an indicator for turning age 14 in 1957 or later. Control variables
are quartic of ages and quartic of birth cohorts. Section \ref{subsec:Estimation_Detail}
describes the empirical specification for estimating the decomposition.

\end{tablenotes}
\end{threeparttable}
\end{table}
\begin{table}[H]
\centering
\begin{threeparttable}
\setlength{\tabcolsep}{1em}

\caption{The Total IV and OLS Weights on the Covariate Groups (Northern Irish
GHS)}
\label{Table:NI_wgt_w}
\begin{centering}
\begin{tabular}{ccccc}
 &  &  &  & \tabularnewline
\hline 
\multirow{2}{*}{Year at 14} & Population & OLS & IV & Subsample\tabularnewline
 & share & weights & weights & OLS\tabularnewline
\hline 
1935--40 & \multirow{2}{*}{0.03} & 0.03 & 0.05 & 0.064\tabularnewline
 &  & (0.01) & (0.04) & (0.024)\tabularnewline
1941--45 & \multirow{2}{*}{0.10} & 0.09 & --0.05 & 0.111\tabularnewline
 &  & (0.02) & (0.08) & (0.011)\tabularnewline
1946--50 & \multirow{2}{*}{0.15} & 0.13 & 0.01 & 0.104\tabularnewline
 &  & (0.02) & (0.07) & (0.007)\tabularnewline
1951--55 & \multirow{2}{*}{0.19} & 0.18 & 0.33 & 0.119\tabularnewline
 &  & (0.03) & (0.12) & (0.008)\tabularnewline
1956--60 & \multirow{2}{*}{0.24} & 0.26 & 0.63 & 0.116\tabularnewline
 &  & (0.04) & (0.13) & (0.007)\tabularnewline
1961--65 & \multirow{2}{*}{0.29} & 0.30 & 0.03 & 0.121\tabularnewline
 &  & (0.04) & (0.08) & (0.006)\tabularnewline
\hline 
\end{tabular}
\par\end{centering}
\begin{tablenotes}
\footnotesize

\item Notes: Standard errors (in parentheses) are robust to heteroskedasticity
and correlation across observations on the same birth cohort and survey
year. Each set of weights sums to one across the whole sample. Subsample
OLS coefficient of years of schooling is obtained with controlling
for quartic terms of ages and birth cohorts. Section \ref{subsec:Estimation_Detail}
describes the empirical specification for estimating the weights.

\end{tablenotes}
\end{threeparttable}
\end{table}
\begin{table}[H]
\caption{Descriptive Statistics (1960--80 Censuses)}
\label{Table:Census_Descriptive}

\centering
\begin{threeparttable}
\vspace{0.5em}
\begin{centering}
\begin{tabular}{llcccccc}
\hline 
\multicolumn{2}{l}{Census year} &  & 1960 &  & \multicolumn{1}{c}{1970} &  & \multicolumn{1}{c}{1980}\tabularnewline
\hline 
\multicolumn{2}{l}{Weekly earnings (\$)} &  & 120.9 &  & 207.9 &  & 437.5\tabularnewline
 &  &  & (67.7) &  & (126.3) &  & (253.1)\tabularnewline
\multicolumn{2}{l}{Years of schooling} &  & 10.55 &  & 11.65 &  & 12.71\tabularnewline
 &  &  & (3.28) &  & (3.27) &  & (3.13)\tabularnewline
\multicolumn{2}{l}{Age} &  & 44.30 &  & 44.48 &  & 44.41\tabularnewline
 &  &  & (2.85) &  & (2.86) &  & (2.91)\tabularnewline
\multicolumn{8}{l}{Years of schooling required by child labor laws at 14}\tabularnewline
 & 6 years or less &  & 0.228 &  & 0.195 &  & 0.057\tabularnewline
 & 7 years &  & 0.368 &  & 0.242 &  & 0.237\tabularnewline
 & 8 years &  & 0.349 &  & 0.496 &  & 0.419\tabularnewline
 & 9 years or more &  & 0.054 &  & 0.066 &  & 0.286\tabularnewline
\hline 
\multicolumn{2}{l}{Sample size} &  & 72,494 &  & 161,435 &  & 378,177\tabularnewline
\hline 
\end{tabular}
\par\end{centering}
\begin{tablenotes}
\footnotesize

\item Notes: Standard deviations are in parentheses, except for indicator
variables.

\end{tablenotes}
\end{threeparttable}
\end{table}

\begin{table}[H]
\caption{First-Stage Coefficients of the CSL instruments}
\label{Table: CSL_FS}

\centering
\begin{threeparttable}
\vspace{0.5em}
\setlength{\tabcolsep}{1em}
\begin{centering}
\begin{tabular}{llccc}
\hline 
 &  &  & Coef. & S.E.\tabularnewline
\hline 
\multicolumn{5}{l}{Schooling required by child labor laws}\tabularnewline
 & 7 years &  & 0.106 & (0.031)\tabularnewline
 & 8 years &  & 0.107 & (0.028)\tabularnewline
 & 9 years or more &  & 0.263 & (0.038)\tabularnewline
\hline 
\multicolumn{2}{l}{Effective F-stat.} &  & \multicolumn{2}{c}{14.88}\tabularnewline
\hline 
\end{tabular}
\par\end{centering}
\begin{tablenotes}
\footnotesize

\item Notes: Standard errors (in parentheses) are robust to heteroskedasticity
and correlation across observations on the same state and year of
birth. The dependent variable is years of schooling. The control variables
are state of birth and year of birth dummies. The effective F-statistics
is based on \citet{olea2013robust}.

\end{tablenotes}
\end{threeparttable}
\end{table}

\begin{table}[H]
\caption{Alternative Specifications for Decomposing the IV--OLS Coefficient
Gap (U.S. Census)}
\label{Table:QOB_decom}

\centering
\begin{threeparttable}
\vspace{0.5em}
\begin{centering}
\begin{tabular}{lllccccccc}
 &  &  &  &  &  &  &  &  & \tabularnewline
\hline 
 &  &  & \multicolumn{3}{c}{Coefficients} &  & \multicolumn{3}{c}{Decomposition}\tabularnewline
\cline{4-6} \cline{5-6} \cline{6-6} \cline{8-10} \cline{9-10} \cline{10-10} 
\multicolumn{2}{l}{Instruments} &  & OLS & IV & IV--OLS &  & $\Delta_{CW}$ & $\Delta_{TW}$ & $\Delta_{ME}$\tabularnewline
\cline{1-2} \cline{2-2} \cline{4-6} \cline{5-6} \cline{6-6} \cline{8-10} \cline{9-10} \cline{10-10} 
1. & Child labor laws &  & 0.067 & 0.084 & 0.017 &  & 0.011 & 0.003 & 0.003\tabularnewline
 & (base specification) &  & (0.0004) & (0.022) & (0.022) &  & (0.004) & (0.003) & (0.021)\tabularnewline
2. & Compulsory  &  & 0.067 & 0.161 & 0.094 &  & 0.007 & 0.004 & 0.084\tabularnewline
 & attendance laws &  & (0.0004) & (0.022) & (0.022) &  & (0.003) & (0.003) & (0.021)\tabularnewline
3. & Required schooling  &  & 0.067 & 0.104 & 0.037 &  & 0.005 & 0.006 & 0.026\tabularnewline
 & \citep{Stephens2014} &  & (0.0004) & (0.014) & (0.014) &  & (0.003) & (0.002) & (0.013)\tabularnewline
4. & Quarter of birth &  & 0.067 & 0.064 & --0.003 &  & 0.000 & --0.002 & --0.002\tabularnewline
 &  &  & (0.0004) & (0.014) & (0.014) &  & (0.0004) & (0.001) & (0.014)\tabularnewline
\hline 
\end{tabular}
\par\end{centering}
\begin{tablenotes}
\footnotesize

\item Notes: Standard errors (in parentheses) are robust to heteroskedasticity
and correlation across observations on the same state and year of
birth. The dependent variable is log weekly earnings. The control
variables are state of birth and year of birth dummies. Section \ref{subsec:Estimation_Detail}
describes the empirical specification for estimating the decomposition.

\end{tablenotes}
\end{threeparttable}
\end{table}

\begin{table}[H]
\centering
\begin{threeparttable}

\caption{The IV and OLS Weights on the Covariate Groups with Alternative Instruments
(U.S. Census)}
\label{Table:QOBL_weights_W}
\begin{centering}
\begin{tabular}{cccccccc}
 &  &  &  &  &  &  & \tabularnewline
\hline 
\multirow{2}{*}{Variable} & \multirow{2}{*}{Group} & Population & OLS & CA IV & RS IV & QOB IV & Subsample\tabularnewline
 &  & share & weights & weights & weights & weights & OLS\tabularnewline
\hline 
 & 1910--19 & \multirow{2}{*}{0.32} & 0.33 & 1.31 & 1.65 & 0.45 & 0.063\tabularnewline
 &  &  & (0.02) & (0.21) & (0.18) & (0.05) & (0.001)\tabularnewline
Year of & 1920--29 & \multirow{2}{*}{0.35} & 0.36 & 1.71 & 0.84 & 0.27 & 0.070\tabularnewline
birth &  &  & (0.02) & (0.24) & (0.17) & (0.04) & (0.001)\tabularnewline
 & 1930--39 & \multirow{2}{*}{0.33} & 0.31 & --2.01 & --1.50 & 0.28 & 0.067\tabularnewline
 &  &  & (0.02) & (0.33) & (0.20) & (0.03) & (0.001)\tabularnewline
\hline 
 & Northeast & \multirow{2}{*}{0.29} & 0.26 & --0.46 & 0.54 & 0.10 & 0.069\tabularnewline
 &  &  & (0.02) & (0.46) & (0.20) & (0.04) & (0.001)\tabularnewline
 & Midwest & \multirow{2}{*}{0.33} & 0.27 & --1.17 & --1.43 & 0.20 & 0.066\tabularnewline
\multirow{2}{*}{\begin{tabular}{@{}c@{}}Region of \\ birth\end{tabular}} &  &  & (0.02) & (0.34) & (0.21) & (0.04) & (0.001)\tabularnewline
 & South & \multirow{2}{*}{0.30} & 0.39 & 3.21 & 2.69 & 0.63 & 0.068\tabularnewline
 &  &  & (0.02) & (0.41) & (0.26) & (0.05) & (0.001)\tabularnewline
 & West & \multirow{2}{*}{0.09} & 0.08 & --0.57 & --0.80 & 0.06 & 0.064\tabularnewline
 &  &  & (0.01) & (0.11) & (0.10) & (0.02) & (0.001)\tabularnewline
\hline 
\end{tabular}
\par\end{centering}
\begin{tablenotes}
\footnotesize

\item Notes: Standard errors (in parentheses) are robust to heteroskedasticity
and correlation across observations on the same state and year of
birth. Each set of weights sums to one across the whole sample. Subsample
OLS coefficient of years of schooling is obtained with the same set
of control variables as regressions in Table \ref{Table:QOB_decom}.
``CA IV'' refers to compulsory attendance instruments, ``RS IV''
refers to required schooling instruments as defined by \citet{Stephens2014},
and ``QOB IV'' refers to quarter-of-birth instruments. Section \ref{subsec:Estimation_Detail}
describes the empirical specification for estimating the weights.

\end{tablenotes}
\end{threeparttable}
\end{table}

\end{refsection}
\end{document}